\documentclass[prb,showpacs,twocolumn,floatfix]{revtex4}
\usepackage{graphicx}
\begin{document}
\def\rhov{{\mbox{\boldmath{$\rho$}}}}
\def\tauv{{\mbox{\boldmath{$\tau$}}}}
\def\Lambdav{{\mbox{\boldmath{$\Lambda$}}}}
\def\sigmav{{\mbox{\boldmath{$\sigma$}}}}
\def\xiv{{\mbox{\boldmath{$\xi$}}}}
\def\chiv{{\mbox{\boldmath{$\chi$}}}}
\def\oh{{\scriptsize 1 \over \scriptsize 2}}
\def\ot{{\scriptsize 1 \over \scriptsize 3}}
\def\of{{\scriptsize 1 \over \scriptsize 4}}
\def\tf{{\scriptsize 3 \over \scriptsize 4}}
\title{The Symmetry of Multiferroics}

\author{A. B. Harris}

\affiliation{ Department of Physics and Astronomy, University of Pennsylvania,
Philadelphia, PA 19104}
%%% ----------------------------------------------------------------------
\date{\today}

\begin{abstract}
This paper represents a detailed instruction manual for constructing the
Landau expansion for magnetoelectric coupling in incommensurate ferroelectric
magnets.  The first step is to describe the magnetic ordering in terms of 
symmetry adapted coordinates which serve as complex valued magnetic
order parameters whose transformation properties are displayed.
In so doing we use the previously proposed technique to exploit
inversion symmetry, since this symmetry had been universally overlooked.
Having order parameters of known symmetry which describe the magnetic
ordering, we are able to construct the trilinear interaction which
couples incommensurate magnetic order to the uniform polarization
in order to treat many of the multiferroic systems so far investigated.
The role of this theory in comparison to microscopic models is discussed.
\end{abstract}
\maketitle

\section{Introduction}

Recently there has been increasing interest in the
interaction between magnetic and electric degrees of
freedom.\cite{FIEBIG} Much interest has centered on 
a family of multiferroics which display a phase transition
in which uniform ferroelectric order appears simultaneously with
incommensurate magnetic ordering.  Early examples of such a system whose
ferroelectric behavior and magnetic structure have been thoroughly
studied are Terbium Manganate, TbMnO$_3$ (TMO).\cite{TMO1,TMO2} and
Nickel Vanadate,  Ni$_3$V$_2$O$_8$
(NVO)\cite{FERRO,JAP,NVOPRB,HANDBOOK}.  A number of other
systems have been shown to have combined magnetic and ferroelectric
transitions,\cite{DMO,CFO,TB25,Y25} but the investigation of their magnetic
structure has been less comprehensive.  Initially this combined
transition was somewhat mysterious, but soon a Landau expansion
was developed\cite{FERRO} to provide a phenomenological explanation of this
phenomenon.  An alternative picture, similar to an earlier
result\cite{CURRENT} based on the concept of a ``spin-current,"
and which we refer to as the ``spiral formulation,''\cite{MOST}
has gained popularity due to its simplicity, but as we will discuss, the
Landau theory is more universally applicable and has a number of advantages.
The purpose of the present paper is to describe the Landau formulation 
in the simplest possible terms and to apply it to a large number
of currently studied multiferroics.  In this way we hope to
demystify this formulation.

It should be noted that this  phenomenon (which we call ``magnetically
induced ferroelectricity") is closely related to the
similar behavior of so-called ``improper ferroelectrics," which
are commonly understood to be the analogous systems in which
uniform magnetic order (ferromagnetism or antiferromagnetism)
drive ferroelectricity.\cite{CHUPIS}  Several decades ago such systems
were studied\cite{AXE} and reviewed\cite{CHUPIS,ROGER} and present many
parallels with the recent developments.

One of the problems one encounters at the outset is how to properly
describe the magnetic structure of systems with complicated unit
cells.  This, of course, is a very old subject, but surprisingly,
as will be documented below,
the full ramifications of symmetry are not widely known.
Accordingly, we feel it necessary to repeat the description of the
symmetry analysis of magnetic structures.  While the first part of
this symmetry analysis is well known to experts, we review it here,
especially because our approach is often far simpler and less technical
than the standard one.  However, either approach lays the groundwork
for incorporating the effects of inversion symmetry, which seem
to have been overlooked until our analysis of 
NVO\cite{FERRO,JAP,HANDBOOK,TMO2,NVOPRB} and TMO.\cite{TMO2}
Inversion symmetry was also addressed by Schweizer with a subsequent
correction.\cite{JS2} Very recently a more formal approach to this problem
has been given by Radaelli and Chapon.\cite{RAD} But, at least
in the simplest cases, the approach initially proposed by us and
used here seems easiest.  We apply this formalism to a number of
currently studied multiferroics, such as MnWO$_4$ (MWO),
TbMn$_2$O$_5$ (TMO25), YMn$_2$O$_5$ (YMO25), and  CuFeO$_2$ (CFO). 
As was the case for 
NVO\cite{FERRO,NVOPRB,JAP,HANDBOOK} and TMO,\cite{TMO2} once one has
in hand the symmetry properties of the magnetic order parameters, one
is then able to construct the trilinear magnetoelectric coupling term
in the free energy which provides a phenomenological explanation of
the combined magnetic and ferroelectric phase transition.

This paper is organized in conformity with the above plan.  In Sec.
II we review a simplified version of the symmetry analysis known
as {\it representation theory}, in which we directly analyze the
symmetry of the inverse susceptibility matrix.  Here we also review the
technique we proposed some time ago\cite{FERRO,NVOPRB,JAP,HANDBOOK,TMO2}
to incorporate the consequences of inversion symmetry.  In Sec. III we
apply this formalism 
to develop magnetic order parameters for a number of multiferroic
systems and in Eq. (\ref{IREQ}) we give a simple example to show
how inversion symmetry influences the symmetry of the allowed
spin distribution.  Then in Sec. IV, we use the symmetry of the
order parameters to construct
a magnetoelectric coupling free energy, whose symmetry
properties are manifested.  In Sec. V we summarize the results of
these calculations and discuss their relation to calculations based
on the spin current model\cite{CURRENT} or the phenomenology of
continuum theory.\cite{MOST}

\section{Review of Representation Theory}

As we shall see, to understand the phenomenology of the magnetoelectric
coupling which gives rise to the combined magnetic and ferroelectric
phase transition, it is essential to characterize and properly
understand the symmetry of the magnetic ordering. In addition, as
we shall see, to fully include symmetry restrictions on possible
magnetic structures that can be accessed via a continuous phase
transition is an extremely powerful aid in the magnetic structure
analysis,  Accordingly in this section we review how symmetry
considerations restrict the possible magnetic structures which can
appear at an ordering transition.  The full symmetry analysis has
previously been presented elsewhere,\cite{FERRO,TMO2,NVOPRB,JAP,HANDBOOK},
but it is useful to repeat it here both to fix the notation and to
give the reader convenient access to this analysis which is so
essential to the present discussion.  To avoid the complexities
of the most general form of this analysis
(called representation theory),\cite{JS2,RAD,JV} we will limit discussion
to systems having some crucial simplifying features.  First, we limit
consideration to systems in which the magnetic ordering is incommensurate.
In the examples we choose $k$ will usually
lie along a symmetry direction of the crystal.  Second,
we only consider systems which have a center of inversion symmetry,
because it is only such systems that have a sharp phase transition
at which long-range ferroelectric order appears. Thirdly, we
restrict attention to crystals having relatively simple symmetry.
(What this means is that except for our discussion of TbMn$_2$O$_5$
we will consider systems where we
do not need the full apparatus of group theory, but can get away with
simply labeling the spin functions which describe magnetic order by
their eigenvalue under various symmetry operations.) By avoiding
the complexities of the most general situations, it is hoped that
this paper will be accessible to more readers.  Finally, as
we will see, it is crucial that the phase transitions we analyze
are either continuous or very nearly so. In many of the examples
we discuss, our simple approach\cite{NVOPRB} is vastly simpler than
that of standard representation theory augmented by representation
theory specialized techniques to explicitly include exploit
inversion symmetry.  

\subsection{Symmetry Analysis of the Magnetic Free Energy}

In this subsection we give a review of the formalism used
previously\cite{FERRO,TMO2} and presented in detail in
Refs. \onlinecite{NVOPRB,HANDBOOK}.
Since we are mainly interested in symmetry properties, we will
describe the magnetic ordering by a version of mean-field theory
in which one writes the magnetic free energy $F_M$ as
\begin{eqnarray}
F_M &=& \frac {1}{2} \sum_{{\bf r}, \alpha ; {\bf r}' \beta}
\chi^{-1}_{\alpha \beta} ({\bf r}, {\bf r}')
S_\alpha ({\bf r}) S_\beta ({\bf r}') \nonumber \\
&& \ + {\cal O} \left( S^4 \right) \ ,
\end{eqnarray}
where $S_\alpha ({\bf r})$ is the thermally averaged $\alpha$-component of
the spin at position ${\bf r}$. In a moment, we will give an explicit
approximation for the inverse susceptibility $\chi$.   We now introduce Fourier transforms
in either of two equivalent formulations.  In the first (which we refer to as
``actual position") one writes
\begin{eqnarray}
S_\alpha ({\bf q} , \tau) &=& N^{-1} \sum_{\bf R} S_\alpha({\bf R}+\tauv)
e^{i {\bf q} \cdot ({\bf R} + \tauv)}
\label{ACTUAL} \end{eqnarray}
whereas in the second (which we refer to as ``unit cell") one writes
\begin{eqnarray}
S_\alpha ({\bf q} , \tau) &=& N^{-1} \sum_{\bf R} S_\alpha({\bf R}+\tauv)
e^{i {\bf q} \cdot {\bf R} } \ ,
\label{UNIT} \end{eqnarray}
where $N$ is the number of unit cells in the system,
$\tauv$ is the location of the $\tau$th site within the unit cell,
and ${\bf R}$ is a lattice vector. 
Note that in Eq. (\ref{ACTUAL}) the phase factor in the Fourier transform
is defined in terms of the actual position of the spin rather than in
terms of the origin of the unit cell, as is done in Eq. (\ref{UNIT}).
In some case (viz. NVO) the results are simpler in the actual position
formulation whereas for others (viz. MWO) the unit cell formulation
is simpler.  We will use whichever formulation is simpler.  In either case
the fact that $S_\alpha$ has to be real indicates that
\begin{eqnarray}
S_\alpha(-{\bf q} , \tau) &=& S_\alpha({\bf q} , \tau)^* \ .
\end{eqnarray}
We thus have
\begin{eqnarray}
F_M &=& \frac{1}{2} \sum_{{\bf q}; \tau, \tau', \alpha , \beta}
\chi^{-1}_{\alpha \beta} ({\bf q}; \tau , \tau' )
S_\alpha ({\bf q},\tau )^* S_\beta ({\bf q}, \tau') \nonumber \\
&& \ + {\cal O} \left( S^4 \right) \ ,
\end{eqnarray}
where (for the ``actual position" formulation)
\begin{eqnarray}
\chi^{-1}_{\alpha \beta} ({\bf q}; \tau , \tau' ) &=&
\sum_{\bf R} \chi^{-1}_{\alpha \beta} (\tau , {\bf R} + \tau')
e^{i {\bf q} \cdot ({\bf R} + \tauv' -\tauv ) } \ .
\end{eqnarray}

To make our discussion
more concrete we cite the simplest approximation
for a system with general anisotropic exchange coupling, so that
the Hamiltonian is
\begin{eqnarray}
{\cal H} &=& \sum_{\alpha , \beta ; {\bf r}, {\bf r}'} J_{\alpha \beta}(
{\bf r} , {\bf r}') s_\alpha ({\bf r}) s_\beta ({\bf r}')
+ \sum_{\alpha {\bf r}} K_\alpha s_\alpha({\bf r})^2 \ ,
\end{eqnarray}
where $s_\alpha ({\bf r})$ is the $\alpha$-component of the spin operator
at ${\bf r}$ and we have included a single ion anisotropy energy
assuming three inequivalent axes, so that the $K_\alpha$ are all different.
One has that
\begin{eqnarray}
\chi^{-1}_{\alpha \beta}( {\bf r}, {\bf r}') &=&
J_{\alpha \beta} ({\bf r}, {\bf r}') + [K_\alpha + ckT] \delta_{\alpha ,\beta}
\delta_{{\bf r},{\bf r}'} \ ,
\end{eqnarray}
where $\delta_{a,b}$ is unity if $a=b$ and is zero otherwise and
$c$ is a spin-dependent constant of order unity, so that
$ckT$ is the entropy associated with a spin $S$. Then
if we have one spin per unit cell, one has
\begin{eqnarray}
\chi^{-1}_{\alpha \beta} ({\bf q}) &=& \delta_{\alpha \beta} \biggl(
2J_1 \left[ \cos (a_\alpha q_x) + \cos (a_\alpha q_y ) \right.
\nonumber \\ && \  \left.
+ \cos (a_\alpha q_z) \right] + akT + K_\alpha \biggr) \ ,
\end{eqnarray}
where $a_\alpha$ is the lattice constant in the
$\alpha$-direction\cite{xyzabc} and we assume that
$K_x<K_y<K_z$.  Graphs of $\chi^{-1}({\bf q})$ are shown in Fig. 1
for both the ferromagnetic ($J_1<0$) and
antiferromagnetic  ($J_1>0$) cases.  For the ferromagnetic case we now
introduce a competing antiferromagnetic next-nearest neighbor (nnn)
interaction $J_2>0$  along the $x$-axis, so that
\begin{eqnarray}
&& \chi^{-1}_{\alpha \alpha} (q_x,q_y=0,q_z=0) = 
\left[ 4J_1 + 2J_1 \cos (a_x q_x) \right. \nonumber \\
&& \ \ \ \left.  + 2J_2 \cos (2 a_x q_x ) + akT + K_\alpha \right] \ ,
\end{eqnarray}
and this is also shown in Fig. 1.  As $T$ is lowered one reaches a critical
temperature where one of the eigenvalues of the inverse susceptibility
matrix becomes zero.  This indicates that the paramagnetic phase is
unstable with respect to order corresponding to the critical eigenvector
associated with the zero eigenvalue.  For the ferromagnet this happens
for zero wavevector and for the antiferromagnet for a zone boundary
wavevector in agreement with our obvious expectation.  For competing
interactions we see that the values of the $J$'s determine a wavevector
at which an eigenvalue of $\chi^{-1}$ is minimal.  This is the
phenomenon called ``wavevector selection,"  and in this case the
selected value of ${\bf q}$ is determined by extremizing
$\chi^{-1}$ to be\cite{NAG}
\begin{eqnarray}
\cos (a_xq) &=& - J_1/(4J_2) \ ,
\end{eqnarray}
providing $J_2 > -J_1/4$. (Otherwise the system is ferromagnetic.)
Note also, that crystal symmetry may select a set of symmetry-related
wavevectors, which comprise what is known as the {\it star} of
${\bf q}$. (For instance, if the system were tetragonal, then crystal
symmetry would imply that one has the same nnn interactions along the
$y$-axis, in which case the system selects a wavevector along the 
$x$-axis and one of equal magnitude along the $y$-axis.

\begin{figure*}
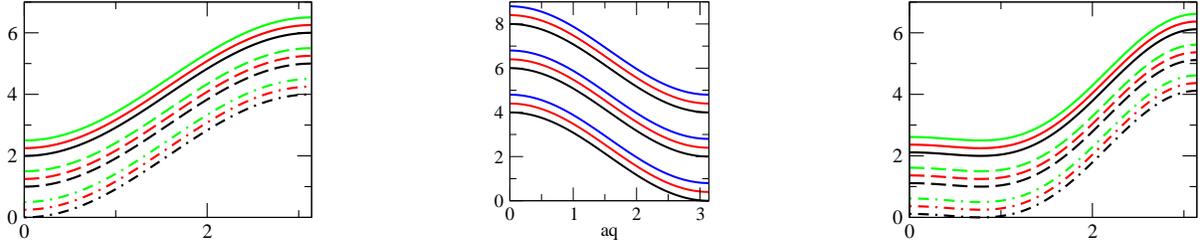

\includegraphics[height=3.2cm]{graph2.eps}
\hspace*{0.85 in}
\includegraphics[height=3.2cm]{new1.eps}
\hspace*{0.85 in}
\includegraphics[height=3.2cm]{graph3.eps}
\vspace{0.2 in}
\caption{\label{CHIQ} Inverse susceptibility $\chi^{-1}(q,0,0)$.
a) Ferromagnetic model ($J_1<0$), b) Antiferromagnetic model ($J_1<0$), and
c) Model with competing interactions (the nn interaction is antiferromagnetic).
In each panel one sees three groups of
curves.  Each group consists of the three curves for $\chi_{\alpha \alpha} (q)$
which depend on the component label $\alpha$ due to the anisotropy.  
The $x$ axis is the easiest axis and the $z$ axis is the hardest. (If the
system is orthorhombic the three axes {\it must} all be inequivalent.
The solid curves are for the highest temperature, the dashed curves are
for an intermediate temperature, and the dash-dot curves are for 
$T=T_c$, the critical temperature for magnetic ordering. Panel c)
illustrates the nontrivial wavevector selection which occurs when one has
competing interactions.}
\end{figure*}

From the above discussion it should be clear that if we assume a
continuous transition so that the transition is associated with
the instability in the terms in the free energy
quadratic in the spin amplitudes, then the nature of the ordered
phase is determined by the critical eigenvector of the inverse
susceptibility, i. e. the eigenvector associated with the
eigenvalue of inverse susceptibility which first goes to
zero as the temperature is reduced. Accordingly, the aim of this
paper analyze how crystal symmetry affects the possible forms
of the critical eigenvector.

\begin{figure}[ht]
\begin{center}
\includegraphics[width=6cm]{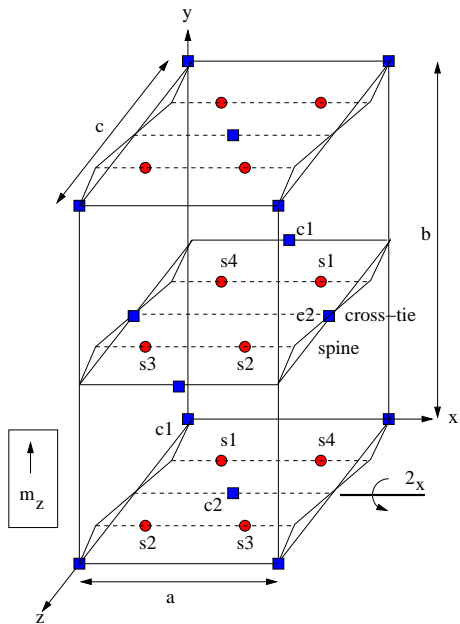}
\caption{(Color online).  Ni sites in the conventional unit cell
of NVO.  The primitive translation vectors ${\bf v}_n$ are
${\bf v}_1 =
(a/2)\hat a + (b/2)\hat b$, ${\bf v}_2 = (a/2)\hat a - (b/2)\hat
b$, and ${\bf v}_3=c \hat c$. The ``cross-tie" sites
(on-line=blue) c1 and c2 lie in a plane with $b=0$. The ``spine"
sites (on-line=red) are labeled s1, s2, s3, and s4 and they may
be visualized as forming chains parallel to the ${\bf a}$-axis.
These chains are in the buckled plane with $b= \pm \delta$, where
$\delta = 0.13 b$ as is indicated.  Cross-tie sites in adjacent planes
(displaced by $(\pm b/2)\hat b$) are indicated by open circles.
Spine sites in adjacent planes are located directly above (or
below) the sites in the plane shown. In the incommensurate phases
the wavevector describing magnetic ordering lies along the {\bf a} axis.
The axis of the two-fold rotation about the $x$-axis is shown.
The glide plane is indicated
by the mirror plane at $z= \tf$ and the arrow above $m_z$ indicates
that a translation of $b/2$ in the $y$-direction is involved.}
\label{NVOSTRFIG}
\end{center}
\end{figure}

When the unit cell contains $n>1$ spins, the  inverse susceptibility
for each wavevector ${\bf q}$ is a $3n \times 3n$ matrix.  The 
ordering transition
occurs when, for some selected wavevector(s), an  eigenvalue first 
becomes zero as the temperature is reduced. In the above simple examples
involving isotropic exchange interactions,
the inverse susceptibility was $3 \times 3$ diagonal matrix, so that
each eigenvector trivially has only one nonzero component.  The critical
eigenvector has spin oriented along the easiest axis, i. e. the one
for which $K_\alpha$ is minimal.  In the present
more general case $n>1$ and arbitrary interactions consistent with crystal
symmetry are allowed. To avoid the technicalities of group theory,
we use as our guiding principle the fact that the free energy, 
being an expansion
in powers of the magnetizations relative to the the paramagnetic state, must
be invariant under all the symmetry operations of the crystal.\cite{IED,LL}
This is the same principle that one uses in discussing the symmetry of 
the electrostatic potential in a crystal. \cite{BS}
We now focus our attention on the critically selected wavevector $q$
which has an eigenvalue which first becomes zero as the temperature
is lowered.  This value of ${\bf q}$ is determined by the interactions
and we will consider it to be an experimentally determined parameter.
Operations which leave the quadratic free energy invariant must leave
invariant the term in the free energy $F_2({\bf q})$ which involves
only the selected wavevector ${\bf q}$, namely
\begin{eqnarray}
F_2({\bf q}) &\equiv & \frac{1}{2} \sum_{ \tau, \tau', \alpha , \beta}
\chi^{-1}_{\alpha \beta} ({\bf q}; \tau , \tau' )
S_\alpha ({\bf q},\tau )^* S_\beta ({\bf q}, \tau') \ .
\label{F2S} \end{eqnarray}
Any symmetry operation takes the original variables before transformation,
$S_\alpha ({\bf q}, {\tau})$, into new ones indicated by primes.
%\cite{WIG}
We write this transformation as
\begin{eqnarray}
S_\alpha'({\bf q},\tauv) = \sum_{\alpha' \tauv} U_{\alpha \tauv; \alpha' \tauv'}
S_{\alpha'}({\bf q},\tauv') \ .
\end{eqnarray}
According to a well known statement of elementary quantum mechanics,
if an operator $T$ commutes with $\chi^{-1}({\bf q})$, then the eigenvectors of
$\chi^{-1}({\bf q})$ are simultaneously eigenvectors of $T$.  (This much
involves a well known analysis.\cite{Bertaut,JS1,Rossat} We will later
consider the effect of inversion, the analysis of which is universally
overlooked).  We will apply this simple condition to a number of
multiferroic systems currently under investigation.  (This approach
can be much more straightforward than the standard one when the operations
which conserve wavevector unavoidably involve translations.)  As a first
example we consider the case of NVO and use the ``actual position" Fourier
transforms.  In Table \ref{NVOSTR} we give
the general positions (this set of positions is the so-called Wyckoff orbit)
for the space group Cmca (\#64 in Ref. \onlinecite{HAHN}) of NVO and this table
defines the operations
of the space group of Cmca.  In Table \ref{NVOMAG} we list the positions
of the two types of  sites occupied by the magnetic (Ni) ions, which are
called ``spine" and ``cross-tie" sites in recognition of 
their distinctive coordination
in the lattice, as can be seen from Fig. \ref{NVOSTRFIG}, where we show 
the conventional unit cell of NVO.  Experiments\cite{NVOPRL,NVOPRB} indicate
that as the temperature is lowered, the system first develops
incommensurate order with ${\bf q}$ along the ${\bf a}$-direction
with $\hat q \approx 0.28$.\cite{RLU}
 
\begin{table}
\vspace{0.2 in}
\begin{tabular} {||c | c ||}
\hline
${\rm E}{\bf r} =(x,y,z)$
& $2_c{\bf r} =(\overline x, \overline y + 1/2, z + 1/2)$ \\
 $2_b {\bf r} = (\overline x, y+1/2, \overline z +1/2)$
& $2_a {\bf r} = (x, \overline y , \overline z)$ \\
${\cal I} {\bf r} =(\overline x,\overline y,\overline z)$
&$m_c{\bf r} =(x, y + 1/2, \overline z + 1/2)$ \\
$m_b{\bf r} = (x, \overline y+1/2, z +1/2)$
&$m_a{\bf r} = (\overline x, y , z)$ \\
\hline \end{tabular}
\caption{\label{NVOSTR}General positions\cite{HAHN,xyz} within
the primitive unit cell for Cmca which describe the symmetry
operations\cite{SYMOP}  of this space group.  $2_\alpha$ is a two-fold
rotation (or screw) axis and $m_\alpha$ is a mirror (or
glide) which takes $r_\alpha$ into $-r_\alpha$.}
\end{table}

\begin{table}
\vspace{0.2 in}
\begin{tabular}{|ccc|}\hline\hline
${\bf r}_{s1}$  &=&  $(0.25, -0.13,0.25)$    \\
${\bf r}_{s2}$  &=&  $(0.25, 0.13, 0.75)$   \\
${\bf r}_{s3}$  &=&  $(0.75, 0.13, 0.75)$   \\
${\bf r}_{s4}$  &=&  $(0.75, -0.13, 0.25)$  \\
${\bf r}_{c1}$  &=&  $(0,    0,    0)$      \\
${\bf r}_{c2}$  &=&  $(0.5,  0,    0.5)$    \\
\hline\hline
\end{tabular}
\caption{\label{NVOMAG} Positions\cite{xyz,SAUER} of ${\rm Ni^{2+}}$
carrying $S$=$1$ within the primitive unit cell illustrated in Fig.
\protect{\ref{NVOSTRFIG}}.
Here ${\bf r}_{sn}$ denotes the position of the $n$th spine site
and ${\bf r}_{cn}$ that of the $n$th cross-tie site.
NVO orders in space group ${\rm Cmca}$, so there are six more atoms
in the conventional orthorhombic unit cell which are obtained by a
translation through $(0.5a,0.5b,0)$.}
\end{table}

The group of operations which conserve wavevector are generated by a)  the
two-fold rotation $2_x$ and b) the glide operation
$m_z$, both of which are defined in Table \ref{NVOSTR}.
We now discuss how the Fourier spin components transform under
various symmetry operations.  Here primed quantities denote the
value of the quantity after transformation. Let ${\cal O}
\equiv {\cal O}_s {\cal O}_r$ be a symmetry operation which
we decompose into operations on the spin ${\cal O}_s$ and on
the position ${\cal O}_r$. The effect of transforming a spin
by such an operator is to replace the spin at the ``final"
position ${\bf R}_f$ by the transformed spin which initially
was at the position ${\cal O}_r^{-1}{\bf R}_f$.  So we write
\begin{eqnarray}
S_\alpha' ({\bf R}_f, \tauv_f ) &=& {\cal O}_s S_\alpha (
{\cal O}_r^{-1}[ {\bf R}_f, \tauv_f]) \nonumber \\
&=& \xi_\alpha ({\cal O}_s) S_\alpha ({\bf R}_i, \tauv_i ) \ ,
\label{RIRFEQ} \end{eqnarray}
where the subscripts ``i" and ``f" denote initial and final
values and $\xi_\alpha({\cal O}_s)$ is the factor introduced
by ${\cal O}_s$ for a pseudovector, namely
\begin{eqnarray}
\xi_x(2_x) &=& 1 \ , \ \ \xi_y(2_x) = \xi_z(2_x)= - 1 \ , \nonumber \\
\xi_x(m_z) &=& \xi_y(m_z)=-1 \ , \ \ \xi_z(m_z) = 1 \ .
\label{XIEQ} \end{eqnarray}
Note that ${\cal O}S_\alpha ({\bf R}, \tauv)$ is not the result of
applying ${\cal O}$ to move and reorient the spin at ${\bf R}+\tauv$,
but instead is the value of the spin at ${\bf R}+\tauv$ after the
spin distribution is acted upon by $\cal O$. 
Thus, for actual position Fourier transforms we have
\begin{eqnarray}
S'_\alpha ({\bf q}, \tauv_f ) &=& N^{-1}  \sum_{\bf R}
S_\alpha'({\bf R}_f, \tauv_f) e^{i {\bf q} \cdot ({\bf R}_f+\tauv_f)}
\nonumber \\ &=&
\xi_\alpha ({\cal O}_s) N^{-1}  \sum_{\bf R}
S_\alpha({\bf R}_i, \tauv_i) e^{i {\bf q} \cdot ({\bf R}_f+\tauv_f)}
\nonumber \\ &=& \xi_\alpha ({\cal O}_s) S_\alpha ({\bf q} , \tauv_i)
e^{i {\bf q} \cdot [{\bf R}_f + \tauv_f - {\bf R}_i - \tauv_i ]} \ .
\label{TRANSAP} \end{eqnarray}
We may write this as
\begin{eqnarray}
{\cal O} S_\alpha ({\bf q},\tauv_f) &=&
\xi_\alpha ({\cal O}_s) S_\alpha ({\bf q} , \tauv_i)
e^{i {\bf q} \cdot [{\bf R}_f + \tauv_f - {\bf R}_i - \tauv_i ]} \ .
\label{TRANSAPO} \end{eqnarray}
This formulation may not be totally intuitive, because one is tempted to
regard the operation ${\cal O}$ acting on a spin at an initial location
and taking it (and perhaps reorienting it) to another location.  Here,
instead, we consider the spin distribution.  The transformed distribution
at a location ${\bf r}$ is related to the distribution at the initial
location ${\cal O}_r^{-1} {\bf r}$.

Similarly, the result for unit cell Fourier transforms is
\begin{eqnarray}
S'_\alpha ({\bf q}, \tauv_f ) &=& \xi_\alpha ({\cal O}_s)
S'_\alpha ({\bf q} , \tauv_i)
e^{i {\bf q} \cdot [{\bf R}_f - {\bf R}_i ]} \ .
\label{TRANSUC} \end{eqnarray}
As before, we may write this as
\begin{eqnarray}
{\cal O} S_\alpha ({\bf q},\tauv_f) &=&
\xi_\alpha ({\cal O}_s) S_\alpha ({\bf q} , \tauv_i)
e^{i {\bf q} \cdot [{\bf R}_f - {\bf R}_i ]} \ .
\label{TRANSUCO} \end{eqnarray}

Under transformation by inversion, $\xi_\alpha ({\cal I})=1$ and
\begin{eqnarray}
S'_\alpha ({\bf q}, \tauv_f )^* &=&
N^{-1} \sum_{\bf R} S_\alpha ({\bf R}_i , \tauv_i)
e^{-i {\bf q} \cdot ({\bf R}_f + \tauv_f)} \nonumber \\
&=& S_\alpha ({\bf q}, \tauv_i)
e^{i {\bf q} \cdot [-{\bf R}_f - \tauv_f - {\bf R}_i - \tauv_i ]}
\nonumber \\ &=& S_\alpha ({\bf q}, \tauv_i)
\label{TRANSIAP} \end{eqnarray}
for actual position Fourier transforms.  For unit cell transforms we get
\begin{eqnarray}
S'_\alpha ({\bf q}, \tauv_f )^* &=& S_\alpha ({\bf q}, \tauv_i)
e^{i {\bf q} \cdot [-{\bf R}_f - {\bf R}_i ]} \nonumber \\
&=& S_\alpha ({\bf q}, \tauv_i) e^{i {\bf q} \cdot [\tauv_f + \tauv_i]} \ .
\label{TRANSIUC} \end{eqnarray}

\begin{table} [ht]
\begin{scriptsize}
\begin{tabular}{|c|c|c|c|c|c|c|c|c|ccc}\hline\hline
Irrep & $\Gamma_1$ & $\Gamma_2$ & $\Gamma_3$ & $\Gamma_4$ \\ \hline
$\lambda(2_x) = $ & $+1$ & $+1$ & $-1$ & $-1$ \vspace{0.05cm}\\
$\lambda(m_z) = $ & $+1$ & $-1$ & $-1$ & $+1$ \vspace{0.05cm}\\
\hline

${\bf S}({\bf q},s1)$&$\begin{array}{c}n^a_s\\n^b_s\\n^c_s\end{array}$ &
$\begin{array}{c}n^a_s\\n^b_s\\n^c_s\end{array}$&
$\begin{array}{c}n^a_s\\n^b_s\\n^c_s\end{array}$&
$\begin{array}{c}n^a_s\\n^b_s\\^c_s\end{array}$\\ \hline

${\bf S}({\bf q},s2)$&$\begin{array}{c}n^a_s\\-n^b_s\\-n^c_s\end{array}$&
$\begin{array}{c}n^a_s\\-n^b_s\\-n^c_s\end{array}$&
$\begin{array}{c}-n^a_s\\n^b_s\\n^c_s\end{array}$&
$\begin{array}{c}-n^a_s\\n^b_s\\n^c_s\end{array}$\\ \hline

${\bf S}({\bf q},s3)$ & $\begin{array}{c} - n^a_s\\ n^b_s\\
- n^c_s\end{array}$ & $\begin{array}{c} n^a_s\\ -n^b_s\\
n^c_s \end{array}$ & $\begin{array}{c}-n^a_s\\
n^b_s\\-n^c_s\end{array}$&
$\begin{array}{c} n^a_s\\ - n^b_s \\ n^c_s\end{array}$\\ \hline

${\bf S}({\bf q},s4)$ & $\begin{array}{c} - n^a_s \\ -n^b_s \\
n^c_s\end{array}$ & $\begin{array}{c} n^a_s\\ n^b_s\\
-n^c_s \end{array}$ & $\begin{array}{c} n^a_s\\ n^b_s\\
-n^c_s \end{array}$ & $\begin{array}{c} -n^a_s \\ -n^b_s\\
n^c_s\end{array}$\\ \hline

${\bf S}({\bf q},c1)$&$\begin{array}{c}n^a_c\\0\\0\end{array}$&
$\begin{array}{c}n^a_c\\0\\0\end{array}$&
$\begin{array}{c}0\\n^b_c\\n^c_c\end{array}$&
$\begin{array}{c}0\\n^b_c\\n^c_c\end{array}$\\ \hline

${\bf S}({\bf q},c2)$ & $\begin{array}{c} -n^a_c\\ 0\\0\end{array}$&
$\begin{array}{c}  n^a_c\\ 0\\0\end{array}$&
$\begin{array}{c} 0\\ n^b_c\\ -n^c_c\end{array}$&
$\begin{array}{c} 0\\ -n^b_c \\ n^c_c\ \end{array}$\\
\hline \hline
\end{tabular}\caption{\label{NVOq} Allowed spin functions (i. e. 
actual position Fourier coefficients) within the unit cell of NVO
for wavevector $(q,0,0)$ which are eigenvectors of
$2_x$ and $m_z$ with the eigenvalues $\lambda$ listed. Irrep stands for
the irreducible representation as labeled in Ref. \onlinecite{NVOPRB}.
The labeling of the sites is as in Table \ref{NVOMAG} and Fig. \ref{NVOSTRFIG}.
Here $n_p^\alpha$ ($p=$s or c, $\alpha=a,b,c$) denotes the complex
quantity $n_p^\alpha ({\bf q})$.}
\end{scriptsize}
\end{table}

Now we apply this formalism to find the actual position Fourier
coefficients which are eigenfunctions of the two operators $2_x$ and
$m_z$.  In so doing note the simplicity of Eq. (\ref{TRANSAP}): since, for
NVO, the operations $2_x$ and $m_z$ do not change the $x$ coordinate,
we simply have
\begin{eqnarray}
S'_\alpha ({\bf q}, \tauv_f ) &=& \xi_\alpha S'_\alpha ({\bf q} , \tauv_i) \ .
\end{eqnarray}
Thus the eigenvalue conditions for $2_x$ acting on the spine sites
(\#1-\#4) are
\begin{eqnarray}
S_\alpha (q,1)' &=& \xi_\alpha (2_x) S_\alpha (q,2) = \lambda(2_x) S_\alpha
(q,1) \nonumber \\
S_\alpha (q,2)' &=& \xi_\alpha (2_x) S_\alpha (q,1) = \lambda(2_x) S_\alpha
(q,2) \nonumber \\
S_\alpha (q,3)' &=& \xi_\alpha (2_x) S_\alpha (q,4) = \lambda(2_x) S_\alpha
(q,3) \nonumber \\
S_\alpha (q,4)' &=& \xi_\alpha (2_x) S_\alpha (q,3) = \lambda(2_x) S_\alpha
(q,4) \ ,
\label{SPINES} \end{eqnarray}
from which we see that $\lambda(2_x)=\pm 1$ and
\begin{eqnarray}
S_\alpha (q,2) &=& [\xi_\alpha (2_x)/\lambda(2_x)] S_\alpha (q,1) \ ,
\nonumber \\
S_\alpha (q,3) &=& [\xi_\alpha (2_x)/\lambda(2_x)] S_\alpha (q,4) \ .
\end{eqnarray}
The eigenvalue conditions for $m_z$ acting on the spine sites are
\begin{eqnarray}
S_\alpha (q,1)' &=& \xi_\alpha (m_z) S_\alpha (q,4) = \lambda(m_z) S_\alpha
(q,1)  \nonumber \\
S_\alpha (q,4)' &=& \xi_\alpha (m_z) S_\alpha (q,1) = \lambda(m_z) S_\alpha
(q,4) \nonumber \\
S_\alpha (q,2)' &=& \xi_\alpha (m_z) S_\alpha (q,3) = \lambda(m_z) S_\alpha
(q,2) \nonumber \\
S_\alpha (q,3)' &=& \xi_\alpha (m_z) S_\alpha (q,2) = \lambda(m_z) S_\alpha
(q,3) \ , 
\end{eqnarray}
from which we see that $\lambda(m_z)=\pm 1$ and
\begin{eqnarray}
S_\alpha (q,4) &=& [\xi_\alpha (m_z)/\lambda(m_z)] S_\alpha (q,1) \ . 
\end{eqnarray}
We thereby construct the eigenvectors for the spine sites
as given in Table \ref{NVOq}. The results for the cross-tie sites
are obtained in the same way and are also given in the table.
Each set of eigenvalues
corresponds, in technical terms, to a single irreducible
representation (irrep).  Since each operator can have either of
two eigenvalues, we have four irreps to consider.  Note that these
spin functions, since they are actually Fourier coefficients,
are complex-valued quantities. [The spin itself is real because
$F(-{\bf q})=F({\bf q})^*$.]  Note that each column of Table
\ref{NVOq} gives the most general form of an
allowed eigenvector for which one has 4 (or 5, depending on the
irrep) independent complex constants.  To further illustrate the
meaning of this table we explicitly write, in Eq. (\ref{SIGMA4}),
below, the spin distribution arising from one irrep, $\Gamma_4$.

So far, the present enalysis reproduces the standard results and indeed
computer programs exist to construct such tables.  But for multiferroics
it may be quicker to obtain and understand how to construct the possible
spin functions by hand rather than to understand how to use the program!
Usually these programs give the results in terms of unit cell
Fourier transforms, which we claim are not as natural a representation
in cases like NVO. In terms of unit cell Fourier transforms the
eigenvalue conditions for $2_x$ acting on the spine sites (\#1-\#4) are
the same as Eq. (\ref{SPINES}) for actual position Fourier transforms
because the operation $2_x$ does not change the unit cell.  However,
for the glide operation $m_z$ this is not the case.  If we start from site
\#1 or site \#2 the translation along the $y$ axis takes the spin to a
final unit cell displaced by $(-a/2)\hat i + (b/2) \hat j$, whereas if we
start from site \#3 or site \#4 the translation along the $y$ axis takes
the spin to a final unit cell displaced by $(a/2)\hat i + (b/2) \hat j$.
Now the eigenvalue conditions for $m_z$ acting on the spine sites
(\#1-\#4) are
\begin{eqnarray}
S_\alpha (q,1)' &=& \xi_\alpha (m_z) S_\alpha (q,4) \eta 
= \lambda(m_z) S_\alpha (q,1)  \nonumber \\
S_\alpha (q,4)' &=& \xi_\alpha (m_z) S_\alpha (q,1) \eta^*
= \lambda(m_z) S_\alpha (q,4) \nonumber \\
S_\alpha (q,2)' &=& \xi_\alpha (m_z) S_\alpha (q,3) \eta
= \lambda(m_z) S_\alpha (q,2) \nonumber \\
S_\alpha (q,3)' &=& \xi_\alpha (m_z) S_\alpha (q,2) \eta^*
= \lambda(m_z) S_\alpha (q,3) \ , 
\end{eqnarray}
where $\eta = \exp(i\pi q)$.  One finds that all entries for $S({\bf q},s3)$,
$S({\bf q},s4)$, and  $S({\bf q},c2)$ now carry the phase factor
$\eta^* = \exp(-i \pi \hat q)$.  But this is just the factor to make the unit
cell result
\begin{eqnarray}
{\bf S}({\bf R}, \tauv) &=& {\bf S}({\bf q}, \tauv )e^{-i{\bf q} \cdot
{\bf R} }
\end{eqnarray}
be the same (to within an overall phase factor) as the actual position result
\begin{eqnarray}
{\bf S}({\bf R}, \tauv) &=& {\bf S}({\bf q}, \tauv )e^{-i{\bf q} \cdot
({\bf R} + \tauv)} \ .
\end{eqnarray}
We should emphasize that in such a simple case as NVO, it is actually not
necessary to invoke any group theoretical concepts to arrive at the
results of Table \ref{NVOq} for the most general spin distribution
consistent with crystal symmetry.
 
More importantly, it is not commonly understood\cite{Bertaut,JS1,Rossat}
that one can also extract information using the symmetry of an operation
(inversion) which need {\it does not} conserve
wavevector.\cite{TMO2,FERRO,JAP,NVOPRB,HANDBOOK,JS2,JV} Since what we
are about to say is unfamiliar, we start from first principles.  The
quadratic free energy may be written as
\begin{eqnarray}
F_2  &=& \sum_{\bf q} \sum_{\tau, \tau' ;\alpha \beta} 
F_{\alpha \beta}^{\tau \tau'}
S_\alpha({\bf q},\tau)^* S_\beta ({\bf q}, \tau')  \ ,
\end{eqnarray}
where we restrict the sum over wavevectors to the star of the wavevector
of interest. One term of this sum is
\begin{eqnarray}
F_2({\bf q}_0)  &=& \sum_{\tau, \tau' ;\alpha \beta}
F_{\alpha \beta}^{\tau \tau'}
S_\alpha({\bf q}_0,\tau)^* S_\beta ({\bf q}_0, \tau')  \ .
\label{QEQ} \end{eqnarray}
It should be clear that the quadratic free energy, $F_2$ is invariant under
all the symmetry operations of the paramagnetic space group (i. e. what
one calls the space group of the crystal).\cite{IED,LL} For centrosymmetric
crystals there are
three classes of such symmetry operations. The first class consists of
those operations which leave $q_0$ invariant and these are the
symmetries taken into account in the usual 
formulation.\cite{Bertaut,JS1,Rossat}
The second class consists of operations which take $q_0$ into
another wavevector of the star (call it ${\bf q}_1$), where
${\bf q}_1 \not= - {\bf q}_0$. Use of these symmetries allows one
to completely characterize the wavefunction at wavevector ${\bf q}_1$
in terms of the wavefunction for ${\bf q}_0$.  These relations
are needed if one is to discuss the possibility of simultaneously
condensing more than one wavevector in the star of ${\bf q}$.\cite{ABHJS}
Finally, the third class consists of spatial inversion (unless the
wavevector and its negative differ by a reciprocal lattice vector,
in which case inversion belongs in class \#1).  The role of inversion
symmetry is almost universally overlooked,\cite{Bertaut,JS1,Rossat}
as is evident from examination of a number of recent papers.
Unlike the operations of class \#1 which takes
$S_n({\bf q})$ into an $S_{n'}({\bf q})$ (for irreps of dimension one
which is true for most cases considered in this paper), inversion takes
$S_n({\bf q})$ into an $S_{n'}(-{\bf q})$. Nevertheless it does take the form
written in Eq. (\ref{QEQ}) into itself and restricts the possible form
of the wavefunctions.  So we now consider the consequences of invariance
of $F_2$ under inversion.\cite{FERRO,HANDBOOK,NVOPRB,TMO2,JAP}  For this
purpose
we write Eq. (\ref{F2S}) in terms of the spin coordinates $m$ of Table
\ref{NVOq}. (The result will, of course, depend on which irrep $\Gamma$
we consider.) In any case, the part of $F_2$ which depends on ${\bf q}_0$
can be written as
\begin{eqnarray}
F_2({\bf q})  &=& \sum_{\tau, \tau' ;\alpha \beta} 
F_{\alpha \beta}^{\tau \tau'}
S_\alpha({\bf q}_0,\tau)^* S_\beta ({\bf q}_0, \tau') \nonumber \\
&=& \sum_{N,\alpha;N',\beta}  G_{N,\alpha ; N', \beta} [n_N^\alpha ]^* 
[n_{N'}^\beta] \nonumber \\
&=& \sum_{N,\alpha;N',\beta}  G_{N,\alpha ; N', \beta} [{\cal I}n_N^\alpha ]^*
[{\cal I}n_{N'}^\beta] \ ,
\label{GEQ} \end{eqnarray}
where $N$ and $N'$ assume the values "s" for spin and "c" for cross-tie and
$\alpha$ and $\beta$ label components.
Now we need to understand the effect of ${\cal I}$ on the spin Fourier
coefficients listed in Table \ref{NVOq}.  Since we use actual position
Fourier coefficients, we apply Eq. (\ref{TRANSIAP}).  For the cross-tie
variables (which sit at a center of inversion symmetry)
inversion takes the spin coordinates of one spine sublattice into 
the complex conjugate of itself:
\begin{eqnarray}
{\cal I} {\bf S}({\bf q},cn) = [{\bf S}({\bf q},cn)]^* \ .
\end{eqnarray}
Thus in terms of the $n$'s this gives
\begin{eqnarray}
{\cal I} n_c^\alpha = [n_c^\alpha ]^* \ ,  \hspace{0.5in} \alpha = x,y,z\ .
\end{eqnarray}

The effect of inversion on the spine variables
again follows from Eq. (\ref{TRANSIAP}). Since inversion
interchanges sublattice \#1 and \#3, we have
\begin{eqnarray}
[{\bf S}({\bf q},s3)]' = [{\bf S}({\bf q},s1)]^* \ . 
\end{eqnarray}
For $\lambda(2_x)=\lambda(m_z)=+1$ (i. e. for irrep $\Gamma_1$),
we substitute the values of the
spin vectors from the first column of Table \ref{NVOq} to get
\begin{eqnarray}
{\cal I} [-n_s^a] &=& [n_s^a]^* \ , \ \ \ \ \ 
{\cal I} [n_s^b] = [n_s^b]^* \ , \nonumber \\
{\cal I} [-n_s^c] &=& [n_s^c]^* \ .
\end{eqnarray}
Note that some components introduce a factor $-1$ under inversion and others do
not.  (Which ones have the minus signs depends on which irrep we consider.)
If we make a change of variable by replacing $n_s^\alpha$ in column \#1
of Table \ref{NVOq}
by $in_s^\alpha$ for those components for which ${\cal I}$ introduces a minus
sign and leave the other components alone, then we may rewrite the first
column of Table \ref{NVOq} in the form given in Table \ref{NVOqq}.
In terms of these new variables one has
\begin{eqnarray}
{\cal I} [n_s^\alpha] = [n_s^\alpha]^* \ .
\label{IEQ} \end{eqnarray}
(It is convenient to define the spin Fourier coefficients so that they all 
transform in the same way under inversion.  Otherwise one would have to keep
track of variables which transform with a plus sign and those which transform
with a minus sign.) Repeating this process for all the other irreps
we write the possible spin functions as those of Table \ref{NVOqq}.
We give an explicit formula for the spin distribution for one irrep in 
Eq. (\ref{SIGMA4}) below.

\begin{table}
\vspace{0.2 in}
\begin{scriptsize}
\begin{tabular}{|c|c|c|c|c|c|c|c|c|ccc}\hline\hline
Irrep$=$ & $\Gamma_1$ & $\Gamma_2$ & $\Gamma_3$ & $\Gamma_4$ \vspace{0.05cm} \\
\hline
$\lambda (2_x)=$ & $+1$ & $+1$ & $-1$ & $-1$ \vspace{0.05cm}\\
$\lambda (m_z)=$ & $+1$ & $-1$ & $-1$ & $+1$ \vspace{0.05cm}\\
\hline

${\bf S}({\bf q},s1)$&$\begin{array}{c}in^a_s\\n^b_s\\in^c_s\end{array}$ &
$\begin{array}{c}n^a_s\\in^b_s\\n^c_s\end{array}$&
$\begin{array}{c}in^a_s\\n^b_s\\in^c_s\end{array}$&
$\begin{array}{c}n^a_s\\in^b_s\\n^c_s\end{array}$\\ \hline

${\bf S}({\bf q},s2)$&$\begin{array}{c}in^a_s\\-n^b_s\\-in^c_s\end{array}$&
$\begin{array}{c}n^a_s\\-in^b_s\\-n^c_s\end{array}$&
$\begin{array}{c}-in^a_s\\n^b_s\\in^c_s\end{array}$&
$\begin{array}{c}-n^a_s\\in^b_s\\n^c_s\end{array}$\\ \hline

${\bf S}({\bf q},s3)$ & $\begin{array}{c} - in^a_s\\ n^b_s\\
- in^c_s\end{array}$ & $\begin{array}{c} n^a_s\\ -in^b_s\\
n^c_s \end{array}$ & $\begin{array}{c}-in^a_s\\
n^b_s\\-in^c_s\end{array}$&
$\begin{array}{c} n^a_s\\ - in^b_s \\ n^c_s\end{array}$\\ \hline

${\bf S}({\bf q},s4)$ & $\begin{array}{c} - in^a_s \\ -n^b_s \\
in^c_s\end{array}$ & $\begin{array}{c} n^a_s\\ in^b_s\\
-n^c_s \end{array}$ & $\begin{array}{c} in^a_s\\ n^b_s\\
-in^c_s \end{array}$ & $\begin{array}{c} -n^a_s \\ -in^b_s\\
n^c_s\end{array}$\\ \hline

${\bf S}({\bf q},c1)$&$\begin{array}{c}n^a_c\\0\\0\end{array}$&
$\begin{array}{c}n^a_c\\0\\0\end{array}$&
$\begin{array}{c}0\\n^b_c\\n^c_c\end{array}$&
$\begin{array}{c}0\\n^b_c\\n^c_c\end{array}$\\ \hline

${\bf S}({\bf q},c2)$ & $\begin{array}{c} -n^a_c\\ 0\\0\end{array}$&
$\begin{array}{c}  n^a_c\\ 0\\0\end{array}$&
$\begin{array}{c} 0\\ n^b_c\\ -n^c_c\end{array}$&
$\begin{array}{c} 0\\ -n^b_c \\ n^c_c\ \end{array}$\\
\hline \hline
\end{tabular}\caption{\label{NVOqq} As Table \ref{NVOq} except that
now the effect of inversion symmetry is taken into account,
as a result of which, apart from an overall phase factor all
the $n$'s in this table can be taken to be real-valued.}
\end{scriptsize}
\end{table}

Now we implement Eq. (\ref{GEQ}), where the spin functions are taken to be
the variables listed in Table \ref{NVOqq}.  First note that the matrix ${\bf G}$
in Eq. (\ref{GEQ}) has to be Hermitian to ensure that $F_2$ be real:
\begin{eqnarray}
G_{M,\alpha; N,\beta} &=& [G_{N,\beta;M,\alpha}]^* \ .
\label{HER} \end{eqnarray}
Then, using Eq. (\ref{IEQ}), we find that Eq. (\ref{GEQ}) is
\begin{eqnarray}
F_2 ({\bf q}_0) &=& \sum_{M,\alpha;N,\beta} [n_M^\alpha]^* G_{M,\alpha;N,\beta} n_N^\beta 
\nonumber \\
&& \ \ = \sum_{M,\alpha;N,\beta} [{\cal I} n_M^\alpha]^* G_{M,\alpha;N,\beta} 
[{\cal I} n_N^\beta]
\nonumber \\
&& \ \ = \sum_{M,\alpha;N,\beta} n_M^\alpha G_{M,\alpha;N,\beta}
[n_N^\beta]^* \nonumber \\
&& \ \ = \sum_{M,\alpha;N,\beta} [n_M^\alpha]^* G_{N,\beta;M,\alpha} [n_N^\beta] \ ,
\end{eqnarray}
where, in the last line, we interchanged the roles of the dummy indices
$M,\alpha$ and $N,\beta$. By comparing the first and last lines, one sees that the 
matrix $G$ is symmetric.  Since this matrix is also Hermitian, all its elements must
be {\it real valued}.  Thus all its eigenvectors can be taken to have
only real-valued components.  But the $m$'s are allowed to be complex valued.
So, the conclusion is that for each irrep, we may write
\begin{eqnarray}
n_N^\alpha (\Gamma) &=& e^{i \phi_\Gamma} [r_N^\alpha (\Gamma)]  \ ,
\end{eqnarray}
where the $r$'s are all real valued and $\phi_\Gamma$ is an overall  phase
which can be chosen arbitrarily for each $\Gamma$.  It is likely that
the phase will be fixed by high-order Umklapp terms in the free energy,
but the effects of such phase locking may be beyond the range of
experiments.

It is worth noting how these results should be (and in a few
cases\cite{FERRO,TMO2,NVOPRB} have been) used in the structure
determinations.  One should choose the best fit to the diffraction data
using, in turn, each irrep (i. e. each set of eigenvalues of $2_x$ and
$m_z$).  Within each representation one parametrizes the spin structure
by choosing the Fourier coefficients as in the relevant
column of Table \ref{NVOqq}.  Note that instead of having 4 or 5 complex
coefficients to describe the six sites within the unit cell (see Table
\ref{NVOq}), one has only 4 or 5 (depending on the representation)
{\it real-valued} coefficients to determine.  The relative phases of
the complex coefficients have all been fixed by invoking inversion
symmetry.  This is clearly a significant step in increasing the precision
of the determination of the magnetic structure from experimental data. 

\subsection{Order Parameters}

We now review how the above symmetry classification influences
the introduction of order parameters which allow the construction
of Landau expansions.\cite{FERRO,NVOPRB}
The form of the order parameter should be
such that it has the potential to describe all ordering which
are allowed by the quadratic free energy $F_2$.  Thus, for an
isotropic Heisenberg model on a cubic lattice, the order parameter
has three components (i. e. it involves a three dimensional irrep)
because although the fourth order terms will
restrict order to occur only along certain directions, as far
as the quadratic terms are concerned, all directions are equivalent.
The analogy here is that the overall phase of the spin function
$\phi(\Gamma)$ is not fixed by the quadratic free energy and
accordingly the order parameter must be a complex variable which
includes such a phase.  One also recognizes that although the amplitude 
of the critical eigenvector is not fixed by the quadratic terms in
the free energy, the ratios of its components are fixed by the
specific form of the inverse susceptibility matrix.  Although we
do not wish to discuss the explicit form of this matrix, what should 
be clear is that the components of the spins which order must
be proportional to the components of the critical eigenvector.  The
actual amplitude of the spin ordering is determined by the competition
between the quadratic and fourth order terms in the free energy.
If $\Gamma_p$ is the irrep which is critical, then just below the
ordering temperature we write
\begin{eqnarray}
n^\alpha_N({\bf q}) &=& \sigmav_p({\bf q}) r^\alpha_N (\Gamma_p ) \ ,
\end{eqnarray}
where the $r$'s are real components of the critical eigenvector 
(coming from irrep $\Gamma_p$) of the
matrix ${\bf G}$ of Eq. (\ref{GEQ}) and are now {\it normalized} by
\begin{eqnarray}
\sum_{\alpha N} [r^\alpha_N]^2 = 1 \ .
\end{eqnarray}
Here the order parameter for irrep $\Gamma({\bf q})$, $\sigmav_p({\bf q})$
is a complex variable,
since it has to incorporate the arbitrary  complex phase $\phi_p$
associated with irrep $\Gamma_p$:
\begin{eqnarray}
\sigmav_p(\pm |{\bf q}|) &=& \sigma_p e^{\pm i \phi_p}\ .
\label{IPHI} \end{eqnarray}
The order parameter transforms as indicated in the tables by
its listed eigenvalues under the symmetry operations
$2_x$ and $m_z$.  Since the components of the critical eigenvector
are dominantly determined by the quadratic terms,\cite{CRITICAL}
one can say that just below the ordering temperature the
description in terms of an order parameter continues to hold but 
\begin{eqnarray}
\sigma_p \sim |T_c -T|^\beta_p \ ,
\end{eqnarray}
where mean-field theory gives $\beta=1/2$ but corrections due to
fluctuation are expected.\cite{RG}

To summarize and illustrate the use of Table \ref{NVOqq} we write an
explicit expression for the magnetizations of the \#1 spine
sublattice and the \#1 cross-tie sublattice for irrep $\Gamma_4$
[$\lambda (2_x)=-1$ and $\lambda (m_z)=+1$].  We use the definition of
the order parameter and sum over both signs of the wavevector to get
\begin{eqnarray}
S_x({\bf r}, s1)&=&  2\sigma_4 r^x_s \cos(qx+\phi_4) \nonumber \\
S_y({\bf r}, s1)&=& 2\sigma_4 r^y_s \sin(qx+\phi_4) \nonumber \\
S_z({\bf r}, s1)&=&  2\sigma_4 r^z_s \cos(qx+\phi_4)  \nonumber \\
S_x({\bf r}, s2)&=& -2\sigma_4 r^x_s \cos(qx+\phi_4) \nonumber \\
S_y({\bf r}, s2)&=& 2\sigma_4 r^y_s \sin(qx+\phi_4) \nonumber \\
S_z({\bf r}, s2)&=&  2\sigma_4 r^z_s \cos(qx+\phi_4)  \nonumber \\
S_x({\bf r}, s3)&=&  2\sigma_4 r^x_s \cos(qx+\phi_4) \nonumber \\
S_y({\bf r}, s3)&=&  -2\sigma_4 r^y_s \sin(qx+\phi_4) \nonumber \\
S_z({\bf r}, s3)&=&  2\sigma_4 r^z_s \cos(qx+\phi_4)  \nonumber \\
S_x({\bf r}, s4)&=& -2\sigma_4 r^x_s \cos(qx+\phi_4) \nonumber \\
S_y({\bf r}, s4)&=&  -2\sigma_4 r^y_s \sin(qx+\phi_4) \nonumber \\
S_z({\bf r}, s4)&=&  2\sigma_4 r^z_s \cos(qx+\phi_4)  \nonumber \\
S_x({\bf r}, c1)&=& 0 \nonumber \\
S_y({\bf r}, c1)&=& 2\sigma_4 r^y_c \cos (qx+\phi_4) \nonumber \\
S_z({\bf r}, c1)&=& 2\sigma_4 r^z_c \cos (qx+\phi_4) \nonumber \\
S_x({\bf r}, c1)&=& 0 \nonumber \\
S_y({\bf r}, c2)&=& -2\sigma_4 r^y_c \cos (qx+\phi_4) \nonumber \\
S_z({\bf r}, c2)&=& 2\sigma_4 r^z_c \cos (qx+\phi_4)
\label{SIGMA4} \end{eqnarray}
and similarly for the other irreps.  Here ${\bf r}\equiv (x,y,z)$ is the actual location of the spin.
Using explicit expressions like the above (or more directly
from Table \ref{NVOqq}), one can verify that
the order parameters have the transformation properties:
\begin{eqnarray}
2_x \sigmav_1({\bf q}) &=& + \sigmav_1({\bf q}) \ , \ \ \ \ \
m_z \sigmav_1({\bf q}) = + \sigmav_1({\bf q}) \ , \nonumber \\
2_x \sigmav_2({\bf q}) &=& + \sigmav_2({\bf q}) \ , \ \ \ \ \
m_z \sigmav_2({\bf q}) =  - \sigmav_2({\bf q}) \ , \nonumber \\
2_x \sigmav_3({\bf q}) &=& - \sigmav_3({\bf q}) \ , \ \ \ \ \
m_z \sigmav_3({\bf q}) = - \sigmav_3({\bf q}) \ , \nonumber \\
2_x \sigmav_4({\bf q}) &=& - \sigmav_4({\bf q}) \ , \ \ \ \ \
m_z \sigmav_4({\bf q}) = + \sigmav_4({\bf q}) 
\label{ROTSYM} \end{eqnarray}
and
\begin{eqnarray}
{\cal I} \sigmav_n({\bf q}) &=&  [\sigmav_n({\bf q})]^*.
\label{IISYM} \end{eqnarray}
Note that even when more than a single irrep is present,
the introduction of order parameters, as done here, provides
a framework within which one can represent the spin distribution
as a linear combination of distributions each having a 
characteristic symmetry, as expressed by Eq. (\ref{ROTSYM}).
When the structure of the unit cell is ignored\cite{MOST}
that information is not readily accessible.
Also note that the phase of each irrep $\Gamma_n$ is defined 
so that when $\phi_n=0$, the wave is inversion-symmetric about 
${\bf r}=0$.  For a single irrep this specification is not important.
However, when one has two irreps, then inversion symmetry is only
maintained if their phases are equal.

In many systems, the initial incommensurate order that first
occurs as the temperature is lowered becomes unstable
as the temperature is further lowered.\cite{NAG}  Typically,
the initial order involves spins oriented along their easy axis
with sinusoidally varying magnitude.  However, the fourth order
terms in the Landau expansion favor fixed length spins.  As
the temperature is lowered the fixed length constraint becomes
progressively more important and at a second, lower, critical
temperature a transition occurs in which transverse components
become nonzero.  Although the situation is more complicated when
there are several spins per unit cell, the result is similar:
the fixed length constraint is best realized when more than a single
irrep has condensed.  So, for NVO and TMO as the temperature is
lowered one encounters a second phase transition in which a second
irrep appears.  Within a low-order Landau expansion this phenomenon is
described by a free energy of the form\cite{NVOPRB}
\begin{eqnarray}
F &=& \frac{1}{2} (T-T_>) \sigma_>^2 + \frac{1}{2} (T-T_<) \sigma_<^2
+ u_> \sigma_>^4 \nonumber \\ && \ \ + u_< \sigma_<^4
+ w \sigma_>^2 \sigma_<^2 \ ,
\label{LANDAU4} \end{eqnarray}
where $T_> > T_<$. This system has been studied in detail by Bruce and Aharony.\cite{AA}
For our purposes, the most important result is that for suitable values
of the parameters ordering in $\sigma_>$ occurs at $T_>$ and at some lower
temperature order in $\sigma_<$ may occur.  The application of this theory
to the present situation is simple: we can (and usually do) have two
magnetic phase transitions in which first one irrep and then at a lower
temperature a second irrep condense.  A question arises as to whether the
condensation of one irrep can induce the condensation of a second irrep.
This is not possible because the two irreps have different symmetry.
But could the presence of two irreps, $\Gamma_>$ and $\Gamma_<$ induce the
appearance of a third irrep $\Gamma_3$ at the temperature at which $\Gamma_<$ first
appears? For that to happen would require that $\Gamma_>^n \otimes \Gamma_<^m\otimes \Gamma_3$
contain the unit representation for some values of $n$ and $m$.
This or any higher combination of representations
is not allowed for the simple four irreps system like NVO.  In more
complex systems one might have to allow for such a phenomenon.

\section{Applications}

In this section we apply the above formalism to a number of
multiferroics of current interest.

\subsection{MnWO$_4$}

MnWO$_4$ (MWO) crystallizes in the space group P2/c (\#14 in Ref.
\onlinecite{HAHN}) whose general positions are given in Table
\ref{WTAB}.  The two magnetic Mn ions per unit cell are at positions
\begin{eqnarray}
\tauv_1 &=& (\oh , y, \of) \ , \ \ \ 
\tauv_2 = (\oh , 1-y, \tf) \ .
\label{SUBL} \end{eqnarray}

\begin{table}
\vspace{0.2 in}
\begin{tabular} {|| c | c ||}
\hline \hline
$E {\bf r} =(x,y,z) \ \ $ & $m_y {\bf r} =(x, \overline y , z+ \oh )\ \ $ \\
${\cal I} {\bf r}= (\overline x , \overline y , \overline z )\ \ $
& $2_y {\bf r}= (\overline x, y, \overline z + \oh )\ \ $ \\
\hline \hline 
\end{tabular}
\caption{\label{WTAB}General Positions for space group P2/c.}
\end{table}

The wavevector of incommensurate magnetic ordering
is\cite{RLU,LAUT} ${\bf q}=(q_x, 1/2,q_z)$ with
$q_x \approx -0.21$ and $q_z \approx 0.46$) and is left invariant
by the identity and $m_y$.  We start by constructing the eigenvectors
of the quadratic free energy (i. e. the inverse susceptibility matrix).
Here we use unit cell Fourier transforms to facilitate comparison
with Ref. \onlinecite{LAUT}.  Below $X$, $Y$, and $Z$
denote integers (in units of lattice constants).  When
\begin{eqnarray}
{\bf R}_f + \tauv_f &=& (X,Y,Z) + \tauv_1 \nonumber \\
&=& (X+\oh , Y+y , Z+ \of) \ ,
\end{eqnarray}
then
\begin{eqnarray}
{\bf R}_i + \tauv_i &=& [m_y]^{-1} ({\bf R}_f + \tauv_f) \nonumber \\
&=& (X + \oh, -Y-y, Z_\of) \nonumber \\
&=& (X, -Y-1,Z-1) + \tauv_2 \ .
\end{eqnarray}
Then Eq. (\ref{TRANSUC}) gives the eigenvalue condition to be
\begin{eqnarray}
S_\alpha' ({\bf q}, \tau_1 ) &=& \xi_\alpha (m_y) S_\alpha ( {\bf q},
\tauv_2) e^{2 \pi i \hat {\bf q} \cdot [(2Y+1)\hat j + \hat k] }\nonumber \\
&=& \xi_\alpha (m_y) S_\alpha ( {\bf q}, \tau_2) e^{\pi i + 2 \pi i \hat q_z}  
\nonumber \\ &=& \lambda S_\alpha ({\bf q},\tau_1) \ ,
\label{EIG1EQ} \end{eqnarray}
where $\xi_x(m_y) =- \xi_y(m_y) =\xi_z(m_y)=-1$.  When
\begin{eqnarray}
{\bf R}_f + \tauv_f &=& (X,Y,Z) + \tauv_2 \nonumber \\
&=& (X+\oh , Y+1-y, Z+ \tf) \ ,
\end{eqnarray}
then 
\begin{eqnarray}
{\bf R}_i + \tauv_i &=& (X+\oh, -Y-1-y, Z+ \of) \nonumber \\
&=& (X ,-Y-1,Z) + \tauv_1 \ ,
\end{eqnarray}
and Eq. (\ref{TRANSUC}) gives the eigenvalue condition to be
\begin{eqnarray}
S_\alpha' ({\bf q}, \tau_2 ) &=& \xi_\alpha (m_y) S_\alpha ( {\bf q},
\tauv_1) e^{2 \pi i \hat {\bf q} \cdot (2Y+1)\hat j }\nonumber \\
&=& \xi_\alpha (m_y) S_\alpha ( {\bf q}, \tau_1) [-1]
= \lambda S_\alpha ({\bf q},\tau_2) \ .
\label{EIG2EQ} \end{eqnarray}
From Eqs. (\ref{EIG1EQ}) and (\ref{EIG2EQ}) we get
$\lambda = \pm e^{i \pi \hat q_z}$ and
\begin{eqnarray}
S_\alpha ({\bf q}, \tau_2) &=& - [\xi_\alpha (m_y)/\lambda] S_\alpha
( {\bf q}, \tau_1) \ .
\end{eqnarray}
So we get the results listed in Table \ref{WREP}.

\begin{table}
\vspace{0.2 in}
\begin{tabular} {|| c || c | c || } \hline \hline
Irrep & $\Gamma_1$ & $\Gamma_2$ \\
$\lambda (m_y) =$ & $e^{i \pi \hat q_z}$ & $-e^{i \pi \hat q_z}$ \\  \hline
${\bf S}({\bf q},1)$  & $\begin{array} {c} a^*n_x \\ a^*n_y \\a^*n_z \end{array}$
& $\begin{array} {c} a^*n_x \\ a^*n_y \\ a^*n_z \end{array}$ \\ 
\hline
${\bf S}({\bf q},2)$  & $\begin{array} {c} an_x \\ -an_y \\ an_z \end{array}$
& $\begin{array} {c} -an_x \\ an_y \\ -an_z \end{array}$ \\ \hline \hline
\end{tabular}
\caption{\label{WREP} Allowed spin eigenfunctions for MWO (apart from
an overall phase factor) before
inversion symmetry is taken into account, where $a=\exp(-i \pi \hat q_z/2)$.
Here the $n({\bf q})$'s are complex and we have taken the liberty to adjust the
overall phase to give a symmetrical looking result.  But these results
are equivalent to Table II of Ref. \onlinecite{LAUT}.}
\end{table}

So far the analysis is essentially the completely standard one.
Now we use the fact that the free energy is invariant under
spatial inversion, even though that operation does not conserve
wavevector.\cite{FERRO,HANDBOOK,NVOPRB,TMO2}
We now determine the effect of inversion on the $n$'s. As will become
apparent use of unit cell Fourier transforms makes this analysis more
complicated than if we had used actual position transforms.
We use Eq. (\ref{TRANSIUC}) to write
\begin{eqnarray}
{\cal I} S( {\bf q},\tau=1) &=&
S({\bf q},\tau=2)^*e^{-2 \pi i \hat {\bf q} \cdot (\hat i + \hat j + \hat k)}
\nonumber \\ &\equiv& b S({\bf q}, 2)^* \ ,
\end{eqnarray}
where $b= - \exp[ -2 \pi i (\hat q_x+\hat q_z)]$.  For $\Gamma_2$ we get
\begin{eqnarray}
{\cal I} [n_x, n_y, n_z] &=& [ - n_x , n_y , - n_z ]^* b \ ,
\end{eqnarray}
which we can write as
\begin{eqnarray}
{\cal I} n_\alpha &=&  b \xi_\alpha (m_y) n_\alpha^* \ . 
\end{eqnarray}
Now the free energy is quadratic in the Fourier spin coefficients, which
are linearly related to the $n$'s.  So the free energy can be written as
\begin{eqnarray}
F_2 &=& {\bf n}^\dagger {\bf G} {\bf n} \ ,
\label{F21EQ} \end{eqnarray}
where ${\bf n} = (n_x,n_y,n_z)$ is a column vector
and ${\bf G}$ is a $3 \times 3$ matrix
which we write as
\begin{eqnarray}
{\bf G} &=& \left[ \begin{array} {c c c} A & \alpha & \beta \\
\alpha^* & B & \gamma \\ \beta^* & \gamma^* & C \\ \end{array}
\right] \ ,
\end{eqnarray}
where, for Hermiticity the Roman letters are real and the Greek ones
complex.  Now we use the fact that also we must have invariance
with respect to inversion, which after all is a crystal symmetry. Thus
\begin{eqnarray}
F_2 &=& [ {\cal I} {\bf n}]^\dagger {\bf G} [{\cal I}{\bf n}] \ .
\end{eqnarray}
This can be written as
\begin{eqnarray}
F_2 &=& \sum_{\alpha \beta} b \xi_\alpha (m_y) n_\alpha
G_{\alpha \beta} b^* a^* \xi_\beta (m_y) n_\beta^* \nonumber  \\ &=&
\sum_{\alpha \beta} \xi_\alpha (m_y) n_\alpha
G_{\alpha \beta} \xi_\beta (m_y) n_\beta^* \ .
\end{eqnarray}
Thus we may write
\begin{eqnarray}
F_2 &=& {\bf n}^{\rm tr} \left[ \begin{array} {c c c}
A & - \alpha & \beta \\ -\alpha^* & B & - \gamma \\ \beta^*
& - \gamma^* & C \\ \end{array} \right] {\bf n}^* \nonumber \\
&=& {\bf n}^\dagger \left[ \begin{array} {c c c}
A & - \alpha^* & \beta^* \\ -\alpha & B & - \gamma^* \\ \beta
& - \gamma & C \\ \end{array} \right] {\bf n} \ ,
\label{F22EQ} \end{eqnarray}
where "tr" indicates transpose (so ${\bf n^{\rm tr}}$ is a row vector).
Since the two expressions for $F_2$, Eqs. (\ref{F21EQ}) and (\ref{F22EQ}),
must be equal we see that
$\alpha=ia$, $\beta=b$, and $\gamma=ic$, where $a$, $b$, and
$c$ must be real. Thus ${\bf G}$ is of the form 
\begin{eqnarray}
{\bf G} &=& \left[ \begin{array} {c c c}  A & ia & b \\
-ia & B & ic \\ b & -ic & C \\ \end{array} \right] \ ,
\end{eqnarray}
where all the letters are real.  This means that the critical eigenvector
describing the long range order has to be of the form
\begin{eqnarray}
(n_x, n_y , n_z ) &=& e^{i \phi} (r, is, t) \ ,
\label{RESULT} \end{eqnarray}
where $r$, $s$, and $t$ are real.  For $\Gamma_2$ we set $e^{i \phi}=-i$.
 For $\Gamma_1$ a similar calculation again yields Eq. (\ref{RESULT}),
but here we set $e^{i \phi}=1$.  (These choices are not essential.
They just make the symmetry more obvious.)  Thus we obtain the final results
given in Table \ref{WWREP}.  Lautenschlager et al\cite{LAUT} say
(just above Table II) ``Depending on the choice of the amplitudes and
phases ..."  What we see here is that inversion symmetry fixes the
phases without the possibility of a choice (just as it did for NVO). 
Note again that we have about half the variables to fix in a structure
determination when we take advantage of inversion invariance to fix
the phase of the complex structure constants.

\begin{table}
\vspace{0.2 in}
\begin{tabular} {|| c || c | c || } \hline \hline
Irrep & $\Gamma_1$ & $\Gamma_2$ \\ \hline
$\lambda (m_y) =$ & $e^{i \pi \hat q_z}$ & $-e^{i \pi \hat q_z}$ \\  \hline \hline
${\bf S}({\bf q},1)$  & $\begin{array} {c} a^*r \\ ia^*s \\ a^*t \end{array}$
& $\begin{array} {c} -ia^*r \\ a^*s \\ -ia^*t \end{array}$ \\ \hline
${\bf S}({\bf q},2)$  & $\begin{array} {c} ar \\ -ias \\ at  \end{array}$
& $\begin{array} {c} iar \\  as \\ iat \end{array}$ \\ \hline \hline
\end{tabular}
\caption{\label{WWREP} As Table \ref{WREP}, except that here inversion
symmetry is taken into account.  Here $r$, $s$, and $t$ are real.
All six components can be multiplied by
an overall phase factor which we have not explicitly written.}
\end{table}

\subsubsection{Order Parameter}

Now we discuss the definition of the order parameter for this system.
For this purpose we replace $r$ by $\sigmav r$, $s$ by $\sigmav s$. etc.,
with the normalization that
\begin{eqnarray}
r^2 + s^2 + t^2 = 1 \ .
\end{eqnarray}
Here the order parameter $\sigmav$ is complex because we always have
the freedom to multiply the wavefunction by a phase factor.  (This phase
factor might be ``locked'' by higher order terms in the free energy, but
we do not consider that phenomenon here.\cite{LOCK})
We record the symmetry properties of the order parameter. With
our choice of phases we have
\begin{eqnarray}
{\cal I} \sigmav_n ({\bf q}) &=& [\sigmav_n ({\bf q})]^* \ , \nonumber \\
m_y \sigmav_n ({\bf q}) &=& \lambda(\Gamma_n) \sigmav_n ({\bf q}) \nonumber \\
m_y \sigmav_n (-{\bf q}) &=& \lambda(\Gamma_n)^* \sigmav_n (-{\bf q}) \ , 
\label{WOPSYM} \end{eqnarray}
where $\sigmav_n({\bf q})$ is the complex-valued order parameter for ordering
of irrep $\Gamma_n$ and $\lambda(\Gamma_n)$ is the eigenvalue of $m_y$ given
in Table \ref{WWREP}.  Now we write an explicit formula for
the spin distribution in terms of the order parameters of the two irreps:
\begin{eqnarray}
{\bf S}({\bf R}, && \tau=1) = 2\sigma_1 \left[
(r_1 \hat i + t_1 \hat k) \cos( {\bf q} \cdot {\bf R} + \phi_1 - \pi q_z/2)
\right.  \nonumber \\ && \ \left. + s_1 \hat j \sin({\bf q} \cdot {\bf R}
+ \phi_1 - \pi q_z/2) \right] \nonumber \\ && \ + 2 \sigma_2 \left[
(-r_2 \hat i - t_2 \hat k) \sin( {\bf q} \cdot {\bf R} + \phi_2 - \pi q_z/2)
\right.
\nonumber \\ && \ \left. + s_2 \hat j \cos({\bf q} \cdot {\bf R} + \phi_2
- \pi q_z/2 ) 
\right] \ ,
\end{eqnarray}
\begin{eqnarray}
{\bf S}({\bf R}, && \tau=2) = 2 \sigma_1 \left[
(r_1 \hat i + t_1 \hat k) \cos( {\bf q} \cdot {\bf R} + \phi_1 + \pi q_z/2)
\right.  \nonumber \\ && \ - \left. s_1 \hat j \sin({\bf q} \cdot {\bf R}
+ \phi_1 + \pi q_z/2 ) \right] \nonumber \\ && + 2\sigma_2 \left[
(r_2 \hat i + t_2 \hat k) \sin( {\bf q} \cdot {\bf R} + \phi_2 + \pi q_z/2)
\right.  \nonumber \\ && \ \left. +s_2 \hat j \cos({\bf q} \cdot {\bf R}
+ \phi_2 + \pi q_z/2) \right] \ .
\end{eqnarray}
One can explicitly verify that these expressions are consistent with
Eq. (\ref{WOPSYM}).
Note that when only one of the order parameters (say $\sigma_n$) is nonzero,
we have inversion symmetry with respect to a redefined origin where $\phi_n=0$.
For each irrep we have to specify three real parameters, $\sigma r_n$,
$\sigma s_n$, and $\sigma t_n$ and one overall phase $\phi_n$ rather than
three complex-valued parameters had we not invoked inversion symmetry.

\subsection{TbMnO$_3$}

\begin{table}
\vspace{0.2 in}
\begin{tabular} {|| c | c ||}
\hline \hline
$E {\bf r} =(x,y,z) \ \ $ 
& $2_x {\bf r} =(x + \oh  , \overline y + \oh , \overline z )\ \ $ \\
$2_z=(\overline x,\overline y , z + \oh )\ \ $
& $2_y=(\overline x + \oh , y + \oh , \overline z+ \oh )\ \ $ \\
${\cal I}{\bf r}=(\overline x,\overline y,\overline z)\ \ $ 
& $m_x{\bf r}=(\overline x + \oh , y + \oh, z )\ \ $ \\
$m_z{\bf r}=(x, y ,\overline z + \oh)\ \ $
& $m_y{\bf r}=(x + \oh , \overline y + \oh , z+ \oh )\ \ $ \\
\hline \hline 
\end{tabular}
\caption{\label{PBNM}General Positions for Pbnm.
Notation as in Table \ref{NVOSTR}.}
\end{table}

\begin{table}
\vspace{0.2 in}
\begin{tabular} {|| c || c | c ||}
\hline \hline
Mn & \ \ $(1)=(0,\oh ,0) \ \ $ & $(2)=(\oh  , 0,0 )\ \ $ \\
& $(3)=(0, \oh , \oh )\ \ $ & $(4)=(\oh , 0 , \oh )\ \ $ \\ \hline
Tb & $(5)=(x,y,\of)$ & \ \ $(6)=(x+\oh,\overline y + \oh, \tf) \ \ $ \\
& $(7)=(\overline x , \overline y , \tf )\ \ $ &
$(8)=(\overline x + \oh , y+\oh , \of )\ \ $ \\ \hline \hline
\end{tabular}
\caption{\label{TMNOPOS}Positions of the Magnetic Ions in the Pbnm Structure
of TbMnO$_3$, with $x=0.9836$ and $y=0.0810$.\cite{BLASCO}}
\end{table}

\begin{figure}[ht]
\begin{center}
\includegraphics[width=8cm]{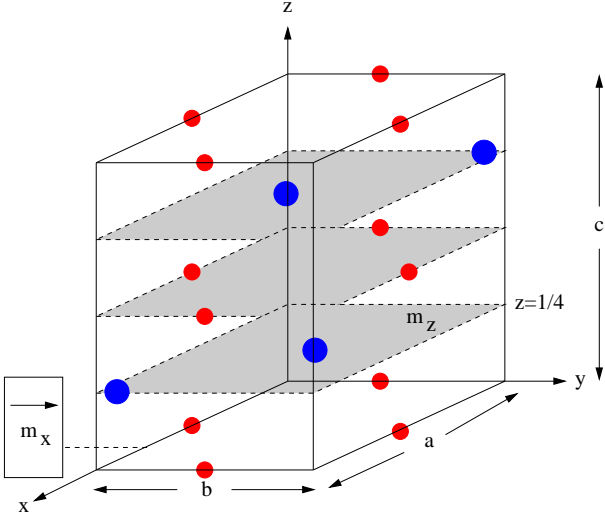}
\caption{(Color online). Mn sites (smaller circles, on-line red) and Tb sites
(larger circles, on-line blue) in the primitive unit cell of TbMnO$_3$.
The Tb sites are in the shaded planes at $z=n\pm \of$ and the
Mn sites are in planes $z=n$ or $z=n+\oh$, where $n$ is an
integer. The incommensurate wavevector is along the ${\bf b}$ axis.
The mirror plane at $z=1/4$ is indicated and the glide plane
$m_x$ is indicated by the mirror plane at $x=3/4$ followed by a 
translation (indicated by the arrow) of $b/2$ along the $y$-axis.}
\label{TMOFIG}
\end{center}
\end{figure}

Here we give the full details of the calculations for TbMnO$_3$
described in Ref. \onlinecite{TMO2}.  The presentation here differs
cosmetically from that in Ref. \onlinecite{JAP}.
The space group of TbMnO$_3$ is Pbnm which is \#62 in
Ref. \onlinecite{HAHN} (although the positions are listed there
for the Pnma setting).  The space group operations for a general
Wyckoff orbit is given in Table \ref{PBNM}. In Table \ref{TMNOPOS}
we list the positions of the Mn and Tb ions within the unit cell
and these are also shown in Fig. \ref{TMOFIG}.

To start we study the operations
that leave invariant the wavevector of the incommensurate phase which
first orders as the temperature is lowered.  Experimentally\cite{KAJI}
this wavevector is found to be $(0,q,0)$, with\cite{RLU}
$q\approx 0.28(2 \pi /b)$.  
These relevant operators (see Table \ref{PBNM}) $m_x$ and $m_z$.  We
follow the approach used for MWO, but use ``actual location" Fourier
transforms.  We set ${\bf R}_f + \tauv_f \equiv {\bf r}$ in order to
use Eq. (\ref{TRANSAP}) and we need to evaluate
\begin{eqnarray}
\Lambda & \equiv & \exp \biggl( 2 \pi i ({\bf q}/b)
\cdot [ {\bf r} - [m_x]^{-1} {\bf r} \biggr)
\nonumber \\ &=&  \exp \biggl( 2 \pi (q/b) \hat j \cdot [ y \hat j  
- [m_x]^{-1} y \hat j ] \biggr) = e^{ i \pi \hat q}  
\end{eqnarray}
and
\begin{eqnarray}
\Lambda' & \equiv & \exp \biggl( 2 \pi i ({\bf q}/b) \cdot [ {\bf r}
- [m_z]^{-1} {\bf r} \biggr)
\nonumber \\ &=&  \exp \biggl( 2 \pi i (q/b) \hat j \cdot [ y \hat j
- [m_x]^{-1} y \hat j ] \biggr) = 1 \ .
\end{eqnarray}
We list, in Table \ref{NINF} the transformation table of sublattice indices
of TMO.

\begin{table}
\begin{tabular} {|| c || c | c | c|} \hline \hline
\ $\tau_i$ \ & $\ \tau_f(m_x)$ \ & \ $\tau_f(m_z)$ \ & \ $\tau_f ({\cal I})$
\ \\ \hline
1 & 2 & 3 & 1 \\
2 & 1 & 4 & 2 \\
3 & 4 & 1 & 3 \\
4 & 3 & 2 & 4 \\
5 & 8 & 5 & 7 \\
6 & 7 & 6 & 8 \\
7 & 6 & 7 & 5 \\
8 & 5 & 8 & 6 \\
\hline \hline
\end{tabular}
\caption{\label{NINF} Transformation table for sublattice indices of TMO
under various operations.}
\end{table}

Therefore the eigenvalue conditions for transformation by $m_x$ are
\begin{eqnarray}
S'_\alpha ({\bf q}, \tau_f ) &=& \xi_\alpha (m_x) S_\alpha ({\bf q},
\tau_i) \Lambda = \lambda(m_x) S_\alpha ({\bf q}, \tau_f) 
\end{eqnarray}
and
\begin{eqnarray}
S'_\alpha ({\bf q}, \tau_f ) &=& \xi_\alpha (m_z) S_\alpha ({\bf q},
\tau_i) = \lambda (m_z) S_\alpha ({\bf q}, \tau_f) \  ,
\end{eqnarray}
where $\xi_x(m_x)= -\xi_y(m_x) = -\xi_z (m_x) =1$ and
$\xi_\alpha (m_z)$ was defined in Eq. (\ref{XIEQ}).
From these equations we see that $\lambda (m_x)$ assumes the values 
$\pm \Lambda$ and $\lambda (m_z)$ the values $\pm 1$.  Then solving the
above equations leads to the results given in Table \ref{TMOq}. (These
results look different than those in Ref. \onlinecite{TMO2} because here 
the Fourier transforms are defined relative to the actual positions, whereas
there they are defined relative to the origin of the unit cell.)
 
\begin{table}
\vspace{0.2 in}
\begin{scriptsize}
\begin{tabular}{||c||c|c|c|c||}\hline\hline
Irrep & $\Gamma_1$ & $\Gamma_2$ & $\Gamma_3$ & $\Gamma_4$ \\ \hline
$\lambda (m_x)=$ & $+\Lambda$ & $-\Lambda$ & $-\Lambda$ & $+\Lambda$ \vspace{0.05cm}\\
$\lambda (m_z)=$ & $+1$ & $-1$ & $+1$ & $-1$ \vspace{0.05cm}\\
\hline

${\bf S}({\bf q},M1)$&$\begin{array}{c}
n^a_M\\- n^b_M\\ - n^c_M \end{array}$ &
$\begin{array}{c} -n^a_M\\  n^b_M\\ n^c_M\end{array}$&
$\begin{array}{c}- n^a_M\\  n^b_M\\  n^c_M\end{array}$&
$\begin{array}{c}  n^a_M\\  - n^b_M\\ - n^c_M\end{array}$
\\ \hline

${\bf S}({\bf q},M2)$&$\begin{array}{c}n^a_M\\n^b_M\\n^c_M\end{array}$&
$\begin{array}{c}n^a_M\\n^b_M\\n^c_M\end{array}$&
$\begin{array}{c}n^a_M\\n^b_M\\n^c_M\end{array}$&
$\begin{array}{c}n^a_M\\n^b_M\\n^c_M\end{array}$\\ \hline

${\bf S}({\bf q},M3)$ & $\begin{array}{c}
-  n^a_M\\  n^b_M\\ -  n^c_s\end{array} $&
$\begin{array}{c}  - n^a_M\\  n^b_M\\ - n^c_M \end{array}$ &
$\begin{array}{c} n^a_M\\ - n^b_M\\  n^c_M\end{array}$&
$\begin{array}{c} n^a_M\\ - n^b_M \\ n^c_M
\end{array}$\\ \hline

${\bf S}({\bf q},M4)$ & $\begin{array}{c} - n^a_M \\ -n^b_M \\
n^c_M\end{array}$ & $\begin{array}{c} n^a_M\\ n^b_M\\
-n^c_M \end{array}$ & $\begin{array}{c} -n^a_M\\ -n^b_M\\
 n^c_M \end{array}$ & $\begin{array}{c} n^a_M \\ n^b_M \\
- n^c_M \end{array}$\\ \hline

${\bf S}({\bf q},T1)$&$\begin{array}{c}0\\0\\ n^c_{T1}\end{array}$&
$\begin{array}{c}n^a_{T1}\\ n^b_{T1} \\0\end{array}$&
$\begin{array}{c}0 \\ 0\\ n^c_{T1} \end{array}$&
$\begin{array}{c} n^a_{T1} \\ n^b_{T1}\\ 0\end{array}$\\ \hline

${\bf S}({\bf q},T2)$ & $\begin{array}{c} 0 \\ 0\\ -  n^c_{T2} \end{array}$&
$\begin{array}{c} -  n^a_{T2}\\  n^b_{T2} \\0\end{array}$&
$\begin{array}{c} 0\\ 0 \\  n^c_{T2}\end{array}$&
$\begin{array}{c} n^a_{T2} \\  - n^b_{T2} \\ 0 \ \end{array}$\\ \hline

${\bf S}({\bf q},T3)$&$\begin{array}{c}0 \\0\\ n^c_{T2} \end{array}$&
$\begin{array}{c} n^a_{T2}\\ n^b_{T2} \\0\end{array}$&
$\begin{array}{c}0\\ 0\\ n^c_{T2}\end{array}$&
$\begin{array}{c} n^a_{T2} \\ n^b_{T2}\\ 0\end{array}$\\ \hline

${\bf S}({\bf q},T4)$ & $\begin{array}{c} 0 \\  0 \\ - n^C_{T1} \end{array}$&
$\begin{array}{c}  - n^a_{T1}\\  n^b_{T1} \\0\end{array}$&
$\begin{array}{c} 0 \\ 0 \\  n^c_{T1}  \end{array}$&
$\begin{array}{c}  n^a_{T1} \\ - n^b_{T1} \\ 0 \ \end{array}$\\
\hline \hline
\end{tabular}\caption{\label{TMOq} Spin functions (i. e. actual
position Fourier coefficients)
within the unit cell of TMO for wavevector $(0,q,0)$ which are eigenvectors of
$m_x$ and $m_z$ with the eigenvalues listed, with
$\Lambda = \exp(-i \pi \hat q)$. All the parameters are complex-valued.
The irreducible representation (irrep) is labeled as in
Ref. \onlinecite{TMO2}.  Inversion symmetry is not yet taken
into account. Note that the two Tb orbits have independent complex
amplitudes.}
\end{scriptsize}
\end{table}

Now, since the crystal is centrosymmetric,
we take symmetry with respect to spatial inversion, ${\cal I}$,
into account. As before, recall that ${\cal I}$ transports the spin to
its spatially inverted position without changing the orientation of
the spin (a pseudovector).  The change of position is equivalent to
changing the sign of the wavevector in the Fourier transform
and this is accomplished by complex conjugation.  Since the Mn ions
sit at centers of inversion symmetry, one has, for the Mn sublattices,
\begin{eqnarray}
{\cal I} {\bf S}({\bf q},n) &=& {\bf S}({\bf q},n)^* \ ,
\label{IMN} \end{eqnarray}
where the second argument specifies the sublattice, as in Table
\ref{TMNOPOS}.
In order to discuss the symmetry of the coordinates we define
$x_1=n^a_M$, $x_2=n^b_M$, $x_3=n^c_M$ and for irreps $\Gamma_1$ and
$\Gamma_3$, $x_4=n^c_{T1}$ and $x_5=n^c_{T2}$, whereas for irreps
$\Gamma_2$ and $\Gamma_4$, $x_4=n^a_{T1}$, $x_5=n^a_{T2}$,
$x_6=n^b_{T1}$, and $x_7=n^b_{T2}$.  Thus Eq. (\ref{IMN}) gives
\begin{eqnarray}
{\cal I} x_n &=& x_n^* \ , \hspace{1 in} n=1, \ 2, \ 3 \ .
\label{IIIEQ} \end{eqnarray}
For the Tb ions ${\cal I}$ interchanges sublattices \#5 and \#7 and
interchanges sublattices \#6 and \#8. So we have
\begin{eqnarray}
{\cal I} {\bf S}({\bf q},5) &=& {\bf S}({\bf q},7)^* \nonumber \\
{\cal I} {\bf S}({\bf q},6) &=& {\bf S}({\bf q},8)^* \ .
\label{ITB} \end{eqnarray}
Therefore we have
\begin{eqnarray}
{\cal I} x_4 = x_5^* \ , \hspace{1 in} {\cal I} x_6 = x_7^* \ .
\label{IVEQ} \end{eqnarray}

Now we use the invariance of the free energy under ${\cal I}$ to write
\begin{eqnarray}
F_2 &=& \sum_{X,\alpha;Y,\beta} S_\alpha({\bf q},X)^* F_{nm} 
S_\beta ({\bf q},Y) \nonumber \\
&=& \sum_{m,n} x_n^* G_{nm} x_m \nonumber \\
&=& \sum_{m,n} [{\cal I} x_n^*] G_{nm} [{\cal I} x_m] \ ,
\label{TBGEQ} \end{eqnarray}
where the matrix ${\bf G}$ is Hermitian.

For irreps $\Gamma_1$ and $\Gamma_3$ the matrix ${\bf G}$ in Eq. (\ref{TBGEQ})
couples five variables.  Equation (\ref{IIIEQ}) implies that the upper left 
$3 \times 3$ submatrix of ${\bf G}$ is real.  Equations (\ref{IIIEQ}) and
(\ref{IVEQ}) imply that $G_{n,4}= G_{5,n}$ and for $n=1,2,3$. We find
that ${\bf G}$   assumes the form
\begin{eqnarray}
{\bf G} &=& \left[ \begin{array} {c c c c c}
a & b & c & \alpha & \alpha^* \\
b & d & e & \beta &  \beta^* \\
c & e & f & \gamma & \gamma^* \\
\alpha^* & \beta^* & \gamma^* & g & \delta \\
\alpha & \beta & \gamma & \delta^* & g  \\ \end{array} \right] \ ,
\label{gMATRIX} \end{eqnarray} 
where the Roman letters are real valued and the Greek are
complex valued.  As shown in the appendix, the
form of this matrix ensures that the
critical eigenvector can be taken to be of the form
\begin{eqnarray}
\psi &=& (n^a_M, n^b_M, n^c_M, n^c_{T1}, {n^c_{T1}}^* ) \equiv 
(r, s, t; \rho, \rho^* ) \ ,
\end{eqnarray}
where the Roman letters are real and the Greek ones complex.
Of course, because the vector can be complex, we should include
an overall  phase factor (which amounts to arbitrarily placing
the origin of the incommensurate structure), so that more generally
\begin{eqnarray}
\psi &=& e^{i \phi} (r, s, t; \rho, \rho^* ) \ .
\end{eqnarray}

For irreps $\Gamma_2$ and $\Gamma_4$ the matrix ${\bf G}$ in
Eq. (\ref{TBGEQ}) couples the seven variables listed just below Eq.
(\ref{IMN}).  Equations (\ref{IIIEQ}) and (\ref{IVEQ}) imply
that $G_{n,4}=G_{5,n}$ and $G_{n,6}=G_{7,n}$ for
$n=1,2,3$.  Therefore ${\bf G}$ assumes the form
\begin{eqnarray}
{\bf G} &=& \left[ \begin{array} {c c c c c c c}
a & b & c & \alpha & \alpha^* & \xi & \xi^* \\
b & d & e & \beta &  \beta^* & \eta & \eta^* \\
c & e & f & \gamma & \gamma^* & \kappa & \kappa^* \\
\alpha^* & \beta^* & \gamma^* & g & \delta & \mu & \nu \\
\alpha & \beta & \gamma & \delta^* & g & \nu^* & \mu^*  \\
\xi^* & \eta^* & \kappa^* & \mu^* & \nu & h & \rho \\
\xi & \eta & \kappa & \nu^* & \mu & \rho^* & h \\ \end{array} \right] \ ,
\label{GMATRIX} \end{eqnarray} 
where Roman letters are real and Greek are complex.  As shown in the
appendix, this form ensures that the eigenvectors are of the form
\begin{eqnarray}
\psi &=& (n^a_M, n^b_M, n^c_M, n^a_{T1}, n^a_{T2}, m^b_{T1}, n^b_{T2} )
\nonumber \\ &=&
e^{i \phi} (r, s, t; \tau, \tau^* , \sigma , \sigma^* ) \ .
\label{GEIGEN} \end{eqnarray}
These results are summarized in Table \ref{TMOqq}. 
Note that the use of inversion symmetry fixes most of the phases and relates
the amplitudes of the two Tb orbits, thereby eliminating almost
half the fitting parameters.\cite{TMO2}

\begin{table}
\vspace{0.2 in}
\begin{scriptsize}
\begin{tabular}{||c||c|c|c|c||}\hline\hline
Irrep & $\Gamma_1$ & $\Gamma_2$ & $\Gamma_3$ & $\Gamma_4$ \\ \hline
$\lambda (m_x)=$ & $+\Lambda$ & $-\Lambda$ & $-\Lambda$ & $+\Lambda$ \vspace{0.05cm}\\
$\lambda (m_z)=$ & $+1$ & $-1$ & $+1$ & $-1$ \vspace{0.05cm}\\
\hline

${\bf S}({\bf q},M1)$&$\begin{array}{c}
 r \\- s \\ - t  \end{array}$ &
$\begin{array}{c} -r \\  s \\  t \end{array}$&
$\begin{array}{c}- r \\  s \\  t \end{array}$&
$\begin{array}{c}  r\\ - s \\ - t \end{array}$
\\ \hline

${\bf S}({\bf q},M2)$&$\begin{array}{c}r \\ s \\ t \end{array}$&
$\begin{array}{c}r \\ s \\ t \end{array}$&
$\begin{array}{c} r \\ s \\ t \end{array}$&
$\begin{array}{c} r \\ s \\ t \end{array}$\\ \hline

${\bf S}({\bf q},M3)$ & $\begin{array}{c}
-  r \\  s \\ -  t \end{array} $&
$\begin{array}{c}  - r \\ s \\ - t  \end{array}$ &
$\begin{array}{c} r \\ - s \\  t \end{array}$&
$\begin{array}{c}  r \\ - s  \\ t  \end{array}$\\ \hline

${\bf S}({\bf q},M4)$ & $\begin{array}{c} - r  \\ - s \\
 t \end{array}$ & $\begin{array}{c} r \\ s \\
- t  \end{array}$ & $\begin{array}{c} -r \\ - s\\
 t  \end{array}$ & $\begin{array}{c} r \\ s  \\
- t  \end{array}$\\ \hline

${\bf S}({\bf q},T1)$&$\begin{array}{c}0\\0\\ \rho \end{array}$&
$\begin{array}{c} \tau \\ \sigma \\0\end{array}$&
$\begin{array}{c}0 \\ 0\\ \rho \end{array}$&
$\begin{array}{c} \tau \\ \sigma \\ 0\end{array}$\\ \hline

${\bf S}({\bf q},T2)$ & $\begin{array}{c} 0 \\ 0\\ -  \rho^* \end{array}$&
$\begin{array}{c}  - \tau^* \\ \sigma^* \\0\end{array}$&
$\begin{array}{c} 0\\ 0 \\  \rho^* \end{array}$&
$\begin{array}{c} \tau^* \\  - \sigma^* \\ 0 \ \end{array}$\\ \hline

${\bf S}({\bf q},T3)$&$\begin{array}{c}0 \\0\\ \rho^* \end{array}$&
$\begin{array}{c} \tau^* \\ \sigma^* \\0\end{array}$&
$\begin{array}{c}0\\ 0\\ \rho^* \end{array}$&
$\begin{array}{c} \tau^* \\ \sigma^* \\ 0\end{array}$\\ \hline

${\bf S}({\bf q},T4)$ & $\begin{array}{c} 0 \\  0 \\ - \rho \end{array}$&
$\begin{array}{c}  - \tau \\ \sigma \\0\end{array}$&
$\begin{array}{c} 0 \\ 0 \\  \rho  \end{array}$&
$\begin{array}{c} \tau \\ - \sigma \\ 0 \ \end{array}$\\
\hline \hline
\end{tabular}
\caption{\label{TMOqq} As Table \ref{TMOq}. 
Apart from an overall phase $\phi_\Gamma$ for each irrep, inversion
symmetry restricts all the manganese Fourier coefficients to be
real and all the Tb coefficients to have the indicated phase relations.}
\end{scriptsize}
\end{table}

\subsubsection{Order Parameters}

We now introduce order parameters $\sigmav_n ({\bf q}) \equiv
\sigma_n e^{i \phi_n}$ for irrep $\Gamma_n$ in
terms of which we can write the spin distribution.
For instance under $\Gamma_3$ one has
\begin{eqnarray}
S_x({\bf r}, M1) &=& -2r \sigma_3 \cos (qy + \phi_3 ) \nonumber \\
S_y({\bf r}, M1) &=& 2s  \sigma_3 \cos (qy + \phi_3 ) \nonumber \\
S_z({\bf r}, M1) &=& 2t  \sigma_3 \cos (qy + \phi_3 ) \nonumber \\
S_x({\bf r}, M2) &=& 2r \sigma_3 \cos (qy + \phi_3 ) \nonumber \\
S_y({\bf r}, M2) &=& 2s  \sigma_3 \cos (qy + \phi_3 ) \nonumber \\
S_z({\bf r}, M2) &=& 2t  \sigma_3 \cos (qy + \phi_3 ) \nonumber \\
S_x({\bf r}, T1) &=& S_y({\bf r}, T1) = 0  \nonumber \\
S_z({\bf r}, T1) &=& 2 \rho \sigma_3 \cos(qy + \phi_3 + \phi_\rho) \nonumber \\
S_x({\bf r}, T2) &=& S_y({\bf r}, T2) = 0  \nonumber \\
S_z({\bf r}, T2) &=& 2 \rho \sigma_3 \cos(qy + \phi_3 - \phi_\rho) \ ,
\label{TMOSPIN} \end{eqnarray}
where we set $\rhov = \rho e^{i \phi_\rho}$ and the parameters are normalized by
\begin{eqnarray}
r^2 + s^2 + t^2 + |\rho |^2 &=& 1 \ .
\end{eqnarray}
In Eq. (\ref{TMOSPIN}) ${\bf r} \equiv (x,y,z)$ is the actual position
of the spin in question.  From Table \ref{TMOq} one can obtain the
symmetry properties of the order parameters for each irrep.  For instance
\begin{eqnarray}
m_x \sigmav_1({\bf q}) &=& + \Lambda \sigmav_1 ({\bf q}) \ , \ \ \ \ \
m_z \sigmav_1({\bf q}) = + \sigmav_1  ({\bf q})\nonumber \\
m_x \sigmav_2({\bf q}) &=& - \Lambda \sigmav_2  ({\bf q})\ , \ \ \ \ \
m_z \sigmav_2({\bf q}) = - \sigmav_2  ({\bf q})\nonumber \\
m_x \sigmav_3({\bf q}) &=& - \Lambda \sigmav_3  ({\bf q})\ , \ \ \ \ \
m_z \sigmav_3({\bf q}) = \sigmav_3  ({\bf q})\nonumber \\
m_x \sigmav_4({\bf q}) &=& + \Lambda \sigmav_4  ({\bf q})\ , \ \ \ \ \
m_z \sigmav_4({\bf q}) = - \sigmav_4 ({\bf q})
\end{eqnarray}
and
\begin{eqnarray}
{\cal I} \sigmav_n ({\bf q})  &=& \sigmav_n^*  ({\bf q}) \ .
\end{eqnarray}

Note that in contrast to the case of NVO, inversion
symmetry does not fix all the phases. However, it again
drastically reduces the number of possible magnetic
structure parameters which have to be determined.
In particular, it is only by using inversion that one
finds that the magnitudes of the Fourier coefficients
of the two distinct Tb sites have to be the same.
Note that if we choose the origin so that $\phi=0$ (which amounts
to renaming the origin so that that becomes true), then
we recover inversion symmetry (taking account that
inversion interchanges terbium sublattice \#3 and \#1).
One can determine that the spin structure is inversion
invariant when one condenses a single representation.

The result of Table 5 applies other manganates provided
their wavevector is also of the form $(0,q_y,0)$. This
includes YMnO$_3$\cite{MUNOZ1} and HoMnO$_3$.\cite{MUNOZ2,BRINKS}
Both these systems order into an incommensurate structure at
about $T_c \approx 42$K.  The Y compound has a second lower-temperature
incommensurate phase, whereas the Ho compound has a lower-temperature
commensurate phase.

\subsection{TbMn$_2$O$_5$}

\begin{figure}[ht]
\begin{center}
\includegraphics[width=8cm]{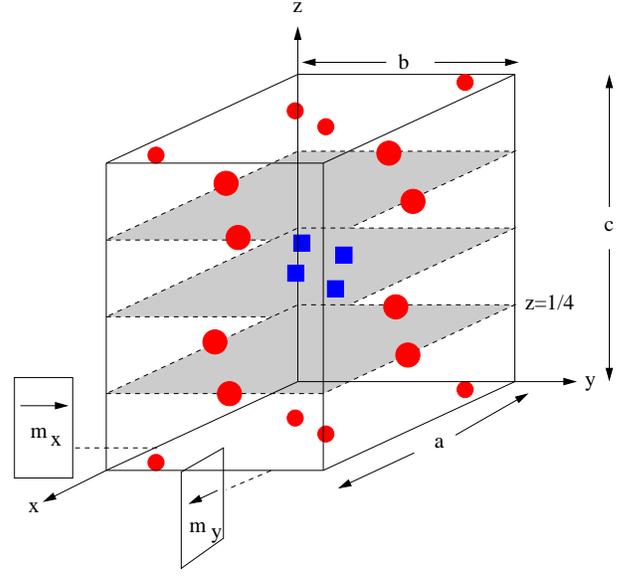}
\includegraphics[width=8cm]{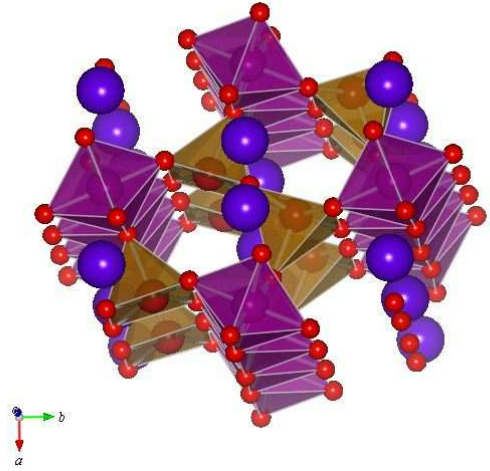}
\caption{(Color online). Two representations of TbMn$_2$O$_5$.
Top: Mn sites (on-line red) with smaller circles
Mn$^{3+}$ and larger circles $^{4+}$ and Tb sites
(squares, on-line blue) in the primitive unit cell
of TbMn$_2$O$_5$.  The Mn$^{+4}$ sites are in the shaded planes at
$z=n\pm \delta$ with $\delta \approx 0.25$ and the Mn$^{+3}$ sites
are in planes $z=n$, where $n$ is an integer.  The Tb ions are in
the planes $z= n + \oh$.  The glide plane $m_x$ is indicated by
the mirror plane at $x=3/4$ followed by a translation (indicated
by the arrow) of $b/2$ along the $y$-axis and similarly for the
glide plane $m_y$. Bottom: Perspective view.  Here the Mn$^{3+}$
are inside oxygen pyramids of small balls
and the Mn$^{4+}$ are inside oxygen octahedra.}
\label{25FIG}
\end{center}
\end{figure}

The space group of TbMn$_2$O$_5$ (TMO25)
is Pbam (\#55 in Ref. \onlinecite{HAHN})
and its general positions are listed in Table \ref{PBAM}.
The positions of the magnetic ions are given in Table \ref{T25TAB}
and are shown in Fig. \ref{25FIG}.

\begin{table}
\vspace{0.2 in}
\begin{tabular} {|| c | c ||}
\hline \hline
$E {\bf r} =(x,y,z) \ \ $ 
& $2_x {\bf r} =(x + \oh  , \overline y + \oh , \overline z )\ \ $ \\
$2_z=(\overline x,\overline y , z )\ \ $
& $2_y=(\overline x + \oh , y + \oh , \overline z )\ \ $ \\
${\cal I}{\bf r}=(\overline x,\overline y,\overline z)\ \ $ 
& $m_x{\bf r}=(\overline x + \oh , y + \oh, z )\ \ $ \\
$m_z{\bf r}=(x, y ,\overline z )\ \ $
& $m_y{\bf r}=(x + \oh , \overline y + \oh , z )\ \ $ \\
\hline \hline 
\end{tabular}
\caption{\label{PBAM}As Table \ref{PBAM}. General Positions for Pbam.}
\end{table}

\begin{table} 
\vspace{0.2 in}
\begin{tabular} {|| c||c| c||} \hline \hline
Mn$^{3+}$ & $(1)=(x,y,0)$ & $(2)=(\overline x, \overline y,0)$ \\
          & $(3)(\overline x + \oh, y+ \oh,0)$ & $(4)=(x+\oh ,\overline y+ \oh,0)$ \\
\hline
Mn$^{4+}$ & $(5)=(\oh,0,z)$ & $(6)=(0, \oh , z)$  \\
          & $(7)=(\oh , 0, \overline z)$ & $(8)=(0, \oh , \overline z )$ \\
\hline
RE        & $(9)=(X,Y,\oh)$ & $(10)=(\overline X, \overline Y, \oh )$ \\
          & $(11)=( \overline X + \oh, Y+\oh, \oh)$ & $(12)=(X+\oh ,
\overline Y + \oh, \oh)$ \\
\hline \hline
\end{tabular}
\caption{\label{T25TAB} Positions of the magnetic ions of TbMn$_2$O$_5$ in 
the Pbam structure.  Here $x=0.09$, $y=-0.15$, $z=0.25$,\cite{BUIS1}
$X=0.14$, and $Y=0.17$.\cite{BUIS2}.  All these values are
taken from the isostructural compound HoMn$_2$O$_5$.}
\end{table}

We will address the situation just below the ordering temperature of
43K.\cite{TB25} We take the ordering wavevector to be \cite{TB25}
to be $(\oh, 0, q)$ with\cite{RLU} $q \approx 0.306$.  (This may be an
approximate value.\cite{BLAKE}) Initially we assume that the
possible spin configurations consistent
with a continuous transition at such a wavevector
are eigenvectors of the operators $m_x$ and $m_y$ which
leave the wavevector invariant. We proceed as for TMO.
We use the unit cell Fourier transforms and write
the eigenvector conditions for transformation by $m_x$ as 
\begin{eqnarray}
S_\alpha({\bf q}, \tau_f)' &=& \xi_\alpha (m_x) S_\alpha ({\bf q},\tau_i)
e^{i {\bf q} ({\bf r}_f - {\bf R}_i)} \nonumber \\ 
&=& \lambda_x S_\alpha ({\bf q}, \tau_f) \ ,
\label{25XEQ} \end{eqnarray}
where $\tau_i$ and ${\bf R}_i$ are respectively the sublattice indices
and unit cell locations before transformation and 
$\tau_f$ and ${\bf R}_f$ are those after transformation.
The eigenvalue equation for transformation by $m_y$ is
\begin{eqnarray}
S_\alpha({\bf q}, \tau_f)' &=& \xi_\alpha (m_y) S_\alpha ({\bf q},\tau_i)
e^{i {\bf q} ({\bf r}_f - {\bf R}_i)} \nonumber \\
&=& \lambda_y S_\alpha ({\bf q}, \tau_f) \ .
\label{25YEQ} \end{eqnarray}

If one attempts to construct spin functions which are simultaneously
eigenfunctions of $m_x$ and $m_y$ one finds that {\it these equations
yield no solution}.  While it is, of course,
true that the operations $m_x$ and $m_y$ take an eigenfunction into an
eigenfunction, it is only for irreps of dimension one that the initial and
final eigenfunctions are the same, as we have assumed.  The present case,
when the wavevector is at the edge of the Brillouin zone is analogous
to the phenomenon of ``sticking" where, for nonsymmorphic space group
(i. e. those having a screw axis or a glide plane) the energy bands
(or phonon spectra) have an almost mysterious degeneracy at the zone
boundary\cite{VH} and the only active irrep has dimension two. This means
that the symmetry operations induce transformations within the subspace
of pairs of eigenfunctions.  We now determine such pairs of eigenfunctions
by a straightforward approach which do not require any knowledge of group
theory.  Here we explicitly consider the symmetries of the matrix
$\chi^{-1}$ for the quadratic terms in the free energy which here
is a $36 \times 36$ dimensional matrix, which we write as
\begin{eqnarray}
\chiv^{-1} &=& \left[ \begin{array} {c c c}
{\bf M}^{(xx)} & {\bf M}^{(xy)} & {\bf M}^{(xz)} \\
{{\bf M}^{(xy)}}^\dagger & {\bf M}^{(yy)} & {\bf M}^{(yz)} \\
{{\bf M}^{(zx)}}^\dagger & {{\bf M}^{(yz)}}^\dagger & {\bf M}^{(zz)} \\
\end{array} \right] \ ,
\label{SUPER} \end{eqnarray}
where ${\bf M}^{(ab)}$ is a 12 dimensional submatrix which describes
coupling between $a$-component and $b$-component spins and is
indexed by sublattice indices $\tau$ and $\tau'$  The symmetries we
invoke are operations of the screw axes, $m_x$ and $m_y$ which
conserve wavevector (to within a reciprocal lattice vector),
and ${\cal I}$, whose effect is usually ignored.  To guide the reader
through the ensuing calculation we summarize the main steps.
We first analyze separately the sectors involving the $x$, $y$, and
$z$ spin components.  We
develop a unitary transformation which takes ${\bf M}^{(\alpha \alpha)}$
into a matrix all of whose elements are real.  This fixes the
phases within the 12 dimensional space of the $\alpha$ spin components
within the unit cell (assuming these relations are not invalidated
by the form of ${\bf M}^{(\alpha \beta)}$, with $\alpha \not= \beta$).
The relative phases between different spin components is fixed by
showing that the unitary transformation introduced above leads to
${\bf M}^{(xy)}$ having all real-valued matrix elements and
${\bf M}^{(xz)}$ and ${\bf M}^{(yz)}$ having all purely imaginary
matrix elements.  The conclusion, then, is that the phases in
the sectors of $x$ and $y$ components are coupled in phase and
the sector of $z$ components are out of phase with the $x$ and $y$ 
components.
    
\begin{table}
\begin{tabular} { ||c || c || c | c || c | c || c | c || c || } \hline \hline
$n_i$ & $m_x$ & \multicolumn{2} {|c||} {$m_y^{\rm a}$} 
& \multicolumn{2} {|c||} {$m_xm_y^{\rm a}$} & \multicolumn{2} {| c ||}
{${\cal I}^{\rm b}$} & \ $m_x m_y{\cal I}$\ \\ \hline
&\ $n_f\ $ \ & $n_f\ $ \ & 
\ $e^{i \phi}$ \ & $n_f$ & \ $e^{i \phi}$ \ &
\ $n_f$ \ & \ $e^{i \phi'}$ & $n_f$ \ \\ \hline
1 & 3 & $4$ & $1$         & 2 & $1$    & $2$  & $1$& 1 \\ 
2 & 4 & $3$ & $1$         & 1 & $1$    & $1$  & $1$ & \ 2 \ \\
3 & 1 & $2$ & \ \ $-1$    & 4 & $-1$    & $4$\ \  & \ \ $-1$ \ \ & $3$ \\
4 & 2 & $1$ & $-1$        & 3 & $-1$    & $3$  & $-1$ & $4$ \\
5 & 6 & $6$ & $-1$        & 5 & $-1$    & $7$  & \ \ $-1$ \ \ & $7$ \\
6 & 5 & $5$ & $1$         & 6 & $1$    & $8$  & $1$ & $8$ \\
7 & 8 & $8$ & $-1$        & 7 & $-1$    & $5$  & $-1$ & $8$ \\
8 & 7 & $7$ & $1$         & 8 & $1$    & $6$  & $1$ & $7$ \\
9 &11 &$12$ & $1$         & 10& $1$    & $10$ & $1$ & $9$ \\
10&12 &$11$ & $1$         & 9 & $1$    & $9$  & $1$ & $10$ \\
11& 9 &$10$ & $-1$        &12 & $-1$    & $12$ & $-1$ & $11$ \\
\ 12\ &10 & $9$ & $-1$        &11 & $-1$    & $11$ & $-1$ & $12$ \\
\hline \end{tabular}
\caption{\label{NINF25}Transformation table for sublattice indices
with associated factors for TMO25 under various operations.
as defined by Eq. (\ref{TRANSUCO}).  For $m_x$, one has
$\exp [i {\bf q} \cdot ({\bf R}_f - {\bf R}_i)]=1$ for all cases
and for $m_x m_y {\cal I}$ the analogous factor is $+1$ in all cases
and this operator relates $S_\alpha ({\bf q}, \tau)$ and
$S_\alpha ({\bf q}, \tau)^*$.  NOTE: This table does not include the
factor of $\xi_\alpha ({\cal O})$ which may be associated with
an operation.}

\noindent a) $\phi = {\bf q} \cdot ({\bf R}_f - {\bf R}_i)$, as
required by Eq. (\ref{TRANSUC}).

\noindent b) $\phi' =  {\bf q} \cdot (\tauv_i + \tauv_f)$, as
required by Eq. (\ref{TRANSIUC}).

\end{table}

\subsubsection{$x$ Components}

As a preliminary, in Table \ref{NINF25}  we list the effect of the
symmetry operations on the sublattice index.  When these symmetries
are used, one finds the $12 \times 12$ submatrix of ${\bf M}^{(xx)}$
which couples only the $x$-components of spins assumes the form
\begin{scriptsize}
\begin{eqnarray}
\left[ \begin{array} {| c c c c | c c | c c | c c c c|}
A & g & h & 0 &  \alpha & \beta & \alpha^* & \beta^* & a & b & c & d \\
g & A & 0 & -h & -\alpha & \beta & -\alpha^* & \beta^* & b & a & -d & -c \\
h  & 0 & A & g & \beta & \alpha & \beta^* & \alpha^* & c & d & a & b \\
0 & -h & g & A & \beta & -\alpha & \beta^* & -\alpha^* & -d & -c & b  & a \\ 
\hline
\alpha^* & -\alpha^* & \beta^* & \beta^* & B & 0  & \epsilon & 0 & \gamma & -\gamma & \delta & \delta \\
\beta^* & \beta^* & \alpha^* & -\alpha^* & 0 & B & 0 & \epsilon  & \delta & \delta & \gamma & -\gamma \\
\hline
\alpha & -\alpha & \beta & \beta & \epsilon^* & 0 & B  & 0 & \gamma^* & -\gamma^* & \delta^* & \delta^* \\
\beta & \beta & \alpha & -\alpha & 0  & \epsilon^* & 0 & B & \delta^* & \delta^* & \gamma^* & -\gamma^* \\
\hline
a & b  & c & -d & \gamma^* & \delta^* & \gamma & \delta & C & e & f & 0 \\
b & a & d & -c & -\gamma^* & \delta^* & -\gamma & \delta & e & C & 0 & -f \\
c & -d & a & b & \delta^* & \gamma^* & \delta & \gamma & f & 0 & C & e \\
d & -c & b & a & \delta^* & -\gamma^* & \delta & -\gamma & 0 & - f & e & C \\
\end{array} \right] \  ,
\label{MATEQ}
\end{eqnarray}
\end{scriptsize}

\noindent
where Roman letters are real quantities and Greek ones complex. (In this matrix the
lines are used to separate different Wyckoff orbits.)  The numbering of 
the rows and columns follows from Table \ref{T25TAB}.  I give a few examples of
how symmetry is used to get this form. Consider the term $T_1$, where 
\begin{eqnarray}
T_1 &=& \chi^{-1}_{1,5} S_x(-{\bf q},1) S_x({\bf q},5) \ .
\end{eqnarray}
Using Table \ref{NINF25} we transform this by $m_x$ into
\begin{eqnarray}
T_1' &=& \chi^{-1}_{1,5} S_x(-{\bf q},3) S_x({\bf q},6) \ ,
\end{eqnarray}
which says that the 1,5 matrix element is equal to the 3,6 matrix element.
(Note that in writing down $T_1'$ we did not need to worry about
$\xi_\alpha$, since this factor comes in squared as unity.)
Likewise if we transform by $m_y$ we get
\begin{eqnarray}
T_1' &=& \chi^{-1}_{15} [-S_x(-{\bf q},4)] [S_x ({\bf q},6)] \ ,
\end{eqnarray}
which says that the 1,5 matrix element is equal to the negative of the
4,6 matrix element.  If we transform by $m_xm_y$ we get
\begin{eqnarray}
T_1' &=& \chi^{-1}_{1,5} [S_x(-{\bf q},2)] [-S_x({\bf q},5)] \ ,
\end{eqnarray}
which says that the 1,5 matrix element is equal to the negative of the
2,5 matrix element.  To illustrate the effect of ${\cal I}$ on $T_1$ we write
\begin{eqnarray}
T_1' &=& \chi^{-1}_{1,5} [S_x({\bf q},2)] [-S_x(-{\bf q},7)] \ ,
\end{eqnarray}
so that the 1,5 element is the negative of the 7,2 element. 
From the form of the matrix in Eq. (\ref{MATEQ}) (or equivalently
referring to Table \ref{TMOTAB} in Appendix B), we see that
we bring this matrix into block diagonal form by introducing the
wavefunctions for $S_x({\bf q},\tau)$,
\begin{scriptsize}
\begin{eqnarray}
\begin{array} {||c || c c c c | c c | c c | c c c c||} \hline
\tau = & 1 & 2 & 3 & 4 & 5 & 6 & 7 & 8 & 9 & 10 & 11 & 12 \\  \hline
\sqrt 2 O^{(x,1)}_{1, \tau} =
& 1 & 0 & 1 & 0 &  0 & 0 & 0 & 0 & 0 & 0 & 0 & 0 \\
\sqrt 2 O^{(x,1)}_{2,\tau} =
& 0 & 1 & 0 & 1 &  0 & 0 & 0 & 0 & 0 & 0 & 0 & 0 \\
2 O^{(x,1)}_{3, \tau} =
& 0 & 0 & 0 & 0 & 1 & 1 & 1 & 1 & 0 & 0 & 0 & 0 \\
2 O^{(x,1)}_{4, \tau} =
& 0 & 0 & 0 & 0 & i & i & -i & -i & 0 & 0 & 0 & 0 \\
\sqrt 2 O^{(x,1)}_{5, \tau} =
& 0 & 0 & 0 & 0 & 0 & 0 & 0 & 0 & 1 & 0 & 1 & 0 \\
\sqrt 2 O^{(x,1)}_{6, \tau} =
& 0 & 0 & 0 & 0 & 0 & 0 & 0 & 0 & 0 & 1 & 0 & 1 \\
\hline \end{array} \ .
\label{SUBSPACE} \end{eqnarray}
\end{scriptsize}
The superscripts $\alpha, n$ on ${\bf O}$ label, respectively, the Cartesian
component and the column of the irrep according to which the wavefunction
transforms.  The subscripts $m,\tau$ label, respectively, the index number
of the wavefunction and the sublattice label.  In this subspace
$\langle {\bf O}^{(x,1)}_n| M^{(xx)} | {\bf O}^{(x,1)}_m \rangle
\equiv \langle n | M^{(xx)} | m \rangle$ is
\begin{scriptsize}
\begin{eqnarray}
\left[ \begin{array} { c | c | c | c | c | c } \hline
A+h & g & \alpha'+\beta' & - \alpha'' - \beta'' & a+c & b+d \\
g & A-h & \beta' -\alpha' &  \alpha'' - \beta'' & b - d & a - c \\
\alpha'+\beta' & \beta' - \alpha' & B+\epsilon' 
& \epsilon" & \delta' + \gamma' & \delta' - \gamma' \\
- \alpha'' - \beta'' & \alpha'' - \beta'' & \epsilon''
& B-\epsilon' & \delta'' + \gamma'' & \delta'' - \gamma'' \\
a + c & b - d & \delta' + \gamma' & \delta'' + \gamma'' & C+f & e \\
b + d & a - c & \delta'-\gamma' & \delta'' - \gamma'' & e & C - f \\
\hline \end{array} \right] \ ,
\label{BLOCK} \end{eqnarray}
\end{scriptsize}
where the coefficients are separated into real and imaginary parts as
$\sqrt 2 \alpha = \alpha' + i \alpha''$, $\sqrt 2 \beta = \beta' + i \beta''$
$\sqrt 2 \gamma = \gamma' + i \gamma''$, and $\sqrt 2 \delta = \delta' + i \delta''$.
There are no nonzero matrix elements between wavefunctions which
transform according to different columns of the irrep.

The partners of these functions can be found from
\begin{eqnarray}
{\bf O}^{(x,2)}_n &=& m_y {\bf O}^{(x,1)}_n \ ,
\end{eqnarray}
so that, using Table \ref{NINF25} and including the factor $\xi_\alpha$, we get
\begin{scriptsize}
\begin{eqnarray}
\begin{array} {||c|| c c c c | c c | c c | c c c c||} \hline
\tau = & 1 & 2 & 3 & 4 & 5 & 6 & 7 & 8 & 9 & 10 & 11 & 12 \\ \hline
\sqrt 2 O^{(x,2)}_{1,\tau} =
& 0 & 1 & 0 & -1 &  0 & 0 & 0 & 0 & 0 & 0 & 0 & 0 \\
\sqrt 2 O^{(x,2)}_{2,\tau} =
& 1 & 0 & -1 & 0 &  0 & 0 & 0 & 0 & 0 & 0 & 0 & 0 \\
2 O^{(x,2)}_{3,\tau} =
& 0 & 0 & 0 & 0 & -1 & 1 & -1 & 1 & 0 & 0 & 0 & 0 \\
2 O^{(x,2)}_{4,\tau} =
& 0 & 0 & 0 & 0 & -i & i & i & -i & 0 & 0 & 0 & 0 \\
\sqrt 2 O^{(x,2)}_{5,\tau}=
& 0 & 0 & 0 & 0 & 0 & 0 & 0 & 0 & 0 & 1 & 0 & -1 \\
\sqrt 2 O^{(x,2)}_{6,\tau} =
& 0 & 0 & 0 & 0 & 0 & 0 & 0 & 0 & 1 & 0 & -1 & 0 \\
\hline \end{array} \ .
\label{SUBSPACE2} \end{eqnarray}
\end{scriptsize}
Within this subspace the matrix $\langle n | M^{(xx)} | m \rangle$
is the same as in Eq. (\ref{BLOCK}) because
\begin{eqnarray}
\langle n| m_y^{-1} M^{(xx)} m_y | m \rangle &=&
\langle n| M^{(xx)} | m \rangle  \ .
\end{eqnarray}

These functions transform as expected for a two dimensional irrep, namely,
\begin{eqnarray}
m_x \left[ \begin{array} {c} O^{(x,1)}_n \\ O^{(x,2)}_n \\ \end{array} \right]
&=& \left[ \begin{array} {c} O^{(x,1)}_n \\ -O^{(x,2)}_n \\ \end{array} \right] \nonumber \\
m_y \left[ \begin{array} {c} O^{(x,1)}_n \\ O^{(x,2)}_n \\ \end{array} \right]
&=& \left[ \begin{array} {c} O^{(x,2)}_n \\ -O^{(x,1)}_n \\ \end{array} \right] \ .
\label{MXTRANS} \end{eqnarray}

We will refer to the transformed coordinates of Eqs. (\ref{SUBSPACE}) 
and (\ref{SUBSPACE2}) as ``symmetry adapted coordinates.''  The fact that
the model-specific matrix that couples them is real, means that
the critical eigenvector is a linear combination of symmetry
adapted coordinates with {\it real} coefficients.
 
\subsubsection{$y$ Components}

The $12 \times 12$ matrix $M^{(yy)}$ coupling $y$ components of spin has exactly
the same form as that given in Eq. (\ref{MATEQ}), although the values
of the constants are unrelated.  This is because
here one has $\xi_y^2=1$ in place of $\xi_x^2=1$.  Therefore the associated
wavefunctions can be expressed just as in Eqs. (\ref{SUBSPACE}) and
(\ref{SUBSPACE2}) except that all the superscripts are changed from $x$ to $y$\
and $\tau$ now labels $S_y({\bf q},\tau)$.  However,
the transformation of the $y$ components rather than the $x$ components, 
requires replacing $\xi_x$ by $\xi_y$ which induces sign changes, so that
\begin{eqnarray}
m_x \left[ \begin{array} {c} O^{(y,1)}_n \\ O^{(y,2)}_n \\ \end{array} \right]
&=& \left[ \begin{array} {c} -O^{(y,1)}_n\\ O^{(y,2)}_n \\ \end{array} \right] \nonumber \\
m_y \left[ \begin{array} {c} O^{(y,1)}_n \\ O^{(y,2)}_n \\ \end{array} \right]
&=& \left[ \begin{array} {c} -O^{(y,2)}_n \\ O^{(y,1)}_n \\ \end{array} \right] \ .
\end{eqnarray}
We want to construct wavefunctions in this sector which transform just
like the $x$ components, so that they can be appropriately combined with
the wavefunctions for the $x$-components.  In view of Eq. (\ref{MXTRANS}) we set
\begin{eqnarray}
O^{(y,1)}_{n,\tau} &=& O^{(x,2)}_{n,\tau} \ , \ \ \  O^{(y,2)}_{n,\tau} =O^{(x,1)}_{n,\tau} \ .
\end{eqnarray}
So the coefficients for ${\bf O}^{(y,1)}_n$ are given by Eq. (\ref{SUBSPACE2}) and those for
${\bf O}^{(y,2)}_n$ by Eq. (\ref{SUBSPACE}).  These wavefunctions are constructed
to transform exactly as those for the $x$ components.

\subsubsection{$z$ Components}

Similarly, we consider the effect of the transformations of the $z$
components. In this case we take account of the factor $\xi_z$ to get 
\begin{eqnarray}
m_x \left[ \begin{array} {c} O^{(z,1)}_n \\ O^{(z,2)}_n \\ \end{array} \right]
&=& \left[ \begin{array} {c} -O^{(z,1)}_n \\ O^{(z,2)}_n \\ \end{array} \right]
\nonumber \\
m_y \left[ \begin{array} {c} O^{(z,1)}_n \\ O^{(z,2)}_n \\ \end{array} \right]
&=& \left[ \begin{array} {c} O^{(z,2)}_n \\ -O^{(z,1)}_n \\ \end{array} \right] \ .
\end{eqnarray}
We now construct wavefunctions in this sector which transform just
like the $x$ components.  In view of Eq. (\ref{MXTRANS}) we set
\begin{eqnarray}
O^{(z,1)}_{n,\tau} &=&O^{(x,2)}_{n,\tau} \ , \ \ \  O^{(z,2)}_{n,\tau} = - O^{(x,1)}_{n,\tau} \ ,
\end{eqnarray}
So the coefficients for ${\bf O}^{(z,1)}_n$ are given by Eq. (\ref{SUBSPACE2}) and those for
${\bf O}^{(z,2)}_n$ are the negatives of those of Eq. (\ref{SUBSPACE}).  These wavefunctions
are constructed to transform exactly as those for the $x$ components.

\subsubsection{The Total Wavefunction and Order Parameters}

Now we analyze the form of ${\bf M}^{(ab)}$ of Eq. (\ref{SUPER}) for
$a \not= b$, using inversion symmetry.  To do this it is convenient
to invoke invariance under the symmetry operation $m_xm_y{\cal I}$
whose effect is given in Table \ref{NINF25}.  We write
\begin{eqnarray}
m_x m_y {\cal I} S_a ({\bf q}, \tau) &=&
\xi_a (m_x) \xi_a (m_y)  \nonumber \\ && \ 
\times S_a ({\bf q}, {\cal R} \tau)^*  \ ,
\end{eqnarray}
where ${\cal R}\tau = \tau$ for $\tau \not= 5,6,7,8$, otherwise
${\cal R} \tau= \tau \pm 2$ within the remaining sector of $\tau$'s
and $a$ (and later $b$) denotes one of $x$, $y$, and $z$.  Thus
\begin{eqnarray}
T & \equiv & S_a ({\bf q}, \tau)^* M^{(ab)}_{\tau \tau'}
S_b ({\bf q}, \tau') \nonumber \\
&=& [ m_x m_y {\cal I} S_a ({\bf q}, \tau)]^* M^{(ab)}_{\tau \tau'}
[ m_x m_y {\cal I} S_b ({\bf q}, \tau')] \nonumber \\
&=& C_{ab} S_a ({\bf q}, {\cal R} \tau) M^{(ab)}_{\tau \tau'}
S_b ({\bf q}, {\cal R} \tau')^* \ ,
\end{eqnarray}
where
\begin{eqnarray}
C_{ab} &=& \xi_a (m_x) \xi_a (m_y) \xi_b (m_x) \xi_b (m_y) \ .
\end{eqnarray}
From this we deduce that
\begin{eqnarray}
M^{(ba)}_{{\cal R}\tau' , {\cal R}\tau} &=& C_{a b} M^{(ab)}_{\tau \tau'} \ ,
\end{eqnarray}
or, since ${\bf M}$ is Hermitian that
\begin{eqnarray}
M^{(ab)}_{\tau \tau'} &=& C_{ab} \left[ M^{(ab)}_{{\cal R}^{-1} \tau ,
{\cal R}^{-1} \tau'}\right]^* \ .
\end{eqnarray}
Now we consider the matrices ${\bf M}^{(ab)}$,
in the symmetry adapted representation where
\begin{eqnarray}
M^{(ab)}_{n,m} &=& \sum_{\tau \tau'}
[O^{ap}_{n \tau}]^* M^{(ab)}_{\tau \tau'} O^{bp}_{m \tau' }
\nonumber \\  &=& \sum_{\tau \tau'} C_{ab} [O^{ap}_{n \tau}]^*
\left[ M^{(ab)}_{{\cal R}^{-1} \tau , {\cal R}^{-1} \tau'}\right]^*
O^{bp}_{m \tau' } \nonumber \\ &=& C_{ab}
\sum_{\tau \tau'} [O^{ap}_{n {\cal R} \tau}]^* \left[
M^{(ab)}_{\tau, \tau'}\right]^*  O^{bp}_{m {\cal R} \tau' } \ .
\end{eqnarray}
There are no matrix elements connecting $p$ and $p' \not= p$ and 
the result is independent of $p$.
One can verify from Eqs. (\ref{SUBSPACE}) and (\ref{SUBSPACE2}) that
\begin{eqnarray}
O^{ap}_{n , {\cal R}\tau} &=& \left[ O^{ap}_{n , \tau} \right]^* \ ,
\end{eqnarray}
so that
\begin{eqnarray}
M^{(ab)}_{n,m} &=& C_{ab} \biggl( \left[ O^{\alpha p}_{n \tau}
\right]^* M^{(ab)}_{\tau, \tau'}  O^{bp}_{m \tau' } \biggr)^*
\nonumber \\ &=& C_{ab} \left[ M^{\alpha \beta}_{nm}  \right]^* \ .
\end{eqnarray}
We have that $C_{xy}=-C_{xz}=-C_{yz}=1$,
so that all the elements of ${\bf M}^{(xy)}$ are real and all the elements of
${\bf M}^{(xz)}$ and ${\bf M}^{(yz)}$ are imaginary.
Thus apart from an over all phase for the eigenfunction of each column, the
phases of all the Fourier coefficients are fixed.
What this means is that the critical eigenvector can be written as
\begin{eqnarray}
\psi &=& \sum_{p=1}^2 \sigmav_p \sum_{n=1}^6 \Biggl(r_{nx} {\bf O}^{(x,p)}_n
+ r_{ny} {\bf O}^{(y,p)}_n \nonumber \\ && \
+ ir_{nz} {\bf O}^{(z,p)}_n \Biggr) \ ,
\end{eqnarray}
where the $r$'s are all real-valued and are normalized by
\begin{eqnarray}
\sum_{n=1}^6 \sum_\alpha [r_{n\alpha }]^2 &=& 1 \ ,
\end{eqnarray}
and $\sigmav_p$ are arbitrary complex numbers.
Thus we have the result of Table \ref{TMO25q}.

\begin{table}
\vspace{0.2 in}
\begin{tabular}{||c||c|c|||| c || c|c ||}\hline\hline
Spin & $\sigmav_1 $ & $\sigmav_2$ & Spin  &$\sigmav_1$ & $\sigmav_2$ \\ \hline

${\bf S}({\bf q},1)$&
$\begin{array}{c} r_{1x} \\  r_{1y} \\ ir_{1z}  \end{array}$ &
$\begin{array}{c} r_{2x} \\  r_{2y} \\ ir_{2z} \end{array}$  &
${\bf S}({\bf q},7)$ & 
$\begin{array}{c} z_x^* \\  -z_y^* \\  iz_z^*  \end{array}$ &
$\begin{array}{c} -z_x^* \\  z_y^* \\  iz_z^* \end{array}$ \\ \hline

${\bf S}({\bf q},2)$&
$\begin{array}{c} r_{2x} \\  r_{2y} \\  -ir_{2z}  \end{array}$ &
$\begin{array}{c} r_{1x} \\  r_{1y} \\  -ir_{1z} \end{array}$ &
${\bf S}({\bf q},8)$ &
$\begin{array}{c} z_x^* \\  z_y^* \\  -iz_z^*  \end{array}$ &
$\begin{array}{c} z_x^* \\  z_y^* \\  iz_z^* \end{array}$ \\ \hline

${\bf S}({\bf q},3)$ &
$\begin{array}{c} r_{1x} \\ -r_{1y} \\  -ir_{1z}  \end{array}$ &
$\begin{array}{c} -r_{2x}\\  r_{2y} \\  ir_{2z} \end{array}$  &
${\bf S}({\bf q},9)$&
$\begin{array}{c} r_{5x} \\  r_{5y} \\  ir_{5z}  \end{array}$ &
$\begin{array}{c} r_{6x} \\  r_{6y} \\  ir_{6z} \end{array}$ \\ \hline

${\bf S}({\bf q},4)$ &
$\begin{array}{c} r_{2x} \\  -r_{2y} \\ i r_{2z}  \end{array}$ &
$\begin{array}{c} -r_{1x}\\  r_{1y} \\  -i r_{1z} \end{array}$  &
${\bf S}({\bf q},10)$&
$\begin{array}{c} r_{6x} \\  r_{6y} \\ -ir_{6z}  \end{array}$ &
$\begin{array}{c} r_{5x} \\  r_{5y} \\  -ir_{5z} \end{array}$ \\ \hline

${\bf S}({\bf q},5)$&
$\begin{array}{c} z_x \\  -z_y \\  iz_z  \end{array}$ &
$\begin{array}{c} -z_x\\  z_y \\  iz_z \end{array}$  &
${\bf S}({\bf q},11)$&
$\begin{array}{c} r_{5x} \\  -r_{5y} \\  -ir_{5z}  \end{array}$ &
$\begin{array}{c} -r_{6x}\\  r_{6y} \\  ir_{6z} \end{array}$ \\ \hline

${\bf S}({\bf q},6)$&
$\begin{array}{c} z_x \\  z_y \\  -iz_z  \end{array}$ &
$\begin{array}{c} z_x\\  z_y \\  iz_z \end{array}$  &
${\bf S}({\bf q},12)$&
$\begin{array}{c} r_{6x} \\  -r_{6y} \\  ir_{6z}  \end{array}$ &
$\begin{array}{c} -r_{5x}\\  r_{5y} \\  -ir_{5z} \end{array}$ \\ \hline
\end{tabular}

\caption{\label{TMO25q} Normalized spin functions (i. e. Fourier
coefficients) within the unit cell of TbMn$_2$O$_5$ for wavevector
$(\oh,0,q)$.  Here $z_\alpha = (r_{3\alpha} + i r_{4 \alpha})/\sqrt 2$.
All the $r$'s are real variables.  The wavefunction listed under
$\sigma_1$ ($\sigma_2$) transforms according to the first (second)
column of the irrep.  The actual spin structure is a linear
combination of the two columns with arbitrary complex coefficients.}
\end{table}

The order parameters are
\begin{eqnarray}
\sigmav_1 & \equiv & \sigma_1 e^{i \phi_1} \ , \ \
\sigmav_2 \equiv \sigma_2 e^{i \phi_2} \ .
\end{eqnarray}
Neither the relative magnitudes of $\sigmav_1$ and $\sigmav_2$ nor their
phases are fixed by the quadratic terms
within the Landau expansion.  Note that the structure
parameters of Table \ref{TMO25q} are determined by the microscopic
interactions which determine the matrix elements in the quadratic
free energy.  (Since these are usually not well known, one has recourse
to a symmetry analysis.)  The direction in $\sigmav_1$-$\sigmav_2$
space which the system assumes, is determined by fourth or higher-order
terms in the Landau expansion.  Since not much is known about these
terms, this direction is reaosonably treated as a parameter to be
extracted from the experimental data.
We use Table \ref{TMO25q} to write the most general spin functions
consistent with crystal symmetry as
\begin{eqnarray}
{\bf S}({\bf R},1) &=& \sigma_1 \left[ (r_{1x}\hat i + r_{1y} \hat j)
\cos ({\bf q} \cdot {\bf R} + \phi_1) \right.
\nonumber \\ && \
\left. + r_{1z} \hat k \sin ({\bf q} \cdot {\bf R} + \phi_1) \right]
\nonumber \\ && \ + \sigma_2 \left[ (r_{2x}\hat i + r_{2y} \hat j]
\cos ({\bf q} \cdot {\bf R} + \phi_2) \right. \nonumber \\
&& \ \left. + r_{2z} \hat k \sin ({\bf q} \cdot {\bf R} + \phi_2) \right] \nonumber \\
{\bf S}({\bf R},2) &=& \sigma_1 \left[ (r_{2x}\hat i + r_{2y} \hat j)
\cos ({\bf q} \cdot {\bf R} + \phi_1) \right.
\nonumber \\ && \
\left. - r_{2z} \hat k \sin ({\bf q} \cdot {\bf R} + \phi_1) \right]
\nonumber \\ && \ + \sigma_2 \left[ (r_{1x}\hat i + r_{1y} \hat j]
\cos ({\bf q} \cdot {\bf R} + \phi_2) \right. \nonumber \\
&& \ \left. - r_{1z} \hat k \sin ({\bf q} \cdot {\bf R} + \phi_2) \right] \nonumber \\
{\bf S}({\bf R},3) &=& \sigma_1 \left[ (r_{1x}\hat i - r_{1y} \hat j)
\cos ({\bf q} \cdot {\bf R} + \phi_1) \right.
\nonumber \\ && \
\left. - r_{1z} \hat k \sin ({\bf q} \cdot {\bf R} + \phi_1) \right]
\nonumber \\ && \ + \sigma_2 \left[ (-r_{2x}\hat i + r_{2y} \hat j]
\cos ({\bf q} \cdot {\bf R} + \phi_2) \right. \nonumber \\
&& \ \left. + r_{2z} \hat k \sin ({\bf q} \cdot {\bf R} + \phi_2) \right] \nonumber \\
{\bf S}({\bf R},4) &=& \sigma_1 \left[ (r_{2x}\hat i - r_{2y} \hat j)
\cos ({\bf q} \cdot {\bf R} + \phi_1) \right.
\nonumber \\ && \
\left. + r_{2z} \hat k \sin ({\bf q} \cdot {\bf R} + \phi_1) \right]
\nonumber \\ && \ + \sigma_2 \left[ (-r_{1x}\hat i + r_{1y} \hat j]
\cos ({\bf q} \cdot {\bf R} + \phi_2) \right. \nonumber \\
&& \ \left. - r_{1z} \hat k \sin ({\bf q} \cdot {\bf R} + \phi_2) \right] \nonumber \\
{\bf S}({\bf R},5) &=& \sigma_1 \left[ (z_x'\hat i - z_y' \hat j - z_z''\hat k)
\cos ({\bf q} \cdot {\bf R} + \phi_1) \right.
\nonumber \\ && \
\left. + (z_x''\hat i - z_y'' \hat j + z_z' \hat k ) \sin ({\bf q} \cdot {\bf R} + \phi_1) \right]
\nonumber \\ && \ + \sigma_2 \left[ (-z_x'\hat i + z_y' \hat j -z_z'' \hat k]
\cos ({\bf q} \cdot {\bf R} + \phi_2) \right. \nonumber \\
&& \ \left. + (-z_x'' \hat i + z_y''\hat j + z_z'  \hat k ) \sin ({\bf q} \cdot {\bf R} + \phi_2)
\right] \nonumber \\
{\bf S}({\bf R}_6) &=& \sigma_1 \left[ (z_x'\hat i + z_y' \hat j + z_z''\hat k)
\cos ({\bf q} \cdot {\bf R} + \phi_1) \right.
\nonumber \\ && \
\left. + (z_x''\hat i + z_y'' \hat j i - z_z' \hat k ) \sin ({\bf q} \cdot {\bf R} + \phi_1) \right]
\nonumber \\ && \ + \sigma_2 \left[ (z_x'\hat i + z_y' \hat j - z_z'' \hat k]
\cos ({\bf q} \cdot {\bf R} + \phi_2) \right. \nonumber \\
&& \ \left. + (z_x'' \hat i + z_y''\hat j + z_z'  \hat k ) \sin ({\bf q} \cdot {\bf R} + \phi_2)
\right] \nonumber \\
{\bf S}({\bf R},7) &=& \sigma_1 \left[ (z_x'\hat i - z_y' \hat j + z_z''\hat k)
\cos ({\bf q} \cdot {\bf R} + \phi_1) \right.
\nonumber \\ && \
\left. + ( - z_x''\hat i + z_y'' \hat j + z_z' \hat k ) \sin ({\bf q} \cdot {\bf R} + \phi_1)
\right] \nonumber \\ && \ + \sigma_2 \left[ ( - z_x'\hat i + z_y \hat j + z_z'' \hat k]
\cos ({\bf q} \cdot {\bf R} + \phi_2) \right. \nonumber \\
&& \ \left. + ( z_x'' \hat i - z_y''\hat j + z_z'  \hat k ) \sin ({\bf q} \cdot {\bf R} + \phi_2)
\right] \nonumber \\
{\bf S}({\bf R}_8) &=& \sigma_1 \left[ (z_x'\hat i + z_y' \hat j - z_z''\hat k)
\cos ({\bf q} \cdot {\bf R} + \phi_1) \right.
\nonumber \\ && \
\left. + ( - z_x''\hat i - z_y'' \hat j - z_z' \hat k ) \sin ({\bf q} \cdot {\bf R} + \phi_1)
\right] \nonumber \\ && \ + \sigma_2 \left[ (z_x'\hat i + z_y \hat j + z_z'' \hat k]
\cos ({\bf q} \cdot {\bf R} + \phi_2) \right. \nonumber \\
&& \ \left. + ( - z_x'' \hat i - z_y''\hat j + z_z'  \hat k ) \sin ({\bf q} \cdot {\bf R} + \phi_2)
\right] \nonumber \\
{\bf S}({\bf R},9) &=& \sigma_1 \left[ (r_{5x}\hat i + r_{5y} \hat j)
\cos ({\bf q} \cdot {\bf R} + \phi_1) \right.
\nonumber \\ && \
\left. + r_{5z} \hat k \sin ({\bf q} \cdot {\bf R} + \phi_1) \right]
\nonumber \\ && \ + \sigma_2 \left[ (r_{6x}\hat i + r_{6y} \hat j]
\cos ({\bf q} \cdot {\bf R} + \phi_2) \right. \nonumber \\
&& \ \left. + r_{6z} \hat k \sin ({\bf q} \cdot {\bf R} + \phi_2) \right] 
\nonumber \\
{\bf S}({\bf R},10) &=& \sigma_1 \left[ (r_{6x}\hat i + r_{6y} \hat j)
\cos ({\bf q} \cdot {\bf R} + \phi_1) \right.
\nonumber \\ && \
\left. - r_{6z} \hat k \sin ({\bf q} \cdot {\bf R} + \phi_1) \right]
\nonumber \\ && \ + \sigma_2 \left[ (r_{5x}\hat i + r_{5y} \hat j]
\cos ({\bf q} \cdot {\bf R} + \phi_2) \right. \nonumber \\
&& \ \left. - r_{5z} \hat k \sin ({\bf q} \cdot {\bf R} + \phi_2) \right]
\nonumber \\
{\bf S}({\bf R},11) &=& \sigma_1 \left[ (r_{5x}\hat i - r_{5y} \hat j)
\cos ({\bf q} \cdot {\bf R} + \phi_1) \right.
\nonumber \\ && \
\left. - r_{5z} \hat k \sin ({\bf q} \cdot {\bf R} + \phi_1) \right]
\nonumber \\ && \ + \sigma_2 \left[ (-r_{6x}\hat i + r_{6y} \hat j]
\cos ({\bf q} \cdot {\bf R} + \phi_2) \right. \nonumber \\
&& \ \left. + r_{6z} \hat k \sin ({\bf q} \cdot {\bf R} + \phi_2) \right] 
\nonumber \\
{\bf S}({\bf R},12) &=& \sigma_1 \left[ (r_{6x}\hat i - r_{6y} \hat j)
\cos ({\bf q} \cdot {\bf R} + \phi_1) \right.
\nonumber \\ && \
\left. + r_{6z} \hat k \sin ({\bf q} \cdot {\bf R} + \phi_1) \right]
\nonumber \\ && \ + \sigma_2 \left[ (-r_{5x}\hat i + r_{5y} \hat j]
\cos ({\bf q} \cdot {\bf R} + \phi_2) \right. \nonumber \\
&& \ \left. - r_{5z} \hat k \sin ({\bf q} \cdot {\bf R} + \phi_2) \right] 
\nonumber \\
\label{25SPINS} \end{eqnarray}
In Table \ref{TMO25q} the position of each spin is ${\bf R}+\tauv_n$,
where the $\tauv$ are listed in Table \ref{T25TAB} and ${\bf R}$ is a Bravais
lattice vector.  The symmetry properties of the order parameters are
\begin{eqnarray}
m_x \left[ \begin{array} {c} \sigmav_1 \\ \sigmav_2 \\ \end{array} \right]
&=& \left[ \begin{array} {c} \sigmav_1 \\ - \sigmav_2 \\ \end{array} \right] \nonumber \\
m_y \left[ \begin{array} {c} \sigmav_1 \\ \sigmav_2 \\ \end{array} \right]
&=& \left[ \begin{array} {c} \sigmav_2 \\ -\sigmav_1 \\ \end{array} \right]
\nonumber \\
{\cal I} \left[ \begin{array} {c} \sigmav_1 \\ \sigmav_2 \\ \end{array} \right]
&=& \left[ \begin{array} {c} \sigmav_2^* \\ \sigmav_1^* \\  \end{array}
\right] \ .
\label{SIGTRANS} \end{eqnarray}

We now check a few representative cases of the above transformation.  If
we apply $m_x$ to $S({\bf q},1)$ we do not change the signs of the
$x$ component but do change the signs of the $y$ and $z$ components.
As a result we get $S({\bf q},3)$ except that $\sigma_y$ has changed sign,
in agreement with the first line of Eq. (\ref{SIGTRANS}).
If we apply $m_y$ to $S({\bf q},1)$ we do not change the sign of
the $y$ component but do change the signs of the $x$ and $z$ components.
As a result we get $S({\bf q},4)$ except that now $\sigmav_1$ is replaced by $\sigmav_2$
and $\sigmav_2$ is replaced by $\sigmav_1$, in agreement with the second line
of Eq. (\ref{SIGTRANS}).  When inversion is applied to $S({\bf q},1)$ we change
the sign of ${\bf R}$ but not the orientation of the spins which are pseudovectors.
We then obtain $S({\bf q},2)$ providing we replace $\sigmav_1$ by $\sigmav_2^*$ and
$\sigmav_2$ by $\sigmav_1^*$, in agreement with the last line of Eq. (\ref{SIGTRANS}). 

\subsubsection{Comparison to Group Theory}

Here I briefly compare the above calculation to the one using the simplest
formulation of representation theory. The first step in the standard
formulation is to find the irreps of the group of the wavevector.  The
easiest way to do this is to introduce a double group having eight
elements (see Appendix B) since we need to take account of the operator
$m_y^2 \equiv -E$. (This is done in Appendix B.)
From this one finds that each Wyckoff orbit and
each spin component can be considered separately (since they do not
transform into one another under the operations we consider).
Then, in every case the only irrep that appears is the two
dimensional one for which we set
\begin{eqnarray}
m_x &=& \left[ \begin{array} {c c } $1$ & $0$ \\ $0$ & $-1$ \\ \end{array}
\right] \
m_y = \left[ \begin{array} {c c } $0$ & $1$ \\ $-1$ & $0$ \\ \end{array}
\right] \
m_x m_y = \left[ \begin{array} {c c } $0$ & $1$ \\ $1$ & $0$ \\ \end{array}
\right] \ .
\label{116EQ} \end{eqnarray}
Indeed, one can verify that the functions in the second (third) column of
Table \ref{TMO25q} comprise a basis vector for column one (two) of this two
dimensional irrep.  One might ask: ``Why have we undertaken the ugly detailed
consideration of the matrix for $F_2$?''  The
point is that within standard representation theory all the variables
in Table \ref{TMO25q} would be independently assigned {\it arbitrary}
phases.  In addition, the amplitudes for the Tb orbits (sublattices \#5, \#6
and sublattices \#7, \#8) would have independent amplitudes.  To get the
results actually shown in Table \ref{TMO25q} one would have to do the
equivalent of analyzing the effect of inversion invariance of the
free energy.  This task would be a very technical
exercise in the arcane aspects of group theory which here we avoid by an
exercise in algebra, which though messy, is basically high school math.
I also warn the reader that canned programs to perform the standard 
representation analysis can not always be relied upon to be correct.
It is worth noting that published papers dealing with TMO25 have not
invoked inversion symmetry.  For instance in Ref. \onlinecite{TB25}
one sees the statement ``As in the incommensurate case[3], each of the 
magnetic atoms in the unit cell is allowed to have an independent SDW,
i. e., its own amplitude and phase," and later on in Ref. \onlinecite{BLAKE}
"all phases were subsequently fixed ... to be rational fractions of
$\pi$."  Use of the present theory would
eliminate most of the phases and would relate the two distinct
Mn$^{4+}$ Wyckoff orbits (just as happened for TMO).

Finally, to see the effect of inversion on a concrete level
I analyze the situation within the
Mn$^{3+}$ orbit and consider only the $x$ components of spin.  The
inverse susceptibility would then be the upper left $4 \times 4$
submatrix shown in Eq. (\ref{MATEQ}). Had we not used inversion
symmetry this submatrix would be the same except that one
would have had
\begin{eqnarray}
M_{14}=-M_{23}=M_{32}=-M_{41} = ir \ ,
\label{IREQ} \end{eqnarray}
where $r$ is real.  One can verify that when $r=0$ the 
eigenvectors can be taken to have only real components, whereas
when $r\not= 0$, the eigenvectors are complex with relative
phases dependent on the value of $r$,

\subsubsection{Comparison to YMn$_2$O$_5$}

YMn$_2$O$_5$ (YMO25) is isostructural to TM025, so its magnetic
structure is relevant to the present discussion.  I will consider
the highest temperature magnetically ordered phase, which appears
between about 20K and 45K.  In this compound Y is nonmagnetic
and in the higher-temperature ordered phase $q_z=1/4$, so the
system is commensurate.  But since the value of $q_z$ is
not special, the symmetry of this state is essentially the
same as that of TMO25.  Throughout this
subsection the structural information is taken from Fig. 2 of
Ref. \onlinecite{Y25}. (The uppermost panel is mislabeled and is
obviously the one we want for the highest temperature ordered phase.)

\begin{figure}[ht]
\begin{center}
\includegraphics[width=8cm]{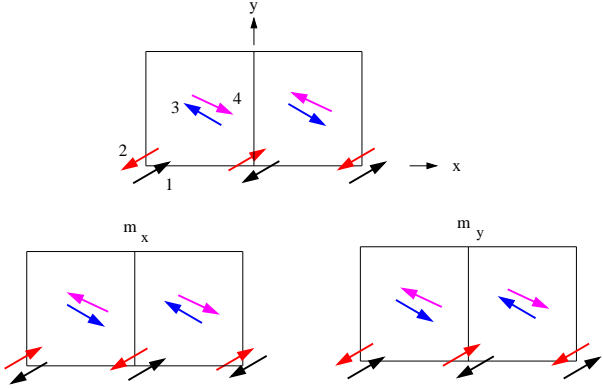}
\caption{(Color online). Top: The spin structure of the
Mn$^{3+}$ ions in YMn$_2$O$_5$
(limited to one {\bf a}-{\bf b} pane), taken from Fig. 2 of Ref.
\onlinecite{Y25}.  The sublattices are labeled in our convention.
Bottom left: the spin structure of after transformation by $m_x$.
Bottom right: spin structure  of TbMn$_2$O$_5$ after transformation
by $m_y$.}
\label{YFIG}
\end{center}
\end{figure}

From Fig. \ref{YFIG} we see that the spin wavefunction is an eigenvector
of $m_x$ with eigenvalue $-1$.  So this structure must be that of
the second column of the irrep.  In accordance with this identification
one sees that the initial wavefunction is orthogonal to the wavefunction
transformed by $m_y$ (since this transformation will produce a wavefunction
associated with the first column). Referring to Eq. (\ref{25SPINS}),
one sees that to describe the pattern of Mn$^{3+}$ spins
one chooses
\begin{eqnarray}
\sigmav_1 &=& 0 \ , \ \ \ \ r_{2x}=-r_{1x}  \approx 0.95 \ , \nonumber \\
&& \ r_{1y} = - r_{2y} \approx 0.3 \ .
\end{eqnarray}
The point we make here is that $\sigmav_1=0$.  Although the values
of these order parameters were not given in Ref. \onlinecite{Y25},
it seems clear that in the lower temperature phase the order
parameters must be comparable in magnitude.

\subsection{CuFeO$_2$}

The space group of CuFeO$_2$ (CFO) is\cite{CFOSTR}
R$\overline 3$m (\#166 in Ref.
\onlinecite{HAHN}) and its general positions are given in Table \ref{R3M}.

\begin{table}
\vspace{0.2 in}
\begin{tabular} {|| c | c | c ||}
\hline \hline
$E {\bf r} =(x,y,z) \ \ $ & $3 {\bf r} =(z, x , y )\ \ $
& $3^2 {\bf r}= (y , z, x)\ \ $ \\
$m_3 {\bf r}= (y, x, z )\ \ $ & $ m_2 {\bf r}=(z, y, x)\ \ $ 
& $m_1{\bf r}=(x, z, y)\ \ $ \\
\hline
${\cal I} {\bf r} =(\overline x, \overline y ,\overline z )\ \ $
& ${\cal I} 3 {\bf r}=(\overline z , \overline x , \overline y )\ \ $
& ${\cal I} 3^2 {\bf r} =(\overline y, \overline z ,\overline x )\ \ $ \\
${\cal I} m_3 {\bf r}=(\overline y , \overline x , \overline z )\ \ $
& ${\cal I} m_2 {\bf r} =(\overline z, \overline y ,\overline x )\ \ $
& ${\cal I} m_1 {\bf r}=(\overline x , \overline z , \overline y )\ \ $ \\
\hline \hline 
\end{tabular}
\caption{\label{R3M}General Positions for R$\overline 3$m, with
respect to rhombohedral axes. Here "3" denotes a three-fold rotation
and $m_n$ labels the three mirror planes which contain the
three-fold axis.}
\end{table}

\begin{figure}[ht]
\begin{center}
\includegraphics[width=6cm]{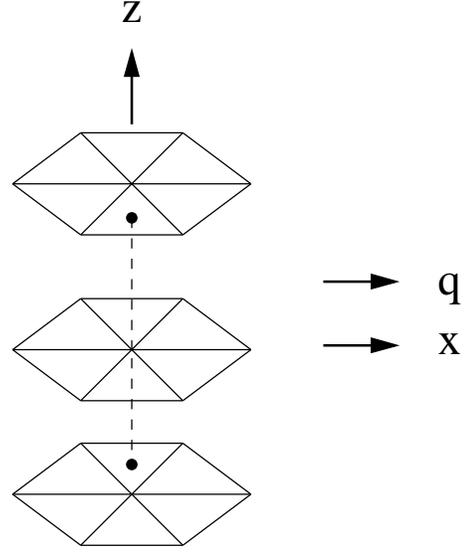}
\caption{The lattice of magnetic Fe ions in CFO.
Here I show sections of three adjacent triangular lattice layers.
The wavevector lies along the $x$-axis. The dashed line indicates that
the central site lies directly above (below) the center of gravity,
indicated by a dot, of a triangle of the layer below (above) it.}
\label{CFOFIG}
\end{center}
\end{figure}

We are interested in structures that can appear for general wavevectors
of the type ${\bf q} \equiv (q,q,0)$ (in crystallographic notation),
in other words
for wavevectors parallel to a nearest neighbor vector of the
triangular plane of Fe ions.  The only operation (other than the identity)
that conserves wavevector is $2_x$ a two-fold rotation about the
axis of the wavevector.  Clearly, the Fourier component $m_z({\bf q})$
obeys
\begin{eqnarray}
2_x m_x ({\bf q}) = m_x({\bf q})
\end{eqnarray}
and we call this irrep \#1.  For irrep \#2 we have
\begin{eqnarray}
2_x m_y({\bf q}) &=& -m_y({\bf q}) \nonumber \\
2_x m_z({\bf q}) &=& -m_z({\bf q}) \ .
\end{eqnarray}
As before inversion fixes the phase of these coefficients to be
the same (within a given irrep), so one has
\begin{eqnarray}
m_x &=& \sigmav_1
\end{eqnarray}
and
\begin{eqnarray}
m_y &=& \sigmav_2 r \ ,
m_z = \sigmav_2 s \ ,
\end{eqnarray}
where $r^2 + s^2 = 1$ and $\sigmav_n(\pm |{\bf q}|)=\sigma_n e^{\pm i \phi_n}$.
Thus, if both irreps are present, we would have
\begin{eqnarray}
m_x({\bf r}) &=& 2 \sigma_1 \cos(qx + \phi_1)\nonumber \\
m_y({\bf r}) &=& 2 \sigma_2 r \cos(qx + \phi_2)\nonumber \\
m_z({\bf r}) &=& 2 \sigma_2 s \cos(qx + \phi_2)\ .
\label{CFOSPIN} \end{eqnarray}
We have the transformation properties 
\begin{eqnarray}
2_x \sigmav_1 &=& \sigmav_1 \ , \ \
2_x \sigmav_2 = - \sigmav_2 \nonumber \\
{\cal I} \sigmav_1 &=& [\sigmav_1]^* \ , \ \
{\cal I} \sigmav_2 = [\sigmav_2]^* \ .
\label{CFOTRANS} \end{eqnarray}
For future reference it is interesting to note that at zero applied
electric and magnetic fields the free energy must
be invariant under taking either $\sigmav_1$ or $\sigmav_2$ into
its negative.  To see this write the free energy as an expansion in 
powers of the order parameters:
\begin{eqnarray}
F &=& \sum_{k,l,m,n} a_{klmn} \sigmav_1^k {\sigmav_1^*}^l
\sigmav_2^m {\sigmav_2^*}^n \ .
\end{eqnarray}
To be time reversal invariant, the total number of powers
($k+l+m+n$) must be even.  Also, to be invariant under $2_x$
the total number of powers of $\sigmav_2$ ($m+n$) has to be
even.  So both $k+l$ and $m+n$ are even.  But wavevector
conservation implies that $l+n=k+m$ (assuming we have
a truly incommensurate phase). Thus $l=k+2\sigma$ and
$n=m-2\sigma$.  Then
\begin{eqnarray}
F &=& \sum_{k,l,m,n} a_{klmn} \sigma_1^{2k+2\sigma}
\sigma_2^{2m-2\sigma} e^{2 i \sigma (\phi_2-\phi_1)} \ .
\end{eqnarray}

\subsection{Discussion}

\subsubsection{Summary of Results}

In Table \ref{PD} we collect the results for various multiferroics.

\begin{table*}
\vspace{0.2 in}
\begin{tabular} {|| c || c | c || c | c | c || c | c ||} \hline \hline
Phase & $T_<$(K)  & $T_>$(K) & ${\bf q}$ & Irreps & Refs. &  FE? & Refs.
\\ \hline
NVO (HTI)   & 6.3  & 9.1 & (q,0,0) & $\Gamma_4$ & \ 
\onlinecite{NVOPRL,NVOPRB} \
& \ \  No \ \   & \onlinecite{FERRO,NVOPRB} \\
NVO (LTI) & 3.9 &  6.3 & (q,0,0) &\ \  $\Gamma_4 + \Gamma_1$\ \  & 
\onlinecite{NVOPRL,NVOPRB} &  $|| \ b$ & \onlinecite{FERRO,NVOPRB}  \\
\hline
TMO (HTI) &  28 & 41 & $(0,q,0)$ &  $\Gamma_3$ & \onlinecite{KAJI,TMO2}
& No & \onlinecite{TMO1} \\
TMO (LTI) &   & 28 & $(0,q,0)$ &  $\Gamma_3$ + $\Gamma_2$ &
\onlinecite{TMO2} &   $ || \ c$ & \onlinecite{TMO1} \\
\hline
TbMn$_2$O$_5$ (HTI) & 38 & 43 & $(\oh ,0,q)^{(\rm a)}$ & $\Gamma^{(\rm b)}$
& \onlinecite{TB25,BLAKE}
& No & \onlinecite{25FERRO} \\
TbMn$_2$O$_5$ (LTI) & 33 & 38 & $(\oh ,0,q)$ & $\Gamma^{(\rm c)}$ 
& \onlinecite{TB25,BLAKE}
& $|| \ \ b $ & \onlinecite{25FERRO} \\
\hline
YMn$_2$O$_5$ (C)$^{(\rm d)}$
& $23$ & $45$ & $(\oh , 0 , \of)$ & $\Gamma^{(\rm b)}$ &
\onlinecite{Y25} & $|| \ b$ & \onlinecite{25FERRO} \\
YMn$_2$O$_5$ (IC) & $$ & $23$ & $(\approx \oh , 0 , q)$ & &
\onlinecite{Y25} & $|| \ b$ & \onlinecite{25FERRO} \\
\hline
CFO$^{\rm (e)}$ (HTI) & 10 & 14 & $(q, q, 0)$ & $\Gamma_1$ &
\onlinecite{CFOJAP} &  No & \onlinecite{CFO} \\
CFO (LTI) & 0? & 10 & $(q, q, 0)$ & $\Gamma_1 + \Gamma_2$ &
\onlinecite{CFO}$^{(\rm f)}$ & $\perp c$ & \onlinecite{CFO}   \\
\hline
MWO & 12.7 & 13.2 & $(q_x, \oh, q_z)$ & $\Gamma_2$& \onlinecite{LAUT}
& No & \onlinecite{MWO1}  \\
MWO & 7.6 & 12.7 & $(q_x, \oh, q_z)$ & $\Gamma_2+\Gamma_1$ & 
\onlinecite{LAUT} &  $|| \  b$  & \onlinecite{MWO1} \\
\hline \hline
\end{tabular}
\caption{\label{PD} Incommensurate Phases of various multiferroics.
Except for CFO each phase is stable for zero applied magnetic field for
$T_< < T < T_>$.  When $T_<=0$ it means that the phase is stable
down to the lowest temperature investigated.  We give the incommensurate
wavevector and the associated irreducible representations in
the notation of our tables. In the column labeled ``FE?" if the
system is ferroelectric we give the direction of the spontaneous
polarization, otherwise the entry is "No."}

\vspace{0.1in} 
\noindent a) At the highest temperature the value of $q_x$ might
not be exactly 1/2.

\noindent b) The irrep is the two dimensional one (see Appendix B).
In the HTI phase only one basis vector is active.

\noindent c) The irrep is the two dimensional one (see Appendix B).
In the LTI phase both basis vectors are active.

\noindent d) This phase commensurate.

\noindent e) Data for CuFeO$_2$ is for $H \approx 8$T.

\noindent f) The magnetic structure was inferred from the existence of
ferroelectricity.
\end{table*}

\subsubsection{Effect of Quartic Terms}

As we now discuss, the quartic terms in the Landau expansion can have
significant qualitative effects.\cite{NVOPRB}  In general, the quartic
terms are the lowest order ones which favor the fixed length spin
constraint, a constraint which is  known to be dominant
at low temperature.\cite{KAPLAN} How this constraint
comes into play depends on what state is selected by the
quadratic terms.  For instance, in the simplest scenario
when one has a ferromagnet or an antiferromagnet, the
instability is such (see Fig. \ref{CHIQ}) that ordering
with uniform spin length takes place.  Thus, as the temperature
is lowered within the ordered phase, the ordering of wavevectors near
$q=0$  for the ferromagnet (near $q=\pi$ for the antiferromagnet)
which would have become unstable if only the quadratic terms
were relevant, is strongly disfavored by the quartic terms.
In the systems considered here the situation is quite different.
For instance, in NVO,\cite{NVOPRL} TMO,\cite{TMO2}
and MWO\cite{LAUT} the
quadratic terms select an incommensurate structure in which
the spins are aligned along an easy axis and their magnitudes
are sinusoidally modulated.  As the temperature is lowered
the quartic terms lead to an instability in which transverse
spin component break the symmetry of the longitudinal
incommensurate phase.  This scenario explains why the 
highest-temperature incommensurate longitudinal phase
becomes unstable to a lower-temperature incommensurate
phase which has both longitudinal and transverse components
which more nearly conserve spin length.

To see this result formally for NVO, TMO, or MWO,
let $\sigmav_>$ ($\sigmav_<$)
be the complex valued order parameter for the higher-temperature
longitudinal (lower-temperature transverse) ordering.
The fourth order terms then lead to the free energy as
\begin{eqnarray}
F &=& a (T-T_>) |\sigmav_>|^2 + b (T-T_<) |\sigmav_<|^2 \nonumber \\
&& \ + A ( |\sigmav_>|^2 + |\sigmav_<|^2)^2
+ B |\sigmav_> \sigmav_<|^2 \nonumber \\
&& + C [ (\sigmav_< \sigmav_>^*)^2
+ ( \sigmav_<^* \sigmav_>)^2]  \ ,
\end{eqnarray}
where $A$, $B$, and $C$ are real. That $C$ is real is a result of
inversion symmetry, which, for these systems leads to
${\cal I} \sigmav_n = \sigmav_n^*$. The high-temperature representation
does allow transverse components and could, in principle, satisfy
the fixed length constraint.  In the usual situation, however,
the exchange couplings are nearly isotropic and this state is
not energetically favored.   If the higher temperature
structure is longitudinal, then $B$ will surely be negative,
whereas if the higher temperature structure conserves spin 
length $B$ will probably be positive.  By properly choosing
the relative phases of the two order parameters the term in $C$
always favors having two irreps.  So the usual scenario
in which the longitudinal phase becomes unstable relative
to transverse ordering is explained (in this phenomenology) by
having $B$ be negative, so that the discussion after
Eq. (\ref{LANDAU4}) applies.

To finish the argument it remains to consider the term in $C$,
which can be written as
\begin{eqnarray}
\delta F_4 &=& 2C \sigma_<^2 \sigma_>^2 \cos (2 \phi_< - 2 \phi_>) \ ,
\label{137} \end{eqnarray}
where again we expressed the order parameters as in Eq. (\ref{IPHI}).
Normally, if two irreps are favored, it is because together they better
satisfy the fixed length constraint.  What that means is that when
spins have substantial length in one irrep, the contribution
to their spin length from the second irrep is small.  In other
words, the two irreps are out of phase and we therefore expect
that to minimize $\delta F_4$ we do not set $\phi_< = \phi_>$, but rather
\begin{eqnarray}
\phi_< = \phi_> \pm \pi /2 \ .
\label{DPHI} \end{eqnarray}
In other words, we expect $C$ in Eq. (\ref{137}) to be positive.
The same reasoning indicates that the fourth order terms will
favor $\phi_2-\phi_1=\pi /2$ in Eq. (\ref{CFOSPIN}) for CFO.

For all of these systems which have two consecutive
continuous transitions one has a family of broken symmetry
states.  At the highest temperature transition one has
spontaneously  broken symmetry which arbitrarily selects
between $\sigma_>$ and $-\sigma_>$.  (This is the simplest scenario
when the wavevector is not truly incommensurate.)  Independently of
which sign is selected for the order parameter $\sigmav_>$,  one similarly 
has a further spontaneous breaking of symmetry to obtain arbitrarily
either $i \sigma_<$ or $-i \sigma_<$. (Here, as mentioned, we
assume a relative phase $\pi/2$ for $\sigmav_<$.  In this
scenario, then, there are four equivalent low temperature phases
corresponding to the choice of signs of the two order parameters.

The cases of TMO25 and YMO25 are different from the above because
they have two order parameters from the same two-dimensional irrep
and which therefore are simultaneously critical.  At quadratic order
one has SU$_2$ symmetry, but this is broken by quartic terms in
the free energy.  The symmetry of these is such that in the
representation of Eq. (\ref{SIGTRANS}) one has
\begin{eqnarray}
F &=& a (T-T_c) \left[ |\sigmav_1|^2 + |\sigmav_2|^2 \right] \nonumber \\
&& \ + A ( |\sigmav_1|^2 + |\sigmav_2|^2)^2
+ B |\sigmav_1 \sigmav_2|^2 \nonumber \\
&& + C \left[ (\sigmav_1 \sigmav_2^*)^2
+ ( \sigmav_1^* \sigmav_2)^2 \right]  \ .
\end{eqnarray}
Terms odd in $\sigmav_2$ are not allowed according to Eq. (\ref{SIGTRANS}).
Also wavevector conservation indicates that two variables must be at
wavevector ${\bf q}$ and two at wavwevector $-{\bf q}$.  Also
$A$, $B$, and $C$ are real.  That $C$ is real is a result
of symmetry under $m_y$, as in Eq. (\ref{SIGTRANS}).
Here the fourth order anisotropy makes itself felt as soon as the
ordered phase is entered.  One can see that the phase difference
between $\sigmav_1$ and $\sigmav_2$ depends on the term proportional
to $C$.  Since the fixed spin length constraint favors this
phase difference to be $\pi /2$, we intuit that $C$ is positive,
so that $\sigmav_2 = \pm i \sigmav_1$. In any case, after minimizing
with respect to the relative phase of the two order parameters, one gets
the sum of the $B$ and $C$ terms as
\begin{eqnarray}
\delta F_4 = (B- |C|) |\sigmav_1 \sigmav_2|^2 \ .
\label{BC4} \end{eqnarray}
If $B-|C|$ is positive, then either $\sigmav_1=0$ or $\sigmav_2=0$.
In the former case the state is odd under $m_x$ and in the latter
case even under $m_x$.  If $B-|C|$ is negative, then the quartic
terms favor
\begin{eqnarray}
|\sigmav_1|= |\sigmav_2|\ .
\label{BCNEG} \end{eqnarray}
It is amusing that the quartic terms in the Landau expansion
dictate that these are the two allowable scenarios unless
one admits to having a multicritical point where $B-|C|=0$.

Now we first consider YMO25 in its
higher temperature commensurate (HTC) ordered phase.  For it additional
fourth order terms occur because $4{\bf q}$ is a reciprocal lattice
vector, but these are not important for the present discussion.
Here the analysis of Ref. \onlinecite{Y25} indicates (see the discussion
of our Fig.  \ref{YFIG}) that only a single order parameter condenses
in the HTC phase.  This indicates that energetics must favor
positive $B-|C|$ in this case.  The question is whether $B-|C|$ is
also positive for TMO25.  As we will see in the next section
one has ferroelectricity unless the magnitudes of the two order
parameters are the same.  For YMO25 the HTC phase is ferroelectric
and the conclusion that only one order parameter is active comports
with this.  However, for TMO25 the situation is not completely clear.
Apparently there is a region such that one has magnetic ordering without
ferroelectricity.\cite{TB25,25FERRO} If this is so, then TMO25 differs
from YMO25 in that its high temperature incommensurate phase has
two equal magnitude order parameters.
 
\section{MAGNETOELECTRIC COUPLING}

Ferroelectricity is induced in these incommensurate magnets by
a coupling which is somewhat similar to that for the so-called
``improper ferroelectrics."\cite {CHUPIS} To see how such a coupling arises
within a phenomenological picture, we imagine expanding the free
energy in powers of the magnetic order parameters which we have
studied in detail in the previous section and also the vector
order parameter for ferroelectricity which is the spontaneous
polarization ${\bf P}$, which, of course, is a zero wavevector
quantity.  If we had noninteracting magnetic and electric systems, then
we would write the noninteracting free energy, $F_{\rm non}$ as
\begin{eqnarray}
F_{\rm non} &=& \oh \sum_\alpha \chi_{E,\alpha}^{-1} P_\alpha^2 \nonumber \\
&& \ + \oh \sum_{\Gamma} a_\Gamma (T-T_\Gamma) |\sigma_\Gamma({\bf q})|^2
+ {\cal O}( \sigma^4 ) \ .
\end{eqnarray}
The first term describes a system which is not close to being unstable
relative to developing a spontaneous polarization (since in the systems
we consider ferroelectricity is {\it induced} by magnetic ordering).
The magnetic terms describe the possibility of having one or more 
phase transitions at which successively  more magnetic order parameters
become nonzero.  As we have mentioned, the scenario of having two
phase transitions in incommensurate magnets is a very common one,\cite{NAG}
and such a scenario is well documented for both NVO\cite{NVOPRL,NVOPRB} and
TMO.\cite{TMO1,TMO2}  
A similar phenomenological description of second harmonic generation
has invoked the necessity of having simultaneously two irreps.\cite{SHG}
Below we will indicate the existence of a term linear in $P$,
schematically of the form $-\lambda M^2P$, where $\lambda$ is
a coupling constant about which not much beyond its symmetry is known.
One sees that when the free energy, including this term, is minimized
with respect to $P$ one obtains the equilibrium value of $P$ as
\begin{eqnarray}
\langle P \rangle &=& \chi_E \lambda M^2 \ .
\end{eqnarray}

\subsection{Symmetry of Magnetoelectric Interaction}

We now consider the free energy of the combined magnetic and electric
degrees of freedom which we write as
\begin{eqnarray}
F &=& F_{\rm non} + F_{\rm int} \ .
\end{eqnarray}
In view of time reversal invariance and wavevector conservation,
the lowest combination of $M(q)$'s
that can appear is proportional to $M_\alpha (-{\bf q}) M_\beta ({\bf q})$.
So generically the term we focus on will be of the form
\begin{eqnarray}
F_{\rm int} &=& \sum_{\alpha \beta \gamma}
c_{\alpha \beta \gamma} M_\alpha ({\bf q}) M_\beta (- {\bf q}) P_\gamma \ ,
\label{INT} \end{eqnarray}
where $\alpha$, $\beta$, and $\gamma$ label Cartesian components.
But, as we have seen in detail, the quantities $M_\alpha ({\bf q})$
are linearly related to the order parameter $\sigma_{\Gamma} ({\bf q})$,
associated with the irrep $\Gamma$. Thus instead of Eq. (\ref{INT}) we write
\begin{eqnarray}
F_{\rm int} &=& \sum_{\Gamma, \Gamma' , \gamma} A_{\Gamma \Gamma' \gamma}
\sigma_{\Gamma}({\bf q}) \sigma_{\Gamma'}({\bf q})^* P_\gamma \ .
\label{SYMINT} \end{eqnarray}
The advantage of this writing the interaction in this form is that it is
expressed in terms of quantities whose symmetry is manifest.  In
particular, the order parameters we have introduced have well specified
symmetries.  For instance it is easy to see that for most of
the systems studied here, magnetism can not induce ferroelectricity
when there is only a single representation present.\cite{FERRO,TMO2}
This follows from the fact that for NVO and TMO, for instance,  
\begin{eqnarray}
{\cal I} |\sigma_n|^2 = |\sigma_n|^2 \ ,
\label{IINVEQ} \end{eqnarray}
as is evident from Eq. (\ref{IISYM}). The interpretation of this is
simple: when one has one representation, it is essentially the same
as having a single incommensurate wave.  But such a single wave will
have inversion symmetry (to as close a tolerance as we wish) with
respect to some lattice point.  This is enough to exclude ferroelectricity.
So the canonical scenario\cite{FERRO,TMO2} is that ferroelectricity
appears, not when the first incommensurate magnetic order parameter
condenses, but rather when a second such order parameter condenses.
Unless the two waves have the same origin, their centers of inversion
symmetry do not coincide and there is no inversion symmetry and
hence ferroelectricity will occur.  One might ask whether or not
the two waves (i. e. two irreps) will be in phase.  The effect,
discussed above, of quartic terms is crucial here.
The quartic terms typically favors the fixed length spin constraint.
To approximately satisfy this constraint, one needs to superpose two
waves which are out of phase.  Indeed the formal result, obtained below,
shows that the spontaneous polarization is proportional to the sine
of the phase difference between the
two irreps.\cite{FERRO} We now consider the various systems in turn.

\subsection{NVO, TMO, and MWO}

We now analyze the canonical magneto-electric interaction in the cases
of NVO, TMO, ad MWO.  These cases are all similar to one another
and in each case the order parameters have been defined so as to
obey Eq. (\ref{IISYM}).  This relation indicates that if we may
choose the origin of the incommensurate system so that the phase
of the order parameter {\it at the origin of a unit cell}
is arbitrarily close to zero.  When this phase is zero,
the spin distribution of this irrep has inversion symmetry relative
to this origin.  In the case when only a single irrep is active,
this symmetry then indicates that the magnetic structure can not
induce a spontaneous polarization.\cite{FERRO}
As mentioned, in the high temperature incommensurate phases of
NVO, TMO, and MWO only one irrep is present, and this argument
indicates that the magneto-electric interaction vanishes
in agreement with the experimental observation\cite{TMO1,FERRO,MWO1}
that this phase is not ferroelectric.

We now turn to the general case when one or more irreps are
present.\cite{FERRO,JAP,NVOPRB,HANDBOOK} We write the
magneto-electric interaction as
\begin{eqnarray}
F_{\rm int} &=& {1 \over 2} \sum_{\gamma \Gamma \Gamma'}
P_\gamma \left[ A_{\Gamma \Gamma' \gamma} 
\sigmav_\Gamma ({\bf q}) \sigmav_{\Gamma'} ({\bf q})^* 
\right. \nonumber \\ && \left. \ + A_{\Gamma' \Gamma \gamma} 
\sigmav_{\Gamma'} ({\bf q}) \sigmav_{\Gamma} ({\bf q})^*  \right] \ .
\label{GENERAL} \end{eqnarray}
For this to yield a real value of $F$ we must have Hermiticity:
$A_{\Gamma \Gamma' \gamma} = A_{\Gamma' \Gamma \gamma}^*$.  
In addition, because this is an expansion relative to the state
in which all order parameters are zero, this interaction has to
be inversion under all operations which leave this ``vacuum''
state invariant.\cite{IED,LL}  In other words this interaction
has to be invariant under inversion (which takes $P_\gamma$ into
$-P_\gamma$).  This condition, together with Eq. (\ref{IISYM})
indicates that $A_{\Gamma \Gamma' \gamma} = -A_{\Gamma' \Gamma \gamma}$.
This condition taken in conjunction with Hermiticity indicates that
$A_{\Gamma \Gamma' \gamma}$ is pure imaginary.  Thus
\begin{eqnarray}
F_{\rm int} &=& {i \over 2} \sum_{\gamma \Gamma \Gamma'}
P_\gamma r_{\Gamma \Gamma' \gamma} 
\left[ \sigmav_\Gamma ({\bf q}) \sigmav_{\Gamma'} ({\bf q})^* \right. 
\nonumber \\ && \ 
\left. - \sigmav_\Gamma ({\bf q})^* \sigmav_{\Gamma'} ({\bf q})  \right] \ .
\end{eqnarray}
Since usually we have only two differenet irreps, we write this as
\begin{eqnarray}
F_{\rm int} &=& \sum_\gamma r_\gamma P_\gamma \sigma_> \sigma_<
\sin(  \phi_> - \phi_< ) \ .
\end{eqnarray}
where $r_\gamma$ is real.
The fact that the result vanishes when the two waves are in phase is
clear because in that case one can find a common origin for both
irreps about which one has inversion symmetry.  In that special case
one has inversion symmetry and no spontaneous polarization can be
induced by magnetism. The above argument applies to all three
systems, NVO,\cite{FERRO} TMO,\cite{TMO2} and MWO.  As we will see
in a moment, it is still possible for inversion symmetry to be broken
and yet induced ferroelectricity not be allowed.

We can also deduce the direction of the spontaneous polarization
by using the transformation properties of the order parameters.
given in Eq. (\ref{ROTSYM}). We start by analyzing
the experimentally relevant cases at low or zero applied
magnetic field.  For NVO the magnetism in
the lower temperature incommensurate phase is described\cite{NVOPRL,NVOPRB}
by the two irreps $\Gamma_4$ and $\Gamma_1$.  One sees from 
Eq. (\ref{ROTSYM}) that
the product $\sigma_1^* \sigma_4$ is even under $m_z$ and
odd under $2_x$.  For the interaction to be an invariant,
$P_\gamma$ has to transform this way also. This implies that
only the ${\bf b}$-component of the spontaneous polarization can 
be nonzero, as observed.\cite{FERRO}
For TMO the lower temperature incommensurate phase
at low magnetic field is described\cite{TMO2} by irreps $\Gamma_3$ and 
$\Gamma_2$.  From Table \ref{TMOqq} we see that
$\sigma_3^*\sigma_2$ is even under $m_x$ and odd under $m_z$,
which indicates that ${\bf P}$ has to be even under $m_x$ and
odd under $m_z$.  This can only happen if ${\bf P}$ lies
along the ${\bf c}$ direction, as observed.\cite{TMO1}

Finally, for MWO, we see that $\sigmav_1 \sigmav_2^*$ is
odd under $m_y$.  This indicates that $P_\gamma$ also has 
to be odd under $m_y$.  In other words ${\bf P}$
can only be oriented along the ${\bf b}$ direction,
again as observed.\cite{MWO1} In this connection one should
note that this conclusion is a result of crystal symmetry,
assuming that the magnetic structure results from
two continuous transitions, so that representation theory
is relevant. This conclusion is at variance with the
argument given by Heyer {\it et al.}\cite{MWO2} who 
``expect a polarization in the plane spanned by the
easy axis and the ${\bf b}$ axis ...," which they 
justify on the basis of the spiral model.\cite{CURRENT,MOST}
It should be noted that their observation that the
spontaneous polarization has a nonzero component along
the ${\bf a}$-axis at zero applied magnetic field 
contradicts the symmetry analysis given here.  The
authors mention that some of the unexpected behavior they
observe might possibly be attributed to a small content
of impurities.  

It is important to realize that the above results are a consequence
of crystal symmetry.  In view of that, it is not sensible to claim
that the fact that a theory gives the result that the polarization
lies along ${\bf b}$ makes it more plausible than some competing
theory.  The point is that any model, if analyzed correctly,
must give the correct orientation for ${\bf P}$.

It is also worth noting that this phenomenology has some
semiquantitative predictions.  To see this, we minimize
$F_{\rm non}+ F_{\rm int}$ with respect to ${\bf P}$ to get
\begin{eqnarray}
P_\gamma = - \chi_{E,\gamma} r_\gamma \sigma_> \sigma_< \sin(\phi_> 
-\phi_<) \ .
\label{PRESEQ} \end{eqnarray}
This result indicates that near the magneto-ferroelectric phase
transition of NVO one has $P \propto \sigma_4 \sigma_1$,\cite{GILL}
or since the high-temperature order parameter $\sigma_4$ is more
or less saturated when the ferroelectric phase is entered,
one has $P \propto \sigma_1$, where $\sigmav_1$ is the order
parameter of the lower temperature incommensurate phase.

As we discussed, in the low temperature incommensurate phase
one will have arbitrary signs of the two order parameters.
However, the presence of a smll electric field will favor
one particular sign of the polarization and hence, by
Eq. (\ref{PRESEQ}) one particular sign for the product
$\sigma_> \sigma_<$. Presumably this could be tested by
a neutron diffraction experiment.

\subsection{TMO25}

The case of TMO25 is somewhat different.  Here we have only
a single irrep.  One expects that as the temperature is
lowered, ordering into an incommensurate state will take place,
but the quadratic terms in the free energy do not select a
direction in $\sigmav_1$-$\sigmav_2$ space.  At present
the data has not been analyzed to say which direction is favored 
at temperature just below the highest ordering temperature.
(For YMO25, as mentioned above, the direction $\sigmav_1=0$ is
favored.) As the temperature is reduced, it is not possible for
another representation to appear because only  one irrep is
involved.  However, ordering according to a second eigenvalue
could occur.  We first analyze the situation assuming that we have
only a single doubly degenerate eigenvalue.  In this case we can have
a spin distribution [as given in Eq. (\ref{25SPINS})] involving the
two order parameters $\sigmav_1$ and $\sigmav_2$ which measure
the amplitude and phase of the ordering of the eigenvector of
the second and third columns of Table \ref{TMO25q}, respectively.
In terms of these order parameters, the magneto-electric coupling
can be written as
\begin{eqnarray}
F_{\rm int} &=& \sum_{nm \gamma} a_{nm \gamma}
\sigma_n^* \sigma_m P\gamma \ ,
\end{eqnarray}
where $\gamma = x, y, z$ and $n,m=1,2$ label 
the columns of the irrep labeled $\sigmav_1$ and
$\sigmav_2$, respectively, in Table \ref{TMO25q}. Since reality
requires that $a_{nm \gamma}=a_{mn \gamma}^*$,
this interaction is of the form
\begin{eqnarray}
F_{\rm int} &=& \sum_\gamma P_\gamma \Biggl[
a_{1\gamma} |\sigmav_1|^2 + a_{2\gamma} |\sigmav_2|^2
\nonumber \\&&  \ +  b_\gamma \sigmav_1 \sigmav_2^* 
+ b_\gamma^* \sigmav_1^* \sigmav_2  \Biggr] \ .
\end{eqnarray}
Now use invariance under inversion, taking note of Eq. (\ref{SIGTRANS}).
One sees that under inversion $\sigmav_1 \sigmav_2^* P_\gamma$ changes
sign, so the only terms which survive lead to the result
\begin{eqnarray}
F_{\rm int} &=& \sum_\gamma r_\gamma P_\gamma
[ |\sigmav_1|^2 - |\sigmav_2|^2 ] \ .
\label{4POL} \end{eqnarray}
As we discussed in connection with Eq. (\ref{BCNEG}) we have two
scenarios depending on whether $B-|C|$ in Eq. (\ref{BC4}) is
positive or negative.  If it is positive, then only one
order parameter is nonzero and we have a nonzero spontaneous 
polarization according to Eq. (\ref{4POL}). In that case, using
Eq. (\ref{SIGTRANS}) we see that $[|\sigmav_1|^2 - |\sigmav_2|^2 ]$
is even under $m_x$ and odd under $m_y$.  For $F_{\rm int}$ to
be invariant under inversion therefore requires that $P_\gamma$
be odd under $m_y$ and even under $m_x$, so ${\bf P}$ has to be
along ${\bf b}$ as is found.\cite{25FERRO} In the other scenario,
when $B-|C|$ is negative, then the right-hand side of Eq. 
(\ref{4POL}) is zero and the state is not ferroelectric.
For TMO25 we are probably in the ferroelectric scenario.
Our analysis therefore suggests that the spin structure of TMO25
should be given by Eq. (\ref{25SPINS}) with only one of the order
parameters nonzero. It would be interesting to analyze the
diffraction data to test this assertion.

\subsection{CFO}

Again we start with the familiar magneto-electric interaction
\begin{eqnarray}
F_{\rm int} &=& \sum_{nm\gamma} A_{nm\gamma}
\sigmav_n \sigmav_m^* P_\gamma  \ ,
\end{eqnarray}
where reality implies that $A_{nm\gamma}=A_{mn\gamma}^*$.
Since we have ${\cal I} \sigmav_n = \sigmav_n^*$, we eliminate
terms with $n=m$: we need two irreps for ferroelectricity.
Indeed, the higher temperature phase with a single order parameter
$\sigmav_1$ is not ferroelectric.\cite{CFO}
Thus the magnetoelectric interaction must be of the form
\begin{eqnarray}
F_{\rm int} &=& \sum_{\gamma} \left[ a_\gamma \sigmav_1 \sigmav_2^* 
+ a_\gamma^* \sigmav_1^* \sigmav_2 \right] P_\gamma \ .
\end{eqnarray}
Inversion symmetry indicates that $a_\gamma =-a_\gamma^*$, so we write
\begin{eqnarray}
F_{\rm int} &=& i \sum_{\gamma} r_\gamma \left[ \sigmav_1 \sigmav_2^* 
- \sigmav_1^* \sigmav_2 \right] P_\gamma \nonumber \\ &=&
2 \sum_\gamma r_\gamma \sigma_1 \sigma_2 \sin (\phi_2 - \phi_1)
P_\gamma \ ,
\end{eqnarray}
where $r_\gamma$ is real.
Now use Eq. (\ref{CFOTRANS}) which gives that $\sigmav_1 \sigmav_2^*$
changes sign under $2_x$. So for the interaction to be invariant under
$2_x$ (as it must be), $P_\gamma$ has to be odd under $2_x$.  This
means that ${\bf P}$ has to be perpendicular to the $x$ axis.
Note that symmetry does not force ${\bf P}$ to lie along the
three-fold axis because the orientation of the incommensurate
wavevector has broken the three-fold symmetry.

In the above analysis we did not mention the fact that the
existence of the ferroelectric phase rquires a magnetic field of
about 8-10T oriented along the three-fold axis. In principle one
should expand the free energy in powers of $H$.  Then
presumably as a function of $H$ one reaches a regime where
first one incommensurate phase orders and then at a lower
temperature the second incommensurate order parameter appears.
Then the phenomenology of the trilinear magnetoelectric
interaction would come into play as analyzed above.  There
is one additional point which merits attention. Namely,
${\bf q}$ could assume a symmetry-related value obtained by one or
two three-fold rotations about the ${\bf c}$ axis.  In the absence 
of any external perturbation to break the three-fold symmetry,
the system would spontaneously break symmetry by arbitrarily
selecting one of the wavevectors in the star.  However, it is
interesting to speculate whether the application of a weak
in-plane magnetic (or electric) field would be enough to
enforce the selection of one of the wavevectors in the star.
If this were so, then the transverse component of the
polarization (which, however, might be small) would be
rotated by the application of such a small external field.

\subsection{High Magnetic Field}

We can also say a word or two about what happens when a 
magnetic field is applied.  In TMO, for instance,
one finds\cite{TMO1} that for applied magnetic fields
above about 10T in either the {\bf a} or {\bf b} directions,
the lower temperature incommensurate phase has a spontaneous
polarization along the ${\bf a}$ axis. Keep in mind that
we want to identify this phase with two irreps and from
the phase diagram we know that the higher temperature
incommensurate phase is maintained into this high field
regime.  So the higher temperature phase is still that
of $\Gamma_3$ at these high fields. Referring to
Table \ref{TMOqq} we see that to get $\sigma_m\sigma_n^*$
to be odd under $m_x$ and even under $m_z$ (in order to get
a polarization along the {\bf a} axis) we can only combine
irrep $\Gamma_1$ with the assumed preexisting $\Gamma_3$.
Therefore it is clear that the magnetic structure has to change at
the same time that direction of spontaneous polarization
changes as a function of applied magnetic field.\cite{HANDBOOK,MOST}
It is also interesting, in this connection to speculate on what
happens if the lower additional irrep had been $\Gamma_4$ so
that $\Gamma_4$ and $\Gamma_3$ would coexist. In that case
$\sigma_4 \sigma_3^*$ is odd under {\it both} $m_x$ and
$m_z$.  These conditions are not consistent with any direction
of polarization, so in this hypothetical case, even though
we have two irreps and break inversion symmetry, a polar
vector (such as the spontaneous polarization) is not
allowed.\cite{COMMENT}
  
For MWO a magnetic field along the ${\bf b}$ axis
of about 10T causes the spontaneous polarization to switch
its direction from along the ${\bf b}$-axis to along
the ${\bf a}$ axis.\cite{MWO1}  We have no phenomenological
explanation of this behavior at present. This behavior seems 
to imply that the wavevector for $H>10$T is no longer of the
form ${\bf q} = (q_x, \oh , q_z)$.

\subsection{Discussion}

What is to be learned from the symmetry analysis of the magnetoelectric
interactions? Perhaps the most important point to keep in mind is
to recognize which results are purely a result of crystal symmetry
and which are model dependent.  For instance, as we have seen,
the direction of the spontaneous polarization is usually a result of
crystal symmetry.  So the fact that a microscopic theory leads to the
observed direction of the polarization does not lend credence to
one model as opposed to another. In a semiquantitative vein, one can
say that symmetry
alone predicts that near the combined magneto-electric phase
transition $P$ will be approximately proportional to the
order parameter raised to the $n$th power, where the value of
$n$ is a result of symmetry. ($n=1$ for NVO or TMO, whereas 
$n=2$ for TMO25).

It also goes without saying that our phenomenological results
are supposed to apply generally, independently of what microscopic
mechanism might be operative for the system in question.  (A number
of such microscopic calculations have appeared 
recently.\cite{CURRENT,TY,SERG1,SERG2,KHOM}) Therefore,
we treat YMO25 and NVO with the same methodology although these
systems are said\cite{TB25} to have different microscopic mechanisms.
A popular phenomenological description is that given by
Mostovoy\cite{MOST} based on a continuum formulation.  However,
this development, although appealing in its simplicity, does
not correctly capture the symmetry of several systems because it
completely  ignores the effect of the different possible symmetries
within the magnetic unit cell.\cite{COMMENT}  Furthermore, it does
not apply to multiferroic systems, such as TMO25 or YMO25, in which
the plane of rotation of the spins is perpendicular to the
wavevector.\cite{Y25} (The spin-current model\cite{CURRENT}
also does not explain ferroelectricity in these systems.)
In addition, a big advantage of the
symmetry analysis presented here concerns small perturbations.
While the structure of NVO and TMO is predominantly a spiral in
the ferroelectric phase, one can speculate on whether there are
small spiral-like components in the nonferroelectric phase.
In other words, could small transverse components
lead to a small (maybe too small for current experiments to
see) spontaneous polarization?  If we take into account the small
magnetic moments induced on the oxygen ions, could these lead to a
small spontaneous polarization in an otherwise nonferroelectric phase?
The answer to these questions is obvious within a symmetry analysis
like that we have given: these induced effects still are governed by
the symmetry of the phase which can only be lowered by a spontaneous
symmetry breaking (which we only expect if we cross a phase boundary).
Therefore all such possible induced effects are taken into account
by our symmetry analysis.

Finally, we note that the form of the magneto-electric interaction
$\sim M^2P$ suggests a microscopic mechanism that has general validity,
although it is not necessarily the dominant mechanism.  This observation
stimulated an investigation of the spin phonon interaction one obtains
by considering the exchange Hamilton
\begin{eqnarray}
{\cal H} &=& \sum_{ij \alpha \beta} J_{\alpha \beta} (i,j)
S_\alpha (i) S_\beta (j)
\end{eqnarray}
when $J_{\alpha \beta}(i,j)$ is expanded to linear order in phonon
displacements.\cite{TY}  After some algebra it was shown that the results
for the direction of the induced spontaneous polarization
(when the spins are ordered appropriately) agrees with the results
of the symmetry arguments used here.  In
addition a first-principles calculation of the phonon modes led to
plausible guesses as to which phonon modes play the key role in
the magneto-electric coupling. But whatever the microscopic model,
the phenomenology presented here should apply.

\section{Conclusion}

In this paper we have shown in detail how one can describe the
symmetry of magnetic and magneto-electric phenomena and have
illustrated the technique by discussing several examples
recently considered in the literature.

The principal results of this work are

$\bullet$
We discussed a method alternative to the traditional one
(called representation analysis) for constructing allowed
spin functions which describe incommensurate magnetic ordering.
In many cases this technique can be especially simple and does
not require an understanding of group theory.

$\bullet$
For systems with a center of inversion symmetry, whether the
simple method mentioned above or the more traditional
traditional representation formalism is used, it is
essential to further include the restrictions imposed by
inversion symmetry, as we pointed out
previously.\cite{FERRO,TMO2,HANDBOOK,JAP,NVOPRB}

$\bullet$
We have illustrated this technique by applying it to systematize
the magnetic structure analysis of several multiferroics many of
which had not been analyzed using inversion symmetry.

$\bullet$
By considering several examples of multiferroics we
further illustrated the general applicability of the
trilinear magneto-electric coupling of the form
$M({\bf q}) M(-{\bf q}) P$, where $M({\bf q})$ is the
magnetization at wavevector ${\bf q}$ and $P$ is the
uniform spontaneous polarization. Usually, when expressed
in terms of order parameters, this interaction requires two
active irreducible representations.

$\bullet$
For TbMn$_2$O$_5$ we analyzed the fourth order terms in
the Landau expansion and predict that the fact that the
system magnetically orders into a ferroelectric phase
indicates that the spin state is described by a single order
parameter according to Eq. (\ref{25SPINS}).  The analysis
of the diffraction data to test this assertion has not
yet been done.

$\bullet$
We briefly discussed the implications of symmetry in
assessing the role of various models proposed for
multiferroics. 

\noindent{\bf ACKNOWLEDGEMENTS}

I acknowledge inspiration and advice from M. Kenzelmann who carried out
several of the group theoretical calculations presented here.  It
should be obvious that this paper owes much to my other collaborators,
especially G. Lawes, T. Yildirim, A. Aharony, O. Entin-Wohlman, C. Broholm,
and A. Ramirez. I thank S.-H. Lee for providing me with the
figure of TbMn$_2$O$_5$ and for insisting that I clarify
various arguments. I thank J. Villain for calling my attention
to some of the history of representation theory.

\begin{appendix}
\section{Form of Eigenvector}
In this appendix we show that the matrix ${\bf G}$ of the form of
Eq. (\ref{GMATRIX}) [and this includes as a subcase the form of
Eq. (\ref{gMATRIX})] has eigenvectors of the form given in Eq.
(\ref{GEIGEN}). Define ${\bf G}' \equiv {\bf U}^{-1} {\bf G} {\bf U}$,
where
\begin{eqnarray}
{\bf U} &=& \left[ \begin{array} {c c c c c c c}
   1   &   0    &    0    &     0     &     0     &     0     &     0      \\
   0   &   1    &    0    &     0     &     0     &     0     &     0      \\
   0   &   0    &    1    &     0     &     0     &     0     &     0      \\
   0   &   0    &    0    & 1/\sqrt 2 & i/\sqrt 2 &     0     &     0      \\
   0   &   0    &    0    & 1/\sqrt 2 &-i/\sqrt 2 &     0     &     0      \\
   0   &   0    &    0    &    0      &    0      & 1/\sqrt 2 & i/\sqrt 2  \\
   0   &   0    &    0    &    0      &    0      & 1/\sqrt 2 &-i/\sqrt 2  \\
\end{array} \right] \ .
\end{eqnarray}
We find that
\begin{scriptsize}
\hspace*{-0.6in} \begin{eqnarray}
\hspace{-0.3 in}
&&{\bf U}^{-1} {\bf G} {\bf U} =\nonumber \\
&& \left[ \begin{array} { c c c c c c c}
  a   &   b   &   c    &
\sqrt 2\alpha'  &   \sqrt 2 \alpha''  &  \sqrt 2 \xi'   &  \sqrt 2 \xi''   \\
  b   &   d   &   e    &
\sqrt 2\beta'   &   \sqrt 2 \beta''   & \sqrt 2 \eta'   & \sqrt 2 \eta''   \\
  c   &   e   &   f    &
\sqrt 2\gamma'  &  \sqrt 2 \gamma''   & \sqrt 2 \kappa' & \sqrt 2 \kappa'' \\
\sqrt 2 \alpha' & \sqrt 2 \beta' & \sqrt 2 \gamma' &
g+ \delta' & \delta''  & \mu' + \nu' & -\mu'' -\nu'' \\
\sqrt 2 \alpha'' & \sqrt 2 \beta'' & \sqrt 2 \gamma'' &
\delta'' & g-\delta'  & \mu'' - \nu'' & \mu' -\nu' \\
\sqrt 2 \xi' & \sqrt 2 \eta' & \sqrt 2 \kappa'
& \mu' + \nu' & \mu'' -\nu''  &h+\rho' & \rho'' \\
\sqrt 2 \xi'' & \sqrt 2 \eta'' & \sqrt 2 \kappa''
& - \mu'' - \nu'' & \mu' -\nu'  &\rho''  &h- \rho' \\
\end{array} \right] \ ,
\end{eqnarray}
\end{scriptsize}
where $\alpha'$ and $\alpha''$ are the real and imaginary parts, 
respectively of $\alpha$ and similarly for the other complex variables.
Note that we have transformed the original matrix into a real symmetric
matrix.  Any eigenvector (which we denote $|R\rangle$) of the transformed
matrix has real-valued components and thus satisfies the equation
\begin{eqnarray}
{\bf U}^{-1} {\bf G} {\bf } {\bf U} |R\rangle &=& \lambda_R |R \rangle,
\end{eqnarray}
from which it follows that
\begin{eqnarray}
[{\bf G}] {\bf U} |R\rangle &=& \lambda_R {\bf U} |R \rangle,
\end{eqnarray}
so that any eigenvector of ${\bf G}$ is of the form 
${\bf U} |R \rangle$, where all components of $|R\rangle$ are real.
If $|R\rangle$ has components $r1, r2, \dots r7$, then 
\begin{eqnarray}
{\bf U} |R \rangle &=& [r1, r2, r3, (r_4+ir_5)/\sqrt 2 , (r_4-ir_5)/\sqrt 2,
\nonumber \\ && \ \ (r_6+ir_7)/\sqrt 2 , (r_6-ir_7)/\sqrt 2 ] \ ,
\end{eqnarray}
which has the form asserted.

\section{Irreps for TMO25}

In this appendix we give the representation analysis for TbMn$_2$O$_5$
for wavevectors of the form $(\oh, 0, q)$, where $q$ has a nonspecial value.
The operators we consider are $E$, $m_x$, $m_y$ and $m_xm_y$, as defined
in Table \ref{PBAM}. Note that $m_y^2(x,y,z)=(x+1,y,z)$, so
that $m_y^2 =-1$ for this wavevector.  Thus, the above set of four
operators do not actually form a group.  Accordingly we consider
the double group which follows by introducing $-E$ defined by
$m_y^2=-E$, $(-E)^2=E$, and $(-E){\cal O}(-E)={\cal O}$.
Since addition has no meaning within a group
we do not discuss additive properties such as $(E)+(-E)=0$.  Then,
if we define $-{\cal O} \equiv (-E) {\cal O}$, we have the character
table given in Table \ref{DOUBLE}.
\begin{table}
\begin{tabular} {| c | c  c c c c|} \hline \hline
Irrep  &  $E$ & $\pm m_x$ & $\pm m_y$ & $\pm m_xm_y$ & $-E$ \\
\hline
$\Gamma_a$ & 1 & 1 & 1 & 1 & 1 \\ 
$\Gamma_b$ & 1 & -1 & 1 & -1 & 1 \\ 
$\Gamma_c$ & 1 & 1 & -1 & -1 & 1 \\ 
$\Gamma_d$ & 1 & -1 & -1 & 1 & 1 \\ \hline
$\Gamma_2$ & 2 & 0 & 0 & 0 & -2 \\ \hline
$G$ & $n$ & 0 & 0 & 0 & $-n$ \\ \hline \hline
\end{tabular}
\caption{\label{DOUBLE} Character table for the double group of the
wavevector.  In the first line we list the five classes of operators for
this group.  In the last line we indicate the characters for the group $G$
which is induced by the $n$-dimensional reducible representation in the
space of the $\alpha$ spin component of spins in a given Wyckoff orbit.}
\end{table}

\begin{table}
\vspace{0.2 in}
\begin{tabular}{||c||c|c|||| c || c|c ||}\hline\hline
Spin & $\sigmav_1 $ & $\sigmav_2 $ & Spin &$\sigmav_1$
& $\sigmav_2$ \\ \hline

${\bf S}({\bf q},1)$&
$\begin{array}{c} r_{1x} \\  r_{1y} \\ r_{1z}  \end{array}$ &
$\begin{array}{c} r_{2x} \\  r_{2y} \\ r_{2z} \end{array}$  &
${\bf S}({\bf q},7)$ & 
$\begin{array}{c} r_{6x} \\  r_{6y} \\  r_{6z}  \end{array}$ &
$\begin{array}{c} -r_{6x} \\  -r_{6y} \\  r_{6z} \end{array}$ \\ \hline

${\bf S}({\bf q},2)$&
$\begin{array}{c} r_{2x} \\  r_{2y} \\  -r_{2z}  \end{array}$ &
$\begin{array}{c} r_{1x} \\  r_{1y} \\  -r_{1z} \end{array}$ &
${\bf S}({\bf q},8)$ &
$\begin{array}{c} r_{6x} \\  -r_{6y}  \\  -r_{6z}  \end{array}$ &
$\begin{array}{c} r_{6x} \\  -r_{6y} \\  r_{6z} \end{array}$ \\ \hline

${\bf S}({\bf q},3)$ &
$\begin{array}{c} r_{1x} \\ -r_{1y} \\  -r_{1z}  \end{array}$ &
$\begin{array}{c} -r_{2x}\\  r_{2y} \\  r_{2z} \end{array}$  &
${\bf S}({\bf q},9)$&
$\begin{array}{c} r_{3x} \\  r_{3y} \\  r_{3z}  \end{array}$ &
$\begin{array}{c} r_{4x} \\  r_{4y} \\  r_{4z} \end{array}$ \\ \hline

${\bf S}({\bf q},4)$ &
$\begin{array}{c} r_{2x} \\  -r_{2y} \\ r_{2z}  \end{array}$ &
$\begin{array}{c} -r_{1x}\\  r_{1y} \\   -r_{1z} \end{array}$  &
${\bf S}({\bf q},10)$&
$\begin{array}{c} r_{4x} \\  r_{4y} \\ -r_{4z}  \end{array}$ &
$\begin{array}{c} r_{3x} \\  r_{3y} \\  -r_{3z} \end{array}$ \\ \hline

${\bf S}({\bf q},5)$&
$\begin{array}{c} r_{5x} \\  r_{5y} \\  r_{5z} \end{array}$ &
$\begin{array}{c} -r_{5x}\\  -r_{5y} \\  r_{5z} \end{array}$  &
${\bf S}({\bf q},11)$&
$\begin{array}{c} r_{3x} \\  -r_{3y} \\  -r_{3z}  \end{array}$ &
$\begin{array}{c} -r_{4x}\\  r_{4y} \\  r_{4z} \end{array}$ \\ \hline

${\bf S}({\bf q},6)$&
$\begin{array}{c} r_{5x} \\  -r_{5y} \\  -r_{5z}  \end{array}$ &
$\begin{array}{c} r_{5x} \\  -r_{5y} \\  r_{5z} \end{array}$  &
${\bf S}({\bf q},12)$&
$\begin{array}{c} r_{4x} \\  -r_{4y} \\  r_{4z}  \end{array}$ &
$\begin{array}{c} -r_{3x}\\  r_{3y} \\  -r_{3z} \end{array}$ \\ \hline
\end{tabular}

\caption{\label{TMOTAB} Spin functions (i. e. unit cell Fourier coefficients)
determined by standard representation analysis without invoking
symmetry operations that relate ${\bf q}$ and $-{\bf q}$. The second and
third columns give the functions
which transform according to the first and second column of the
two dimensional irrep. These coefficients are all complex parameters.}
\end{table}

The Mn$^{4+}$ Wyckoff orbits contain two atoms and all the other
orbits contain four atoms. In either case we may consider separately
an orbit and a single component, $x$, $y$, or $z$ of spin.  So the
corresponding spin functions form a basis set of $n$ vectors, where
$n=2$ for the single spin components of Mn$^{4+}$ and $n=4$, otherwise.
In each case, the operations involving $m_x$ and/or $m_y$ interchange
sites and therefore have zero diagonal elements. Their character,
which is their trace within 
this space of $n$ vectors is therefore zero.  On the other hand
$E$ and $-E$ give diagonal elements of $+1$ and $-1$, respectively.
So their character (or trace) is $\pm n$ and we have the last line
of the table for this reducible representation $G$.

In this character table we also list (in the last line) the characters
of these operations within the vector space of wavefunctions of a given
spin component over a Wyckoff orbit of $n$ sites. Comparing this last
line of the table to the character of the irreps we see that ${\bf G}$
contains only the irrep $\Gamma_2$ and it contains this irrep $n/2$
times.  This means that for the system of three spin
components over 12 sites, we have 36 complex components and these
function generate a reducible representation which contains
$\Gamma_2$ 18 times. If there were no other symmetries to
consider, this result would imply that to determine the structure
one would have to fix the 18 complex-valued parameters.  The two
dimensional representation can be realized by Eq. (\ref{116EQ}).
The basis vectors which transform as the first
and second columns, respectively of the two dimensional
representation are given in Table \ref{TMOTAB}.
One can check the entries of this table by verifying that the
effect of $m_x$ and $m_y$ on the vectors of this table are
in conformity with Eq. (\ref{116EQ}).

However, after taking account of
inversion symmetry we have only 18 real-valued structural parameters
of Table \ref{TMO25q} to determine.

\end{appendix}


\begin{thebibliography} {99}
\bibitem{FIEBIG}
M. Fiebig, J. Phys. D: Appl. Phys. {\bf 38}, R123 (2005).
\bibitem{TMO1} %1
T. Kimura, T. Goto, H. Shintani, K. Ishizka, T. Arima, and Y. Tokura, 
Nature {\bf 426}, 55 (2003).
\bibitem{TMO2} %2
M. Kenzelmann, A. B. Harris, S. Jonas, C. Broholm, J. Schafer, S. B. Kim,
C. L. Zhang, S.-W. Cheong, O. Vajk, and J. W. Lynn, Phys. Rev. Lett. {\bf 95}, 087206 (2005).
\bibitem{FERRO} %3
G. Lawes, A. B. Harris, T. Kimura, N. Rogado, R. J. Cava,
A. Aharony, O. Entin-Wohlman, T. Yildirim, M. Kenzelmann,
C. Broholm, and A. P. Ramirez, Phys. Rev. Lett. {\bf 95}, 087205 (2005)
\bibitem{JAP} %4
A. B. Harris, J. Appl. Phys. {\bf 99}, 08E303 (2006).
\bibitem{NVOPRB} %5
M. Kenzelmann, A. B. Harris, A. Aharony, O. Entin-Wohlman, T. Yildirim, 
Q. Huang,
S. Park, G. Lawes, C. Broholm, N. Rogado, R. J. Cava, K. H. Kim, G. Jorge, and
A. P. Ramirez, Phys. Rev. B {\bf 74}, 014429 (2006).
\bibitem{HANDBOOK} %6
A. B. Harris and G. Lawes, "Ferroelectricity in Incommensurate Magnets,"
in "The Handbook of Magnetism and Advanced Magnetic Materials," Ed.
H. Kronmuller and S. Parkin, Wiley 2006. cond-mat/0508617.
\bibitem{DMO} %7
T. Goto, T. Kimura, G. Lawes, A. P. Ramirez, and Y. Tokura, Phys. Rev.
Lett. {\bf 92}, 257201 (2004).
\bibitem{CFO} %8
T. Kimura, J. C. Lashley, and A. P. Ramirez, Phys. Rev. B {\bf 73},
220401(R) (2006).
\bibitem{TB25} %9
L. C. Chapon, G. R. Blake, M. J. Guttmann, S. Park, N. Hur, P. G. Radaelli,
and S.-W. Cheong, Phys. Rev. Lett. {\bf 93}, 177402 (2004).
\bibitem{Y25}  % 10: Yttrium Mn_2O_5
L. C. Chapon, P. G. Radaelli, G. R. Blake, S. Park, 
and S.-W. Cheong, Phys. Rev. Lett. {\bf 96}, 097601 (2006).
\bibitem{CURRENT} %11: spin-current
H. Katsura, N. Nagaosa, and A. V. Balatsky, Phys. Rev. Lett. {\bf 95}
057205 (2005).
\bibitem{MOST} %12
M. Mostovoy, Phys. Rev. Lett. {\bf 96}, 067601 (2006).
\bibitem{CHUPIS} %13
G. A. Smolenskii and I. E. Chupis, Sov. Phys. Usp. {\bf 25}, 475 (1982).
\bibitem{AXE} %14
B. Dorner, J. Axe, and G. Shirane, Phys. Rev. B {\bf 6}, 1950 (1972).
\bibitem{ROGER} %15
R. A. Cowley, Advances in Physics {\bf 29}, 1 (1980).
\bibitem{JS2} %16
J. Schweizer, C. R. Physique {\bf 6}, 375 (2005).
Erratum, to be published in C. R. Physique.
\bibitem{RAD} %17
P. G. Radaelli and L. C. Chapon, cond-mat 0609087.
\bibitem{JV} %18
J. Villain, J. Schweizer, and A. B. Harris, to be published.
\bibitem{xyzabc} %19 etc
For orthorhombic crystals we will interchangeably refer to axes as
either $x$, $y$, $z$, or ${\bf a}$, ${\bf b}$. ${\bf c}$.
\bibitem{NAG} %19
T. Nagamiya, in {\it Solid State Physics}, ed. F. Seitz and D. Turnbull
(Academic, New York, 1967), Vol. 20, p346.
\bibitem{IED} %20
I. E. Dzialoshinskii, Sov. Phys. JETP {\bf 5}, 1259 (1957).
\bibitem{LL} %21
L. D. Landau and E. M. Lifshitz, {\it Statistical Physics}
(Pergamon, London, 1958).
\bibitem{BS} %22
K. D. Bowers and J. Owen, Rep. Prg. Phys. {\bf 1}, 304 (1955).
%\bibitem{WIG} %23
%We interpret a transformation as a transformation on the coordinate system
%in the sense of E. P. Wigner, {\it Group Theory},
% Expanded and Improved Edition,
%(Academic, New York, 1959), p1, as contrasted to a transformation which
%replaces the old vector by a new vector (as is discussed by Wigner on p2.)
\bibitem{Bertaut} %24
E. F. Bertaut, Journal de Physique, Collloque C1, {\bf 32} 462 (1971). 
\bibitem{JS1} %25
J. Schweizer, J. Phys. {\bf IV}, 11 (2001) Pr9.
\bibitem{Rossat} %26
J. Rossat-Mignod, in {\it Methods of Experimental Physics}, Chap. 20:
{\it Magnetic Structures}, ed. K. Skold and D. L. Price, Vol. {\bf 23},
p69 (Academic Press, 1987).
\bibitem{HAHN} %27
A. J. C. Wilson, {\it International Tables for Crystallography} (Kluwer
Academic, Dordrecht, 1995) Vol. A.
\bibitem{xyz} %28
Coordinates of positions within the unit cell are given as fractions of the
appropriate lattice constants.
\bibitem{SAUER} %29
E. E. Sauerbrei, F. Faggiani, and C. Calvo, Acta Crystallogr. B {\bf 29}
2304 (1973).
\bibitem{SYMOP} %30
Often these symmetry operations are labeled $h_n$, referring
Refs. \onlinecite{HAHN} or \onlinecite{KOV}. Rather than use
these symbols with no mnemonic value, we 
introduce the notation $m_\alpha$ for a mirror 
(or glide) plane which changes the sign of the $\alpha$ coordinate
and $2_\alpha$ for a two-fold rotation (or screw) axis parallel to
the $\alpha$ direction.
\bibitem{KOV} %31
O. V. Kovalev, {\it Representations of Crystallographic Space Groups:
Irreducible Representations, Induced representations and Corepresentations}
(Gordon and Breahc, Amsterdam, 1993).
\bibitem{NVOPRL} %32
G.Lawes, M. Kenzelmann,  N. Rogado, K. H. Kim, G. Jorge, R. J. Cava,
A. Aharony, O. Entin-Wohlman, A. B. Harris, T. Yildirim, Q. Z. Huang, S. Park,
C. Broholm, and A. P. Ramirez, Phys. Rev. Lett. {\bf 93}, 247201 (2004).
\bibitem{RLU} %33
We use the convention that $\hat q$ denotes a wavevector
in reciprocal lattice units
(rlu) so that $q_x$ in real units is $2 \pi \hat q_x/a$.  When
$q$ is multiplied by a distance obviously one has to either
take $q$ in real units or take distances in inverse rlu's.
\bibitem{ABHJS} %34
A. B. Harris and J. Schweizer, Phys. Rev. B {\bf 74} ppp (2006).
\bibitem{CRITICAL} %35
The fourth order terms can also renormalize the quadratic terms,
especially if one is far from the critical point. But asymptotically
close to the transition this effect is negligible. 
\bibitem{RG} %36
H. Kawamura, Phys. Rev. B {\bf 38}, 4916 (1988).
\bibitem{AA} %37
A. D. Bruce and A. Aharony, Phys. Rev. B {\bf 11}, 478 (1975).
\bibitem{LAUT} %38
G. Lautenschlager, H. Weitzel, T. Vogt, R. Hock,A. B\'ohm, M. Bonnet,
and H. Fuess, Phys. Rev. B {\bf 48}, 6087 (1993).
\bibitem{MWO1} %39
K. Taniguchi, N. Abe, T. Takenobu, Y. Iwasa, and T. Arima,
Phys. Rev. Lett. {\bf 97}, 097203 (2006).
\bibitem{BLASCO} %41 Tb: x=0.9836(1), y=0.0810(1)
J. Blasco, C. Ritter, J. Garcia, J. M. de Teresa, J. P\'erez-Cacho,
and M. R. Ibarra, Phys. Rev. B {\bf 62}, 5609 (2000).
\bibitem{KAJI} %42
R. Kajimoto, H. Yoshizawa, H. Shintani, T. Kimura, and Y. Tokura,
Phys. Rev. B {\bf 70} 012401 (2004); {\bf 70}, 219904(E) (2004).
\bibitem{LOCK} %43
Symmetry constraints on lock-in is discussed by Y. Park, K. Cho, and
H.-G. Kim, J. Appl. Phys.  {\bf 83}, 4628 (1998).
\bibitem{MUNOZ1} % YMO
A. Munoz, J. A. Alonso, M. T. Casais, M. J. Martinez-Lope,
J. L. Martinez, and M. T. Fernandez-Diaz, J. Phys. Condens. Matter
{\bf 14}, 3285 (2002).
\bibitem{MUNOZ2} % HMO
A. Munoz, M. T. Casais, J. A. Alonso, M. J. Martinez-Lope,
J. L. Martinez, and M. T. Fernandez-Diaz, Inorg. Chem. {\bf 40},
1020 (2001).
\bibitem{BRINKS}
H. W. Brinks, J. Rodriguez-Carvajal, H. Fjellvag, A. Kjekshus, and
B. C. Hauback, Phys. Rev. B {\bf 63}, 094411 (2001).
\bibitem{BUIS1} %44  Mn(+3) and Mn(+4) coords
G. Buisson, Phys. Stat. Sol. {\bf 16}, 533 (1973).
\bibitem{BUIS2} %45  RE coords
G. Buisson, Phys. Stat. Sol. {\bf 17}, 191 (1973).
\bibitem{BLAKE} %Symmetry analysis for 25
G. R. Blake, L. C. Chapon, P. G. Radaelli, S. Park N. Hur, S.-W. Cheong, and
J. Rodriguez-Carvajal, Phys. Rev. B {\bf 71}, 214402 (2005).
\bibitem{VH} %47
V. Heine, {\it Group Theory in Quantum Mechanics}, (Pergamon, New York, 1960)
p284. For an example, see A. J. Berlinsky and C. F. Coll, III,
Phys. Rev. B {\bf 5}, 1587 (1972).
\bibitem{CFOJAP} %48
S. Mitsuda, H. Yoshizawa, N. Yamaguchi, and M. Mekata, J. Phys.
Soc. Jpn. {\bf 60}, 1885 (1991).
%\bibitem{SVIS03} %49
%L. E. Svistov, A. I. Smirnov, L. A. Prozorova, O. A. Petrenko,
%L. N. Damianets, and A. Ya. Shapiro, Phys. Rev. B {\bf 67}, 094434 (2003).
%\bibitem{SVIS06} %50
%L. E. Svistov, A. I. Smirnov, L. A. Prozorova, O. A. Petrenko,
%A. Micheler, N. B\'ottgen, A. Ya. Shapiro, and L. N. Damianets,
%Phys. Rev. B {\bf 67}, 094434 (2003).
\bibitem{PHASE} %51
When we ``fix" the phase $\phi$ of a variable, it means that
the variable is of the form $re^{i \phi}$, where $r$ is
real, but not necessarily positive.
%\bibitem{RFMO} %52
%RFMO
\bibitem{CFOSTR}
C. T. Prewitt, R. D. Shannon, and D. B. Rogers, Inorg. Chem.
{\bf 10}, 791 (1971).
%\bibitem{THESIS} %53
%G. Gasparovic, Ph. D. Thesis, Johns Hopklins University, 2004.
%\bibitem{RFMONAT} %54
%M. Kenzelmann, G. Lawes, A. B. Harris, G. Gasparovic, C. Broholm,
%A. P. Ramirez, G. A. Jorge, M. Jaime, S. Park, Q. Huang, A. Ya. Shapiro,
%and L. A. Damianets, to be published.
\bibitem{KAPLAN} %55
T. Kaplan, Phil. Mag., in press. 
\bibitem{SHG} %56
D. Frohlich, St Leute, V. V. Pavlov, and R. V. Pisarev,
Phys. Rev. Lett. {\bf 81}, 3239 (1998).
\bibitem{MWO2} %40
O. Heyer, N. Hollmann, I. Klassen, S. Jodlauk, L. Bohat\'y,
P. Becker, J. A. Mydosh, T. Lorenz, and D. Khomskii,
cond-mat/0608498.
\bibitem{GILL} %57
It is true that the coefficient $r_\gamma$ can depend on temperature
and it would be of the form $r_\gamma(T) \approx r_\gamma (T_c) + r_1 (T-T_c)$.
There is no reason to think that $r_\gamma(T_c)$ would be unusually small,
and we therefore have the desired result as long as we are not far from
the critical temperature, $T_c$.
\bibitem{25FERRO} %58
I. Kagomiya, S. Matsumoto, K. Kohn, Y. Fukuda, T. Shoubu, H. Kimura,
Y. Noda, and N. Ikeda, Ferroelectrics {\bf 286}, 167 (2003.
\bibitem{TY} %60
A. B, Harris, T. Yildirim,  A. Aharony, and O. Entin-Wohlman, Phys. Rev.
B {\bf 73} 184433 (2006).
\bibitem{SERG1}
I. A. Sergienko and E. Dagotto, Phys. Rev. B {\bf 73}, 094434 (2006).
\bibitem{SERG2}
I. A. Sergienko, C. Sen, and E. Dagotto, cond-mat/0608075.
\bibitem{KHOM}
D. I. Khomskii, J. Magn. Magn. Mat. (2004).
\bibitem{COMMENT} %59
M. Kenzelmann and A. B. Harris, to be published.
%\bibitem{TINKHAM}
%M. Tinkham, {\it Group Theory and Quantum Mechanics}, (McGraw-Hill,
%New York, 1964). 
\end{thebibliography}
\end{document}